\documentclass[amsmath,
               amssymb,
               aps,
               prb,
               reprint,
               superscriptaddress,
               floatfix]{revtex4-2}

\usepackage{color}
\usepackage{times}
\usepackage{graphicx} 
\usepackage{subfigure}
\usepackage{array,dcolumn} 
\usepackage{mathrsfs}
\usepackage{amsmath,amssymb,amsfonts}
\usepackage{bm} 
\usepackage{latexsym}
\usepackage{textcomp}
\usepackage{pifont}
\usepackage{gensymb}
\usepackage{epstopdf}
\usepackage{hyperref}
\newcommand {\del} {\partial}

\newcommand {\p} {\prime}
\newcommand {\s} {\sigma}
\newcommand {\beq} { \begin{equation} }
\newcommand {\eeq} { \end{equation} }
\newcommand {\bea} { \begin{eqnarray} }
\newcommand {\eea} { \end{eqnarray} }

\begin{document}
\author{Shinjan  Mandal}
 \affiliation{Center for Condensed Matter Theory, Department of Physics, Indian Institute of Science, Bangalore 560012, India}
 \author{Shrihari Soundararajan}
  \affiliation{Center for Condensed Matter Theory, Department of Physics, Indian Institute of Science, Bangalore 560012, India}
  \affiliation{Mechanical Engineering Department, University of California, Riverside, CA 92521, USA}
\author{Manish Jain}
 \affiliation{Center for Condensed Matter Theory, Department of Physics, Indian Institute of Science, Bangalore 560012, India}
\author{H. R. Krishnamurthy}
 \affiliation{Center for Condensed Matter Theory, Department of Physics, Indian Institute of Science, Bangalore 560012, India}
 \affiliation{International Centre for Theoretical Sciences, Tata Institute of Fundamental Research, Bengaluru 560089, India}

\title
{Possibilities for enhanced electron-phonon interactions and high-$T_c$ superconductivity in engineered bimetallic nano-structured superlattices}

\begin{abstract}
We explore theoretically the properties of engineered bimetallic nano-structured superlattices where an array of nano-clusters of a simple (single band) metal are embedded periodically inside another simple metal with a different work function. The exploration is done using a simplified tight-binding model with Coulomb interactions included, as well as density functional theory. Taking arrays of ``Ag'' clusters of fixed sizes and configurations (when unrelaxed) embedded periodically in an ``Au'' matrix as an example, we show that a significant enhancement of electron-phonon interactions ensues, implying possibilities for high-$T_c$ superconductivity. The enhancement stems from a strong coupling, via Coulomb interactions, between the dipolar charge distribution that forms at the Au-Ag interfaces and the breathing and other modes of vibration of the light Ag atoms caged inside the heavier Au matrix. The interface dipoles form because of the interplay between the mismatch of the local potential seen by the conduction electrons localised in Wannier orbitals at the Ag and Au sites (the Ag sites being slightly repulsive relative to the Au sites) and the (long-range) Coulomb repulsion between electrons occupying these Wannier orbitals. We also discuss the DC transport in such systems.
\end{abstract}

\maketitle

\section{Introduction}

Sometime ago, intriguing and astounding experimental findings suggestive of room temperature superconductivity were reported \cite {Au_Ag_unpub1, Au_Ag_unpub2, Saha_et_al-2022}  in systems of {\em silver} nano-particles dispersed in a {\em gold} matrix. These findings did not find wide acceptance in the community, partly due to the fact that the samples once prepared degraded with time, making them difficult to share, and the difficulties involved in controlled production of new samples meant that other groups were not able to reproduce these findings. However, encouraging new recent results  reporting improved control in preparing such systems \cite {Maji_et_al-2023}, and more interestingly, unprecedented enhancements of electron-phonon interactions in them as reflected in resistivity and point contact spectroscopy measurements  \cite {AG_el_ph_enh-2024} , suggest that a revival of interest in these systems is very much warranted. 

In this paper we explore the properties of such Ag-Au nano-structured systems theoretically, using a simplified semi-phenomenological tight-binding  model (TBM) {\em with Coulomb interactions included}, as well as with density functional theory (DFT),  as regards their electrons; and a simplified semi-phenomenological force constant model (FCM) for the phonons. For arrays of ``Ag'' nano-clusters of fixed sizes and configurations (when unrelaxed) embedded periodically in a ``Au'' matrix leading to a superlattice, we show that a significant enhancement of electron-phonon interactions can indeed arise.  The mechanism originates from the formation of dipolar charge redistribution at the Au-Ag interfaces, and their strong coupling, via Coulomb interactions, with the breathing and other modes of the lighter Ag atoms caged inside the heavier Au matrix. These ``interface dipoles" form because of the interplay between the mismatch of the local potential seen the by the conduction electrons localised in Wannier orbitals at the Ag and Au sites, (with the Ag sites being slightly repulsive relative to the Au sites as is suggested by the fact that the work function of silver is smaller than that of gold), and the (long-range) Coulomb interactions between the electrons occupying these Wannier orbitals. 

Electron-phonon interactions have been studied since the 1930's \cite{Ziman-60, Allen-2000, Giustino-2017}, and it is well known that in pure Ag or Au, the interactions are weak enough that Ag and Au do not show superconductivity down to the lowest temperatures measured. The dramatic enhancements that have been observed experimentally in systems with Ag clusters embedded in an Au matrix are therefore striking and mysterious.  We resolve this mystery in this paper, and show that the enhancement is almost inevitable in nano-structured lattices involving metals with differing work functions.

The Au-Ag nano-structured superlattice that we consider is constructed as follows. We use $\mathbf{R}$ to denote the (Bravais) lattice vectors of the superlattice. Each unit cell of this superlattice, i.e.,  the {\em supercell}, contains a basis of  $N_c$  Au/Ag atoms labelled by an index $j$, with a fraction $f_{Ag}$ of Ag atoms clustered at its centre surrounded by the atoms of Au which constitute the rest. We denote the position vectors of these basis atoms by $\mathbf{d}_j^{0}$ prior to {\em relaxation}, and  $\mathbf{d}_j$ after {\em relaxation} (see subsection (3) below). The lattice spacings of pure silver and gold are very close (both are FCC structures, with conventional unit cell sizes of 0.40860 and 0.40786 nm respectively\cite{Au-Ag-lat}). Hence, for simplicity, we choose  $\mathbf{d}_j^{0}$ such that the resulting lattice is the same for both pure gold and silver. We will refer to this as the  {\em pure} lattice, with lattice vectors denoted by $\mathbf{R}_0$ (~$\Leftrightarrow (\mathbf{d}^0_j + \mathbf{R}) $ ).  For all our calculations we use periodic boundary conditions with $\mathcal{N}$ supercells, i.e., $\mathcal{N} \times N_c$ total number of sites, and total volume $\mathcal{V}$ (or area $\mathcal{A}$ in 2-D) in our system Correspondingly, a grid of wave-vectors $\mathbf{k}$ with $\mathcal{N}$ uniformly distributed points within the small Brillouin zone (SBZ) appropriate to the superlattice labels the Bloch functions of the electrons in the system.

\begin{figure}
    \centering
    \includegraphics[width=0.48\textwidth]{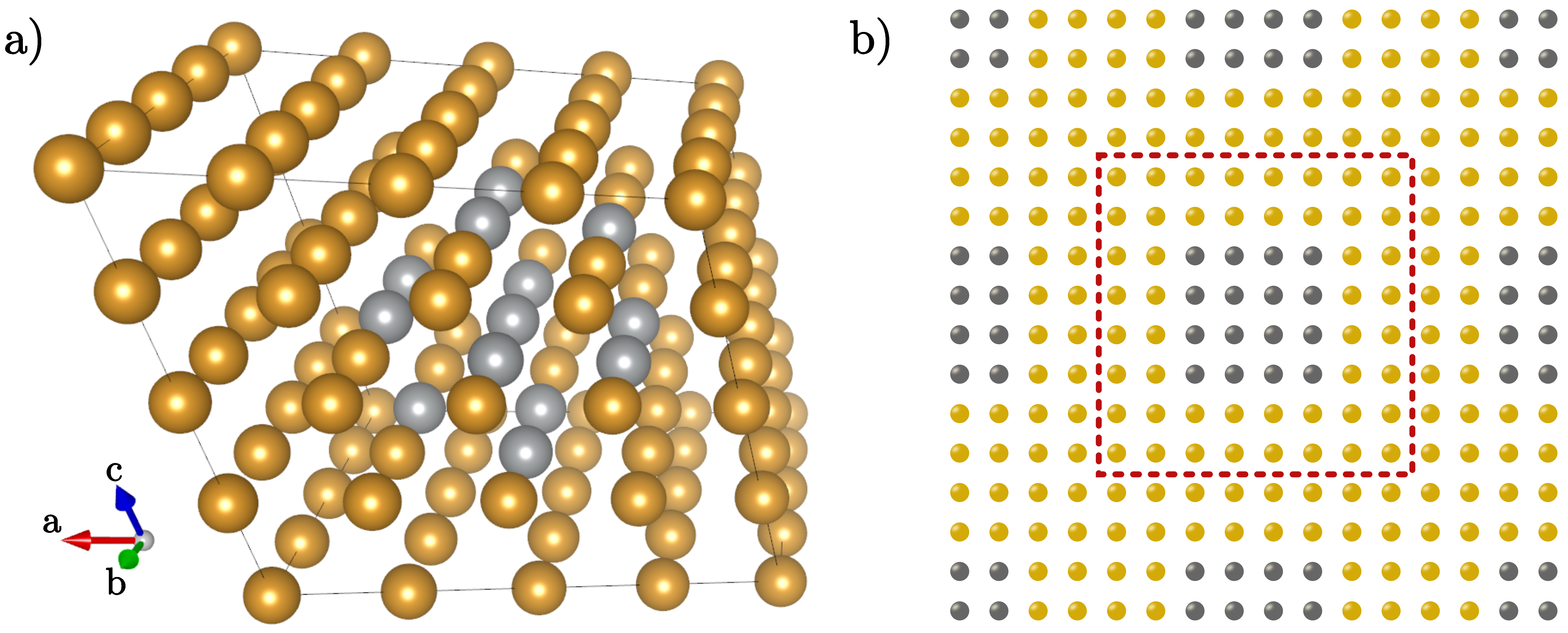}
    \caption{a) The 3-D FCC-SL structure with 13 Ag atoms embedded within 51 Au atoms used for our DFT calculations. b) The 2-D S-SL system used for our phenomenological modelling. The unit cell, marked with red dotted line, consists of a total of 64 atoms, with a $(4\times4)$ cell of silver embedded within an outer shell of $48$ Au atoms.}
    \label{fig:fig1}
\end{figure}

For the specific calculations we report in this paper we use two instances of the above, both with  $N_c = 64$, as illustrated in Fig. 1. The first (Fig. 1a) is an idealised version of the 3-D Ag-Au nano-structures discussed in refs. \cite{Maji_et_al-2023, AG_el_ph_enh-2024}:  a periodic array of 13-site close packed Ag clusters embedded (prior to relaxation) in a Au matrix of 51 pure FCC sites with a conventional cubic unit cell of side $a_0 = 0.41 nm$, forming 64 site 3-D supercells, corresponding to a first neighbour distance $d_1 = a_0 / \sqrt{2} = 0.29 nm$, and  $f_{Ag} = 0.203$. We will refer to this as the FCC superlattice (FCC-SL). The second (Fig. 1b) is a square super-lattice (S-SL), where a periodic square array of $(4 \times 4)$ clusters of ``Ag" sites with lattice spacing $d_1$ are surrounded by 48 ``Au" sites forming a superlattice of 64-site 2-D supercells, corresponding to $f_{Ag} = 0.250$. 

Fig. 2 shows the DFT band structure and the Density of states (DOS) of the  3-D Au-Ag FCC-SL specified above. More details regarding the calculations are provided in Appendix A.

\begin{figure}[b]
    \centering
    \includegraphics [width=0.48\textwidth] {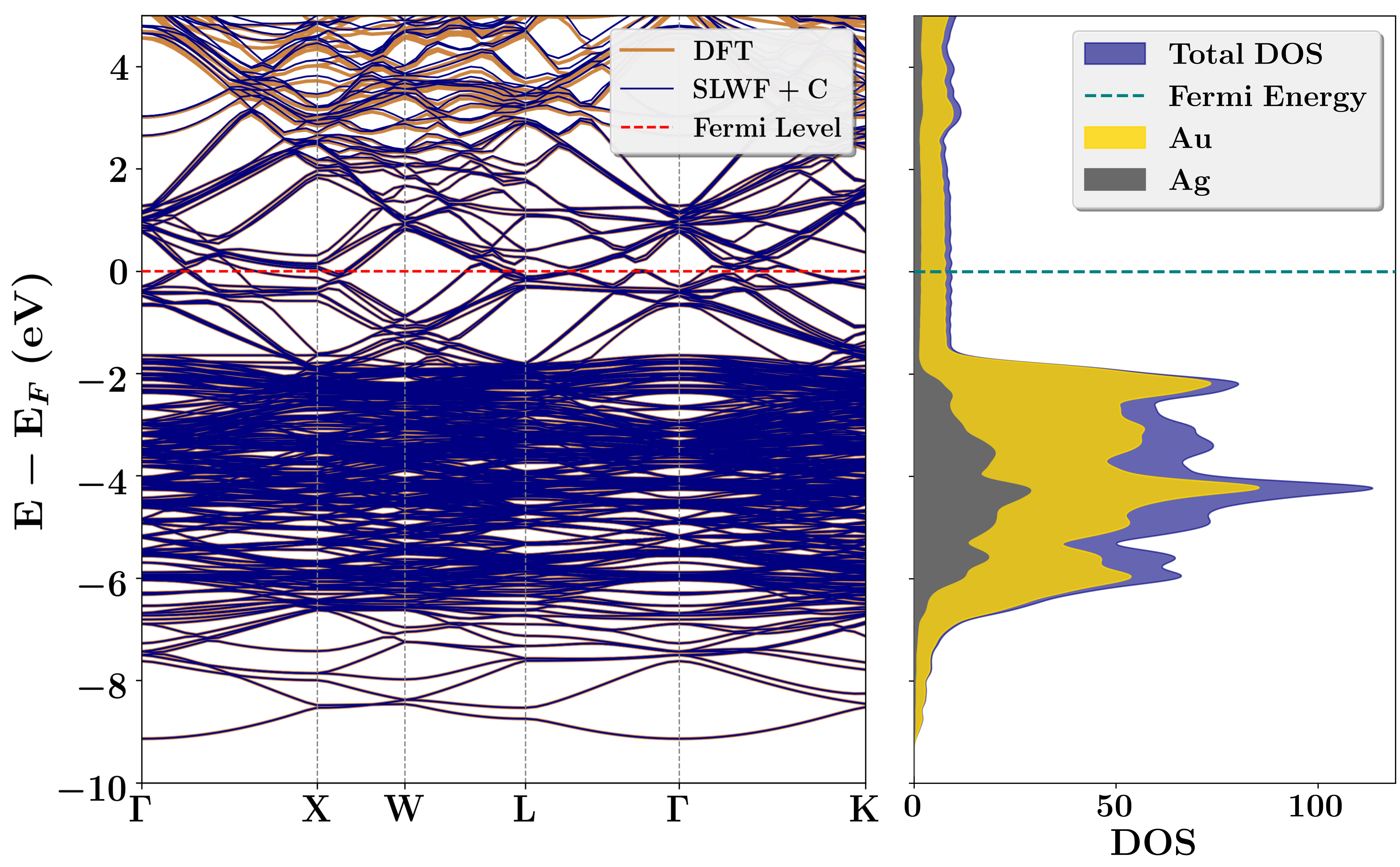}
    \caption { The DFT electronic band structure for the 64 atom Au-Ag FCC-SL nanostructured superlattice obtained via Wannier interpolation compared to the bands obtained from DFT calculations. The Wannier projected DOS are also shown. More details can be found in Appendix A. }
    \label {fig:fig2}
\end{figure}

The bands near the Fermi level in Fig. 2 can be thought of as resulting from band folding and hybridisation of the corresponding bands of Ag and Au, or equivalently, as arising from the hybridisation of of the Wannier orbitals of these bands localised at the Ag and Au sites. By identifying the latter (see Appendix A for details) we have calculated the net charge localised on the Ag and Au sites, and find that there is a deficit of electrons (corresponding to a positive charge) of 0.022 on the central Ag site and 0.043 on the other 12 Ag sites of the Ag cluster; and an excess of electrons  (corresponding to negative charges) on the Au sites, ranging from 0.023 for the Au sites at the interface with the Ag cluster, to 0.013 - 0.003  for the Au sites farther away from the interface. This demonstrates, for a realistic setting, that the nano-structuring indeed leads to the formation of the interface dipoles mentioned earlier. Additionally, as is also visually noticeable in the band-structure in Fig. 2, the hybridisation mentioned above leads to ``flattening'' of the conduction band states near the Fermi level, as reflected in a reduction of the velocities of the electrons in these states, leading to a reduction in the {\em transport} DOS of the system by as much as 1/3 \cite{vel-reduct}. 

A detailed ab-initio exploration of the physical properties even of the above superlattice, let alone superlattices with larger Ag clusters and Ag-Au supercells, involving Density Functional Perturbation Theory calculations of the phonon band structure and the electron-phonon matrix elements, poses significant computational challenges. We believe that  the essential physics involved, as well as  a reasonable account of the conduction electrons, phonons and the electron phonon interactions in such systems, can be obtained within the framework of a semi-phenomenological TBM for the electrons, and a simple FCM for the phonons, with the Coulomb interactions being treated within the Hartree approximation. Using such a framework, below we demonstrate that the formation of the interface dipoles is an {\em inevitable consequence of the nano-structuring with dissimilar metals}, and discuss their physical consequences. 

In particular, we show that the following additional features emerge,  due to  the dependence of the tight-binding hopping amplitudes and the long range Coulomb interactions on the positions of the atoms: (1). There is a lattice ``relaxation'' leading to the displacement of atoms from $\mathbf{d}^0_j$ to new equilibrium positions $\mathbf{d}_j$ within the supercell; (2). Additional contributions to the force constants of the system arise, especially a stiffening of the bonds at the Au-Ag interface, and a consequent increase or decrease in the frequencies of the phonon modes involving the interface atoms;  (3) Finally, and most importantly,  {\em new contributions to the electron-phonon (el-ph) interactions arise, leading to their substantial enhancement} due to the long range Coulomb interactions and the interface dipoles. We discuss how these ingredients  can together lead to the observed unprecedented enhancements of the resistivity \cite {AG_el_ph_enh-2024} mentioned above.

In the rest of the main part of the paper (sections II - V) below, we specify in greater detail our modelling, outline the ideas and approximations involved in our calculations, specify the equations governing the most important of our results,  discuss some typical results from our numerical calculations, and some conclusions that follow from these. The details of our theoretical treatment, and additional results, are set out in the appendices.  

\section{Electrons in the Nano-structured Superlattice}

The simplified semi-phenomenological TBM that we use for our calculations is based on the assumption that the low energy electronic excitations involve only the s-like conduction band states of silver and gold, and can be expressed entirely in terms of the set of operators $\boldsymbol{c}_{j \mathbf{R} \s}$ and  $\boldsymbol{c}_{j \mathbf{R} \s}^\dag$ which respectively destroy and create electrons with spin $\s$ in the {\em Wannier} orbitals \cite{Ashcroft_Mermin-76} of this band localised at the Au/Ag sites at $\mathbf{R} + \mathbf{d}_j$. Additional simplifying assumptions we make are:
\begin{enumerate}
	\item
	The amplitude $\mathfrak{t}_{j \mathbf{R} ; j^\prime \mathbf{R}^\prime}$ for electrons to hop between sites $j \mathbf{R}$ and $j^\prime \mathbf{R}^\prime$ is assumed to be the same whether the sites are occupied by silver or gold atoms,  and a function only of the distance between the two sites, $d_{j \mathbf{R} ; j^\prime \mathbf{R}^\prime} = |\mathbf{d}_{j \mathbf{R} ; j^\prime \mathbf{R}^\prime}|$,  where  $\mathbf{d}_{j \mathbf{R} ; j^\prime \mathbf{R}^\prime} \equiv  (\mathbf{R} + \mathbf{d}_j) - (\mathbf{R}^\prime + \mathbf{d}_{j^\prime})$ is the inter-site vector to $j \mathbf{R}$ from $j^\prime \mathbf{R}^\prime$ . We model it using an exponentially decaying Slater-Koster form \cite{Slater_Koster-54, Harrison-89}, $\mathfrak{t}_{j \mathbf{R} ; j^\prime \mathbf{R}^\prime} = \mathfrak{t} (d_{j \mathbf{R} ; j^\prime \mathbf{R}^\prime}) =  \mathfrak{t}_0  \exp [- (d_{j \mathbf{R} ; j^\prime \mathbf{R}^\prime} - d_1) / \xi_0 ]$, where $d_1$  is the first neighbour distance, with the first neighbour hopping $\mathfrak{t}_0$ and $\xi_0/d_1$ being treated as adjustable parameters. We set $\mathfrak{t}_{j \mathbf{R} ; j^\prime \mathbf{R}^\prime}$ to zero if $(j \mathbf{R}) = (j^\prime \mathbf{R}^\prime)$, or if $d_{j \mathbf{R} ; j^\prime \mathbf{R}^\prime} > d_c $ where $d_c$ is an appropriately chosen cutoff distance beyond which the hopping amplitude are assumed to be negligible. 
	\item
	The {\em dominant} parts of the Coulomb interaction terms involving the conduction electrons are parameterised by an onsite repulsion energy $U_0$ and a long-range inter-site repulsion energy between electrons at sites  $j \mathbf{R}$ and $j^\prime \mathbf{R}^\prime$ which scales inversely as the distance between these sites, again without distinction as to  whether the sites involved host silver or gold atoms. We write this as $V(d_{j \mathbf{R} ; j^\prime \mathbf{R}^\prime}) = V_0 / d_{j \mathbf{R} ; j^\prime \mathbf{R}^\prime}$.
	\item
	The {\em{only}} term in the electronic Hamiltonian of our simplified model that distinguishes between silver and gold sites is a {\em local} site energy $\epsilon_j$ which does not depend on $\mathbf{R}$ because of the assumed superlattice periodicity. We assume that it is zero for all gold sites and the same positive number $\epsilon_0$ for all the silver sites, taking into account the fact that the work-function for silver is smaller than that of the gold by 0.6 - 0.9 eV , depending on the orientation of the surface \cite{workfunction}. As we expect our theory to be applicable for other choices of the metals making up the nano-structured superlattice, we have carried out our calculations, and report results, for a range of values of $\epsilon_0$ and $V_0/d_1$.
\end{enumerate}

Using these ingredients, we write the electronic Hamiltonian as the sum of one-electron terms (accounting for hopping processes, local site energies and the chemical potential), Coulomb energy terms, and the core energy (in which we include the electronic energy from all the completely filled core bands other than the s-like conduction band, as well as the bulk of the nuclear repulsion energy).

\begin{equation}
    \boldsymbol{H}_{el}  =  \boldsymbol{H}_{1-el} + \boldsymbol{H}_{coul} + E_{core} \label{model_ham_eqn}
\end{equation}
\begin{multline}
        \boldsymbol{H}_{1-el}  =    - \sum^{\quad \p}_{j, \mathbf{R}; j^\prime, \mathbf{R}^\prime} \mathfrak{t}_{j \mathbf{R}; j^\prime \mathbf{R}^\prime} \sum_{\s}  \boldsymbol{c}_{j \mathbf{R} \s}^\dag \boldsymbol{c}_{j^\prime \mathbf{R}^\prime \s}  \\ +  \, \sum_{j, \mathbf{R}} \, (\epsilon_j -\mu) \, (\boldsymbol{n}_{j \mathbf{R}} - 1)  \label{H_hop_eqn}
\end{multline}
\begin{multline}
       \boldsymbol{H}_{ce}  =    U_0  \sum_{j, \mathbf{R}}   (\boldsymbol{n}_{j \mathbf{R} \uparrow} -\frac{1}{2})    (\boldsymbol{n}_{j \mathbf{R} \downarrow} - \frac{1}{2}) \; \\ + \, \frac {V_0}{2} \sum^{\quad \p}_{j, \mathbf{R}; j^\prime, \mathbf{R}^\prime}  \frac{1} {d_{j \mathbf{R}; j^\prime \mathbf{R}^\prime}}  \, (\boldsymbol{n}_{j \mathbf{R} } -1)  \,   (\boldsymbol{n}_{j^\prime \mathbf{R}^\prime} - 1)
       \label{H_ce_eqn}  
\end{multline}

Here $\boldsymbol{n}_{j \mathbf{R} \s} \equiv  \boldsymbol{c}_{j \mathbf{R} \s}^\dag \boldsymbol{c}_{j \mathbf{R} \s}$ and $\boldsymbol{n}_{j \mathbf{R} } \equiv \sum_\s \boldsymbol{n}_{j \mathbf{R} \s}$ are respectively the spin resolved and the total number operators at the site $j \mathbf{R}$ , and the superscript $\prime$ over the summations  above denotes the constraint $j \mathbf{R} \neq j^\prime \mathbf{R}^\prime$, i.e., the two sites involved {\em have to be different}. Note that we have included a part of the Coulomb interaction involving the nuclear charges (corresponding to singly charged ions) in the onsite and long range Coulomb energy terms, such that at the Hartree mean-field level they do not contribute to the total energy when the Wannier orbital occupancy is exactly one (half for each spin orientation) at every site, as is the case for pure gold or silver. The chemical potential $\mu$ is to be determined by the ``half filling condition'' that the total number of electrons in the system is exactly $\mathcal{N} \times N_c$ , i.e., one per site. For the specific calculations we report in this paper, mostly for the S-SL, we use 
$\mathfrak{t}_0$ as the basic unit of energy (by arbitrarily setting it equal to 1 eV), with the other energy parameters ($\epsilon_0, U_0, V_0/d_1$, and $\mu$) also specified in eV (although they are really being specified in units of $\mathfrak{t}_0$, and need to be scaled appropriately if a different $\mathfrak{t}$ is used); and explore the properties of the model system for various choices of their values as well as the value of $\xi_0$.

As is well known, the above Hamiltonian is difficult to solve except when $U_0$ and $V_0/d_1$ are small enough compared to the bandwidth of the conduction band, which we denote as $\mathfrak{W}$  ($ \mathfrak{W}_0 \sim 8 \mathfrak{t}_0$ for the pure square lattice and  $\sim 16  \mathfrak{t}_0$  for the pure FCC lattice) to be treatable by perturbation theory. For the purposes of this paper, where our goal is to establish generic features of nano-structured bi-metallic superlattices,  we confine ourselves to treating the interaction terms within the {\em restricted} Hartree approximation, where we do not allow for the superlattice-translation or spin symmetry to be broken. The effective electronic Hamiltonian within the Hartree approximation, $\boldsymbol{H}_{el;H}$  is essentially the same (apart from an energy shift - see Appendix B for details) as  $\boldsymbol{H}_{1-el}$ except for the replacement $\epsilon_j \rightarrow \epsilon_j + \Phi_j$.  Here $\Phi_j$ is the Hartree potential at the site $j\mathbf{R}$ given by   
\beq
\Phi_j \equiv  \frac{1}{2} \, U_0 \, \delta n_j + V_0 \sum^{\quad \p}_{j^\prime, \mathbf{R}^\prime} \, \frac{\delta n_{j^\prime}} {d_{j \mathbf{R}; j^\prime \mathbf{R}^\prime}} \, ,
\label{Phi_j_eqn}
\eeq
$\delta n_j \equiv \langle (\boldsymbol{n}_{j \mathbf{R} } -1) \rangle$ being the excess occupancy (of the Wannier orbital) at the site $j \mathbf{R}$, and  $\Phi_j$ and $\delta n_j$ are to be determined self consistently. Note that by assumption (of unbroken superlattice-translation symmetry), $\delta n_j$ and $\Phi_j$ are independent of $\mathbf{R}$.  $~\boldsymbol{H}_{el;H}$ is straightforwardly diagonalised into the simple form $\boldsymbol{H}_{el;H} = \sum_{\mathbf{k}, m, \sigma} \varepsilon_{\mathbf{k} m}  \tilde{\boldsymbol{c}}^\dag_{\mathbf{k} m \sigma} \tilde{\boldsymbol{c}}_{\mathbf{k} m \sigma}$ by the transformations

\begin{align}
    \boldsymbol{c}_{j \mathbf{R} \sigma} &= \frac {1} {\sqrt{\mathcal{N}}} \sum_{\mathbf{k}} e^{i \mathbf{k} \cdot (\mathbf{R} + \mathbf{d}_j)} \tilde{\boldsymbol{c}}_{j \mathbf{k} \sigma} \nonumber \\
    &=  \frac {1} {\sqrt{\mathcal{N}}} \sum_{\mathbf{k}, m} e^{i \mathbf{k} \cdot (\mathbf{R} + \mathbf{d}_j)} \varphi_{j; \mathbf{k} m} \tilde{\boldsymbol{c}}_{\mathbf{k} m \sigma} \, 
\label{c_ctilde_transform_eqn}
\end{align}
Here $\varphi_{j; \mathbf{k} m} \,$ and $\varepsilon_{\mathbf{k} m} \, $ with both being labeled by the wave-vector $~\mathbf{k}$  and the integer band index $m$ which distinguishes the $N_c$ different bands, are respectively the eigenvectors and eigenvalues of the Fourier transform of the $N_c \times N_c$ (first quantised) one-electron lattice Hamiltonian matrix; i.e., 
\bea
\sum_{j^\prime} \tilde{h}_{j j^\prime} (\mathbf{k}) \varphi_{j^\prime; \mathbf{k} m} &= &\varepsilon_{\mathbf{k} m} \varphi_{j; \mathbf{k} m}  \nonumber \\ 
\tilde{h}_{j j^\prime} (\mathbf{k}) \equiv(\epsilon_j -\mu + \Phi_j) \delta_{j j^\prime} &-& \sum^{\quad \p}_{\mathbf{R}^\prime}  \, \left[ e^{ - i \mathbf{k} \cdot \mathbf{d}_{j \mathbf{R}; j^\prime \mathbf{R}^\prime} } \, \mathfrak{t}_{j \mathbf{R}; j^\prime \mathbf{R}^\prime} \right] \, \notag \\ & &
\label{k_space_ham_eqn}
\eea
Because of the periodic boundary conditions $\mathbf{k}$ takes $\mathcal{N}$ discrete values inside the superlattice Brillouin zone (SBZ) whose volume in reciprocal space is $1/N_c$ of the volume of the  unit-cell Brilloun zone (UBZ) of the pure lattice,  
If the system temperature is $T$, the excess occupancies $\{\delta n_j \}$ are determined in terms of the eigenvectors $\varphi_{j; \mathbf{k} m}$, and the thermal occupancies of the band levels $\varepsilon_{\mathbf{k} m}$ (which are already {\em relative} to the chemical potential) governed by the Fermi function  $n^-_F$ as
\bea
\delta n_j = \left( \frac {2} {\mathcal{N}} \sum_{\mathbf{k}, m} n^-_F(\varepsilon_{\mathbf{k} m}) |\varphi_{j; \mathbf{k} m}|^2 \right) - 1 ; \nonumber \\ \; n^-_F(\epsilon) \equiv \frac {1} {e^{ \beta \epsilon} + 1} , \, \ \ \ \ \  \beta \equiv 1/(k_B T) \, .
\label{nj_eqn}
\eea
Eq.s~(\ref{Phi_j_eqn}), (\ref{k_space_ham_eqn}) and (\ref{nj_eqn}), together with the half-filling constraint which requires $ \sum_j \delta n_j = 0 $, constitute the self-consistency conditions \cite{Hartree-wo-relax} that determine the $N_c$ different excess occupancies $\{\delta n_j \}$, the corresponding Hartree potentials $\{ \Phi_j \}$, and the chemical potential $\mu$. 

In case of pure gold or silver, it is easy to see that both $\delta n_j$ and $\Phi_j$ are self-consistently zero for all $j$, i.e., each site has exactly one electron, and the Hartree potential is zero. Then the band dispersions $\{ \varepsilon_{\mathbf{k} m} \}$ and the eigenvectors $\varphi_{j; \mathbf{k} m}$ for the superlattice are merely those obtainable by {\em folding} the single pure Au (or Ag) s-band with wave-vectors $\mathbf{k}_0$ in the pure lattice UBZ down to the much smaller SBZs of the FCC-SL or S-SL. For a more detailed discussion, see Appendix B.

However, when clusters of Ag atoms are embedded in the Au matrix, the results, obtained from a numerical solution of the self consistent equations (see Appendix B for details), are dramatically different, as shown in Fig.s 3 and 4, both obtained for the S-SL, and  corresponding respectively to the case {\em without} and {\em with} interactions. 

Fig. 3a shows that the repulsive local potential $\epsilon_0$ on the Ag sites causes the folded bands mentioned above to hybridise and split, leading to band-gaps where the folded bands for the pure lattice would have intersected. More importantly, it causes the buildup of a substantial  deficit of electrons ($\delta n_j < 0$) on the Ag sites (~ 42 - 52 \% for the case shown, corresponding to $\epsilon_0 = 1.0$ eV, $\xi_0 = 1.00 d_1$, and $d_c = 8 d_1$), and a consequent excess of electrons ($~\delta n_j > 0$) of order 10 to 17 \%  on the Au sites to maintain overall charge neutrality. See Fig 3b. 

\begin{figure}
    \centering
    \includegraphics[width=0.48\textwidth]{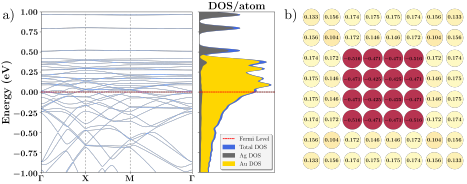}
    \caption { a) The electronic band structure in the TBM model on the S-SL without any Coulomb interactions for $\epsilon_0 = 1.0$ eV, $\xi_0 = 1.0 d_1$, and $d_c = 8 d_1$. b) The electron occupancy excess $(\delta n_j)$ at each atomic site is depicted using colour-coding. In addition, the value of $\delta n_j$ at each site is given within the corresponding circle.}
    \label{fig:fig3}
\end{figure}

When interactions are taken into account, the onsite Coulomb repulsion $U_0$ suppresses this charge disproportionation, so as to reduce the  the corresponding Hartree energy cost  given by $(U_0 / 4) \sum_{j, \mathbf{R}} (\delta n_j)^2$. However, the long range Coulomb repulsion controlled by $V_0$,  with its Hartree energy being given by $(V_0 / 2) \sum^{\quad \prime}_{j, \mathbf{R}; j^\prime, \mathbf{R}^\prime}  (\delta n_j \, \delta n_{j^\prime}) /  d_{j \mathbf{R}; j^\prime \mathbf{R}^\prime}$, causes the charge disproportionation  to be concentrated on the sites at the Ag-Au interface, with the interfacial Ag sites acquiring the most negative  $\delta n_j$ values and interfacial Au sites the largest positive $\delta n_j$ values, as this is clearly energetically favourable. Consequently, the overall suppression caused by $V_0$ is smaller.  Thus we have here the formation of what we will refer to henceforth as {\em interface dipoles} (even though the opposing charges right at the interface are not strictly equal in magnitude). Fig. 4, showing the results from our self-consistent calculations for the S-SL for $ U_0 = 2.0$ eV and $V_0/d_1 = 1.0$ eV \cite{V0-values},  and the same values as above for the other parameters, illustrates the above features.  

\begin{figure}
    \centering
    \includegraphics[width=0.48\textwidth]{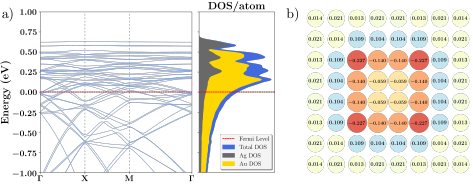}
    \caption{a) The electronic band structure in the S-SL model with Coulomb interactions included, for $U_0=2.0, V_0/d_1 = 1.0, \epsilon_0 = 1.0$ (all in eV), $\xi_0 = 1.0 d_1$ and $d_c = 8 d_1$. b) The electron occupancy excess  $(\delta n_j)$ at each atomic site for the above set of parameters, depicted using colour-coding. In addition, the value of $\delta n_j$ at each site is given within the corresponding circle.}
    \label{fig:fig4}
\end{figure}

A good overall measure of the charge disproportionation is $\delta n_{Au}$, the {\em average} electron occupancy excess per atom on the Au sites. (The corresponding quantity on the Ag sites is $\delta n_{Ag} = - [(1-f_{Ag})/f_{Ag}] \delta n_{Au}$.)  Fig. 5 shows the variation of  $\delta n_{Au}$ with $\epsilon_0$ for the noninteracting case as well as for  $V_0/d_1$ equal to 1.0 eV, 2.0 eV, 3.0 eV and 4.0 eV \cite{V0-values}, with the other parameters held fixed at the values mentioned earlier, in the context of Fig.4. In the noninteracting case, $\delta n_{Au}$ rises rapidly with $\epsilon_0$ at first, but the increase slows down for larger  $\epsilon_0$.  Increasing $V_0/d_1$  leads to increased suppression of $\delta n_{Au}$, as remarked earlier, and its increase with 
$\epsilon_0$ becomes nearly linear in the interacting case for the values of the parameters we have explored, with a slope that decreases with increasing $V_0/d_1$.  

\begin{figure}[b]
    \centering
    \includegraphics[width=0.48\textwidth]{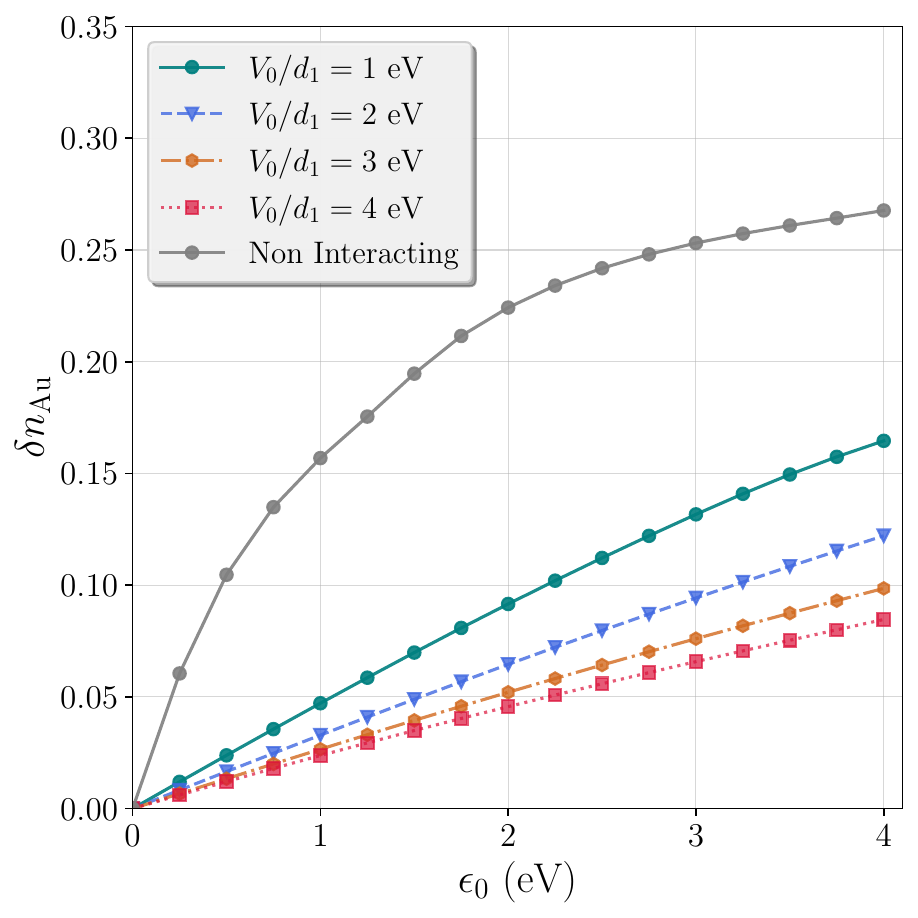}
    \caption{The variation of the average excess electron occupancy per site at the Au sites, $ \delta n_{j} $, as a function of $\epsilon_0$ for the noninteracting case as well as for several values of $V_0/d_1$. The other parameters are held fixed at the values mentioned  in the context of Fig.4.}
    \label{fig:fig5}
\end{figure}

Another feature of the (Hartree renormalised) band structure of the nano-structured superlattices, visually noticeable in Fig.4, is the slight ``flattening'' of the bands near the Fermi energy, which corresponds to a reduction in the velocities of the electrons in the (Hartree renormalised) band states near the chemical potential (Fermi level at zero temperature), where the $\alpha^{th}$ component of the velocity for the Bloch-state labelled by ($\mathbf{k} m$) is given by  $v^\alpha_{\mathbf{k} m} \equiv  \hbar ^{-1}  \del \varepsilon_{\mathbf{k}  m} /  \del  k^\alpha$. We have calculated three quantitative averaged measures of this flattening:  the one-electron density of states (DOS)  
$\mathcal{D}(0)$, the {\em transport DOS}, $\mathcal{D}^{(tr)}(0)$, (both per supercell per spin,)  and the mean squared velocity which is the ratio of the above two, all at the chemical potential, given (in the thermodynamic, i.e., large $\mathcal{N}$, limit) by  
\bea
\mathcal{D}(0) &\equiv&  \frac{1}{\mathcal{N}}  \sum_{\mathbf{k}, m} \delta (\varepsilon_{\mathbf{k} m} ) \, ;  \; \nonumber \\ \mathcal{D}^{(tr)}(0) &\equiv&  \frac{1}{\mathcal{N} D}  \sum_{\mathbf{k}, m}  (\mathbf{v}_{\mathbf{k}, m} \cdot \mathbf{v}_{\mathbf{k}, m}) \delta (\varepsilon_{\mathbf{k} m} ) \, ; \; \nonumber \\
\langle v^2 \rangle &\equiv& \frac{\mathcal{D}^{(tr)}(0)}{\mathcal{D}(0))} 
\label{dos_tr_dos_eqn}
\eea
In practical calculations, involving discrete $\mathbf{k}$-grids, these correspond essentially to counting the total number of states per unit cell in a small energy window  around the Fermi energy in case of $\mathcal{D}(0)$, and weighting this counting with the  {\em squared velocity} in case of  $\mathcal{D}^{(tr)}(0)$. See Appendix B for details of the calculations. 

 \begin{figure}
    \centering
    \includegraphics[width=0.48\textwidth]{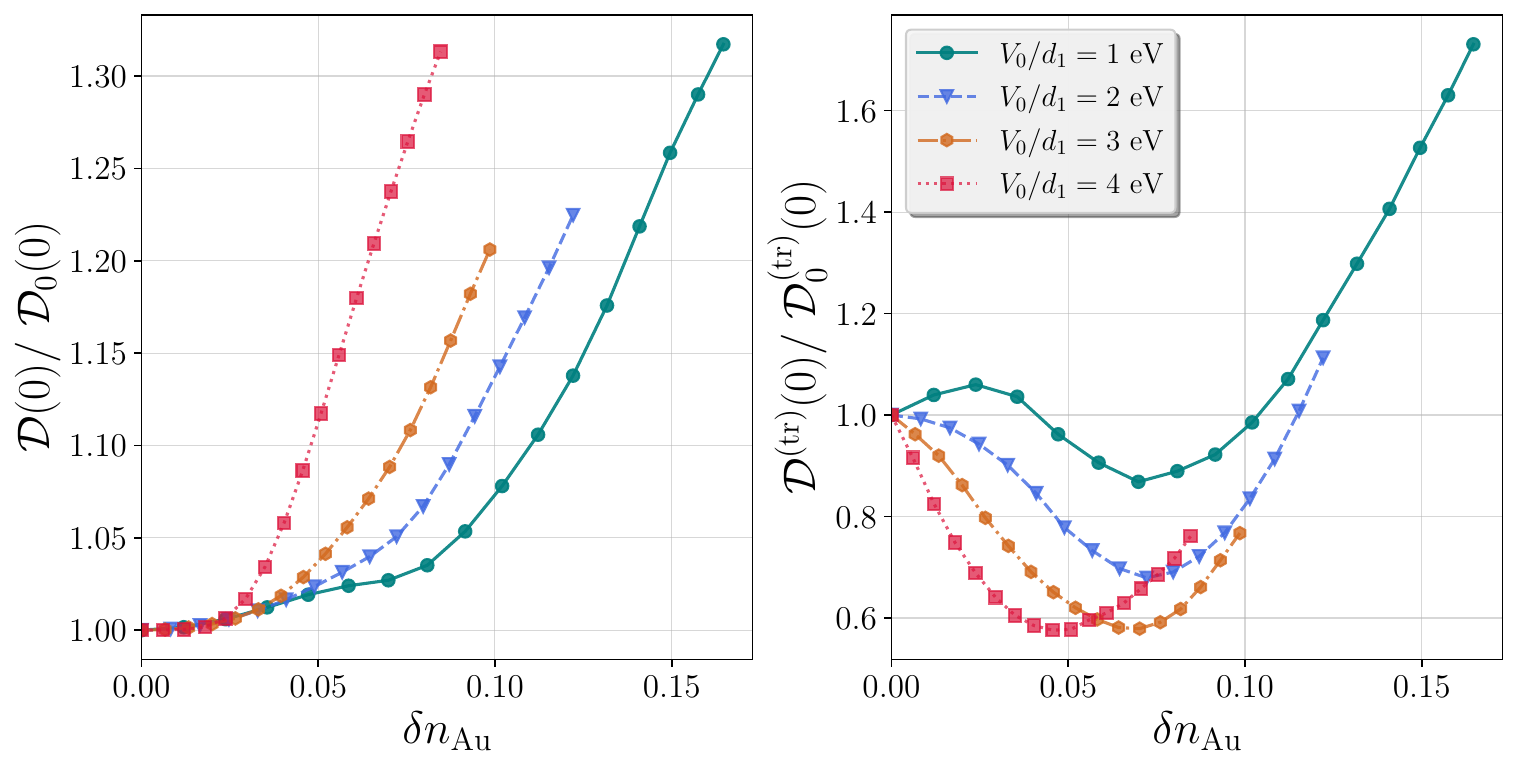}
    \caption{Plots of the ratios of the electronic DOS $ \mathcal{D}(0)$, (left panel,) and the transport DOS, $\mathcal{D}^{(tr)}(0)$, (right panel,) for the S-SL system to those for the pure square lattice, as functions of  $\delta n_{Au}$ for several values of $V_0/d_1$. (The other parameters are held fixed at the values mentioned in the context of Fig.4). }
    \label{fig:fig6}
\end{figure}

Fig. 6 shows our results for $\mathcal{D}(0) / \mathcal{D}_0 (0)$ (left panel) and 
$\mathcal{D}^{(tr)}(0) / \mathcal{D}_0^{(tr)}(0)$ (right panel), where the subscript 0 refers to the pure lattice (corresponding to $\epsilon_0 = 0$), as functions of $\delta n_{Au}$ for the same set of values of $V_0/d_1$ and the other parameters as in the case of Fig. 5. The DOS increases with  $\delta n_{Au}$ as it is proportional (in the thermodynamic limit, and for $T = 0$) to sums of integrals of $|\mathbf{v}_{\mathbf{k}, m}|^{-1}$ over the Fermi surfaces of all the partially filled bands\cite{Ashcroft_Mermin-76}. The suppression of 
$\mathcal{D}^{(tr)}(0)$ with  larger $V_0/d_1$ suggests that the suppression of the velocities in this case is large enough to compensate for the increase in $\mathcal{D}(0)$, as is confirmed by the plots of $\langle v^2 \rangle / \langle v^2 \rangle_0$ shown in Fig. 11 in Appendix B, but this is not so for smaller values  $V_0/d_1$ and larger $\delta n_{Au}$.  The suppression  of $\mathcal{D}^{(tr)}(0)$ is somewhat smaller overall for the 2D S-SL case compared to its suppression ($\sim 1/3$) obtained from the DFT-LDA calculations for  the  3D FCC-SL mentioned earlier.  
 
More results for the parameter values we have explored which exhibit the features arising from the nano-structuring mentioned above can be found in Appendix B.  

\section{Phonons in Nano-structured Superlattices}

$E_{core}$, the core energy term in Eq. (\ref{model_ham_eqn}),  is a c-number as far as the conduction electrons are concerned; but via its dependence on the positions of all the atoms, it is the source of the primary component of the potential energy $\boldsymbol{V}$ for the atomic displacements away from equilibrium. The leading, harmonic, terms in $\boldsymbol{V}$ for displacing the atoms from $\{ \mathbf{R}+\mathbf{d}_j \}$ to $\{ \mathbf{R}+\mathbf{d}_j + \mathbf{u}_{j \mathbf{R}} \}$ can be written in two equivalent forms \cite{Ashcroft_Mermin-76}
\bea
\boldsymbol{V}_{ph} \; &=& \; \frac{1}{4}  \sum_{j \mathbf{R} \alpha ; j^\prime \mathbf{R}^\prime \alpha^\prime}  K^{\alpha \alpha^\prime}_{j j^\prime } (\mathbf{R}-\mathbf{R}^\prime)  \; u^{\alpha}_{j \mathbf{R}; j^\prime \mathbf{R}^\prime}  \; u^{\alpha^\prime}_{j \mathbf{R}; j^\prime \mathbf{R}^\prime} \; \nonumber \\ &=&  \; \frac{1}{2} \sum_{j \mathbf{R} \alpha ; j^\prime \mathbf{R}^\prime \alpha^\prime}  \kappa^{\alpha \alpha^\prime}_{j j^\prime } (\mathbf{R}-\mathbf{R}^\prime)  \; u^{\alpha}_{j \mathbf{R}}  \; u^{\alpha^\prime}_{j^\prime \mathbf{R}^\prime},
\label{E_core_expn_eqn}
\eea
where $\mathbf{u}_{j \mathbf{R}; j^\prime \mathbf{R}^\prime} \equiv \mathbf{u}_{j \mathbf{R}} - \mathbf{u}_{j^\prime \mathbf{R}^\prime} $ is the relative displacement  between the pair of sites involved,  the superscripts $\alpha$, $\alpha^\prime$ denote the appropriate Cartesian components of the vectors or tensors involved. The ``spring constant tensors''  $K$ and the ``force constant tensors'' $\kappa$ are related via the equation 
\begin{multline}
    \kappa^{\alpha \alpha^\prime}_{j j^\prime } (\mathbf{R} - \mathbf{R}^\prime) =   - \;  K^{\alpha \alpha^\prime}_{j j^\prime} (\mathbf{R}-\mathbf{R}^\prime) \\
 + \delta_{j j^\prime} \; \delta_{\mathbf{R},\mathbf{R}^\prime} \;  \sum_{\mathbf{R}^{\prime \prime}, j^{\prime \prime}}  K^{\alpha \alpha^\prime}_{j j^{\prime \prime}}  (\mathbf{R} -\mathbf{R}^{\prime \prime} )
\end{multline}
As is well known \cite{Ashcroft_Mermin-76}, the phonon Hamiltonian for a $D$-dimensional superlattice can be completely decoupled into  $D \mathcal{N} N_c$ {\em independent} harmonic oscillators, corresponding to the  phonon modes characterised by   $\mathcal{N}$ wave vectors $\{\mathbf{q}\}$, an integer mode number label $s$ which takes values from 1 to $D N_c$, frequencies  $\{ \omega_{\mathbf{q} s} \} $, and polarisation vector components $\{ \eta^{\alpha}_{j; \mathbf{q} s} \}$. Here $\{ (\omega_{\mathbf{q} s})^2 \} $ and $\{ \eta^{\alpha}_{j; \mathbf{q} s} \}$ are the eigenvalues and eigenvectors respectively of the $D N_c \times D N_c$ {\em dynamical matrix} $\mathcal{D}^{\alpha \alpha^\prime}_{j j^\prime } (\mathbf{q})$ given by
\bea
\mathcal{D}^{\alpha \alpha^\prime}_{j j^\prime } (\mathbf{q})  &=&  \frac{1} { \sqrt{M_j M_{j^\prime}} } \sum_{\mathbf{R}^\prime}  \left[    e^{ i \mathbf{q} \cdot \mathbf{d}_{j \mathbf{R}; j^\prime \mathbf{R}^\prime} } \;  \kappa^{\alpha \alpha^\prime}_{j j^\prime} (\mathbf{R}-\mathbf{R}^\prime) \right] \, \nonumber \\ &\equiv& \, \frac{1} { \sqrt{M_j M_{j^\prime}} } \, \tilde{\kappa}^{\alpha \alpha^\prime}_{j j^\prime} (\mathbf{q}) \, .
\label{kappa_to_dyn_matrix_eqn}
\eea
with $M_j$ being the mass of the atom at the site $(j \mathbf{R})$ \cite{no-R-dep}. 

\begin{figure}
    \centering
    \includegraphics[width=0.48\textwidth]{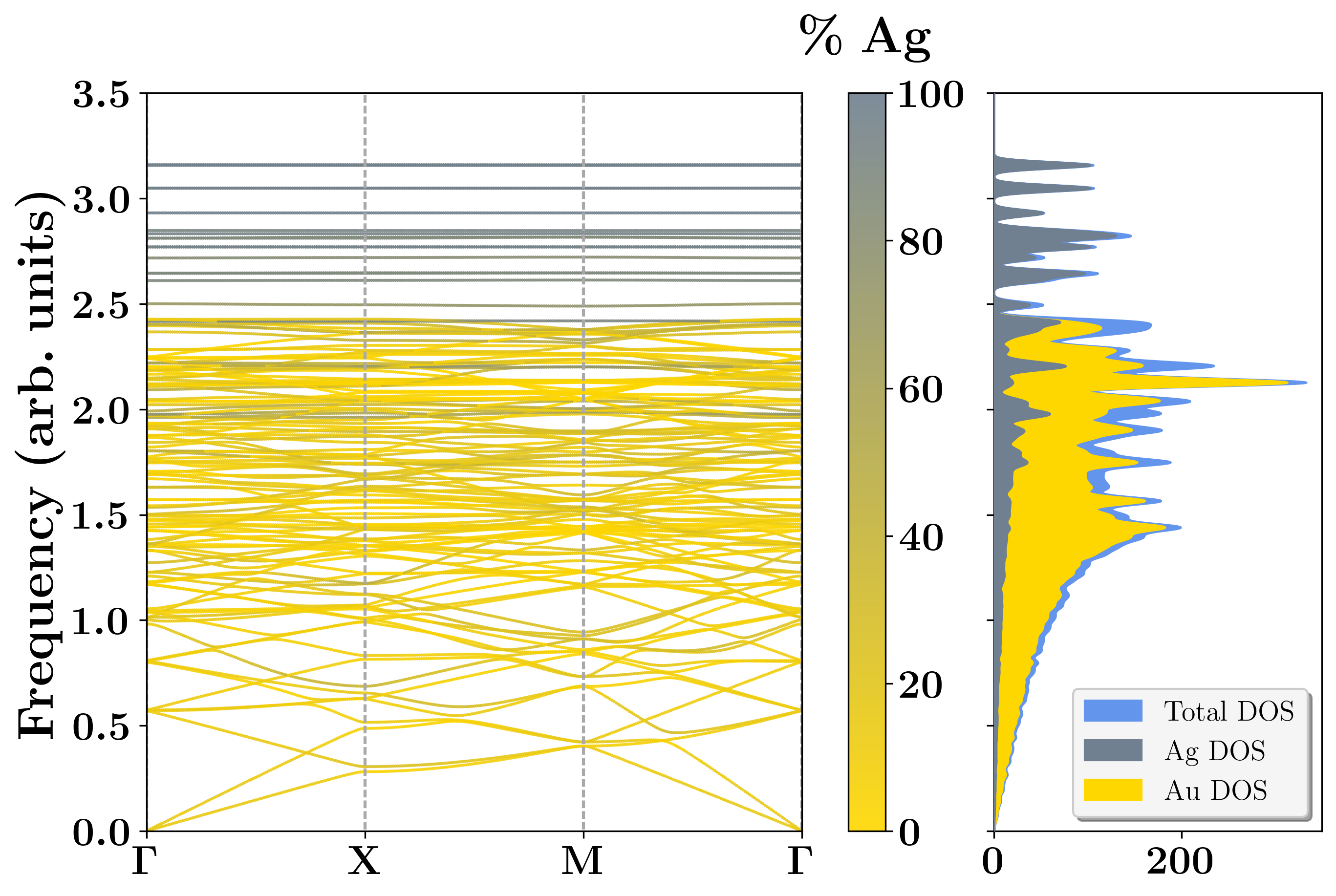}
    \caption{The phonon bands for the S-SL FCM obtained by assuming that only the first and second neighbour bond stretching spring constants, $K_1^{(s)}$ and $K_2^{(s)}$, are non zero, and in units where $K_1^{(s)}$ and $K_2^{(s)}$  have been set equal to 1.0 and 0.5 respectively, and $M^{(Au)}$, the mass of the Au atom, has also been set equal to 1.0. The values of the phonon frequencies in real units can be obtained by multiplication with the appropriate value of $\sqrt{(K_1^{(s)}/M^{(Au)})}$. The DOS projected onto the Au and the Ag atoms are also shown. Note the dramatic appearance of the ``flat'' bands of Ag phonons.}
    \label{fig:fig7}
\end{figure}

In phenomenological force-constant modelling (FCM) of lattice dynamics it is common practice to invoke the additional assumption that the contribution to the harmonic energy cost for displacements involving the pair of atoms at sites  $j \mathbf{R}$ and $j^\prime \mathbf{R}^\prime$  is {\em axially symmetric}  with respect to the inter-site vector $\mathbf{d}_{j \mathbf{R}; j^\prime \mathbf{R}^\prime} \equiv (\mathbf{R}^\prime + \mathbf{d}_{j^\prime}) - (\mathbf{R} + \mathbf{d}_j)$, involving only  two spring constants $K^{(s)}_{j j^\prime} (\mathbf{R}-\mathbf{R}^\prime)$ and $K^{(b)}_{j j^\prime} (\mathbf{R}-\mathbf{R}^\prime)$ corresponding respectively to bond-stretching and bond-bending\cite{axial-symmetry}; hence, 
\begin{multline}
K^{\alpha \alpha^\prime}_{j j^\prime } (\mathbf{R}-\mathbf{R}^\prime)  =  2 K^{(s)}_{j j^\prime}(\mathbf{R}-\mathbf{R}^\prime) \frac{ d^{\alpha}_{j \mathbf{R}; j^\prime \mathbf{R}^\prime} d^{\alpha^\prime} _{j \mathbf{R}; j^\prime \mathbf{R}^\prime}} {|\mathbf{d}_{j \mathbf{R}; j^\prime \mathbf{R}^\prime}|^2}  \\ + 2 K^{(b)}_{j j^\prime}(\mathbf{R}-\mathbf{R}^\prime) \left[   \delta_{\alpha, \alpha^\prime} -  \frac{ d^{\alpha}_{j \mathbf{R}; j^\prime \mathbf{R}^\prime} d^{ \alpha^\prime} _{j \mathbf{R}; j^\prime \mathbf{R}^\prime}} {|\mathbf{d}_{j \mathbf{R}; j^\prime \mathbf{R}^\prime}|^2} \right] ,
\label{stretch_bend_force_const_eqn}
\end{multline}
from which, proceeding as discussed above, one can determine all the $D N_c$ phonon frequencies and polarisation vectors as functions of $\mathbf{q}$.

In case of pure silver or gold, as has been shown \cite{Drexel-72, Lynn_et_al-73}, a reasonably good account of the observed phonon spectra (determined from neutron scattering data) is obtainable by assuming that the spring constants $K^{(s)}$ and $K^{(b)}$  are nonzero only for just the first neighbour atom pairs [Root Mean Squared (RMS) error $ \sim 4.1 \%$],  or slightly better, just the first, second and third neighbour atom pairs (RMS error $ \sim 3.3 \%$).  Furthermore, both are roughly the same for Ag and Au (to within $ \sim 60 \%$), with a substantial part of their frequency differences coming  from the difference in their atomic masses,  and  $K^{(s)} \gg K^{(b)}$. Hence, for illustrative purposes, for the S-SL we assume that only the first and second neighbour bond-stretching spring constants are non zero, and choose spring-constant units such that $K^{(s)}_1 = 1$ and $K^{(s)}_2 = 0.5$ irrespective of whether Au-Au or Ag-Ag or Au-Ag atom-pairs are involved,  and mass units such that $M^{(Au)} = 1$ and  $M^{(Ag)} = 0.54$.  Depending upon how one chooses $K^{(s)}_1$, the phonon frequencies in ``real units'' can be obtained by using the multiplicative ``conversion factor'' given by the appropriate value of $\sqrt{(K_1^{(s)}/M^{(Au)})}$\cite{real-units}.    

As in the electronic case, the phonon bands for pure Ag or pure Au treated as a superlattice are folded versions of the acoustic branches one would have obtained  for the pure lattices with 1 atom per unit cell - see Appendix C, in particular Fig. 12, for details. Furthermore, for the simplified FCM we are using, with the same force constants for both Au and Ag, the bands for pure Ag are the same as those of pure Au except for being scaled up by a factor of $\sqrt{(M^{(Au)}/M^{(Ag)})} = 1.36$. However, in case of the nano-structured Au-Ag superlattice, the results are again {\em dramatically different},  and are shown in Fig. 7 for the S-SL. The highest frequency modes almost entirely involve vibrations of the Ag atoms, as is clear from the projected DOS shown in Fig.7. Furthermore, their flatness (with respect to $\mathbf{q}$) suggests that the corresponding modes are almost completely localised within the respective Ag clusters, with the lighter Ag atoms rattling inside the ``cages''  formed by the the heavier Au atoms. More details are provided Appendix C.

\section{Electron-Phonon interaction effects in the nano-structured Superlattice}

The dependence of the hopping amplitude in Eq. (\ref{H_hop_eqn}) and the long range Coulomb energy in Eq. (\ref{H_ce_eqn}) on the positions of the atoms leads to three characteristic consequences {\em due to the nanostructuring}. The details are provided in Appendix D. Here we summarise our main findings.

The first is that there will be a relaxation of the positions of the basis of atoms within each supercell, leading to their displacements from $\mathbf{d}^0_j$ to new equilibrium positions $\mathbf{d}_j$. In the limit when the displacements due to the relaxation are small enough that {\em anharmonic forces} arising from $E_{core}$ can be ignored, the equation for the relaxation is given by 
\begin{multline}
    \sum_{j^\prime, \alpha^\prime}  \tilde{\kappa}^{\alpha \alpha^\prime}_{j j^\prime} (\mathbf{0}) \,  (d^{\alpha^\prime}_{j^\prime} - d^{0 \alpha^\prime}_{j^\prime}) \, = \,  \sum^{\quad \p}_{j^\prime,  \mathbf{R}^\prime} \bigg[  V_0 \frac{\delta n_j  \, \delta n_{j^\prime}} {(d_{j \mathbf{R}; j^\prime \mathbf{R}^\prime})^2} \\ - \mathfrak{t} (d_{j \mathbf{R}; j^\prime \mathbf{R}^\prime}) \frac{( C_{j \mathbf{R}; j^\prime \mathbf{R}^\prime} + C_{j^\prime \mathbf{R}^\prime ; j \mathbf{R}}) } {\xi_0}  \bigg] \,  \frac{d^{\alpha} _{j \mathbf{R}; j^\prime \mathbf{R}^\prime} } {d_{j \mathbf{R}; j^\prime \mathbf{R}^\prime}}  \, ,
\label{lat_relax_eqn}
\end{multline}
where the matrix $\tilde{\kappa}$ was defined in Eq. (\ref{kappa_to_dyn_matrix_eqn}), and  $C_{j \mathbf{R}; j^\prime \mathbf{R}^\prime} \equiv  \sum_\s \langle \boldsymbol{c}_{j \mathbf{R} \s}^\dag \boldsymbol{c}_{j^\prime \mathbf{R}^\prime \s} \rangle$. Based on energetics, we expect the Au-Ag bonds at the  interface to contract, and Ag-Ag and Au-Au bonds to expand.

Second, as a consequence of the basis relaxation mentioned above, additional contributions to the spring constants of the system will arise, especially near the Au-Ag interface, leading either to their stiffening or softening. There are four different contributions: two of these, which come from the expectation values of the changes in the Hamiltonian to second order in the lattice displacements, are of the axially symmetric form; the other two, involving changes in $C_{j \mathbf{R}; j^\prime \mathbf{R}^\prime}$ and $\delta n_j$ to first order in the lattice displacements, are not. Based on the fact that for most materials, thermal expansion is accompanied by a softening of phonons (most noticeable for optical modes), we expect that the contraction of Au-Ag bonds at the interface will cause the corresponding spring constants  to  increase, while the expansion of the Au-Au and Ag-Ag bonds will cause the corresponding spring constants to decrease. Consequently, we expect the frequencies of the nearly localised modes involving the Ag atoms at the interface to increase.

The detailed derivation of both the consequences mentioned above can be found in Appendix D.  We note that if the basis relaxation is substantial, the final structure results from a higher level of self-consistency involving (the Hartree self consistent) electronic structure, the basis relaxation, and if required, the full force-fields generated by $E_{core}$ that include anharmonic effects. Its calculation would thus be immensely more difficult than the already challenging calculations we report  in this paper, which are for the unrelaxed S-SL. Hence we postpone the study of the above two consequences to future work.

The third consequence, which is of direct relevance to the new experimental findings reported in ref. \cite {AG_el_ph_enh-2024}, and is the main focus of this paper, is that {\em entirely new contributions} to the electron-phonon ({\em el-ph}) interactions can arise due to the nano-structuring involving dissimilar metals, leading to their substantial enhancement. Consider the {\em el-ph} interaction Hamiltonian corresponding to the scattering of a conduction electron from the state $ |\mathbf{k}, m \rangle$ to the state  $ |(\mathbf{k}+\mathbf{q})  m^\prime \rangle$ by absorbing a phonon with labels $ (\mathbf{q}, s )$ (or emitting a phonon $ (- \mathbf{q}, s )$ written in the standard (second-quantised) form: 
\begin{multline}
    \boldsymbol{H}_{el-ph} = \frac {1} {\sqrt{\mathcal{N}}} \sum_{\mathbf{k}, m, m^\prime, \s} \sum_{\mathbf{q}, s}  \mathfrak{g}_{m^\prime m; s} (\mathbf{k}, \mathbf{q}) \, \times 
    \\ \tilde{\boldsymbol{c}}^\dag_{(\mathbf{k}+\mathbf{q})  m^\prime \s} \tilde{\boldsymbol{c}}_{\mathbf{k} m \s} \,(\boldsymbol{a}_{\mathbf{q} s} + \boldsymbol{a}^\dag_{-\mathbf{q} s}) ,
\label{el_ph_ham_std_form_eqn}
\end{multline}
We can show (see Appendix D  for details)  that the {\em el-ph} coupling matrix element $\mathfrak{g}_{m^\prime m; s} (\mathbf{k}, \mathbf{q})$ is the sum of two distinct contributions, arising respectively from the hopping and Coulomb energy terms in the Hamiltonian, and are given by 
\begin{widetext}
\beq
\mathfrak{g}^{(hop)}_{m^\prime m; s} (\mathbf{k}, \mathbf{q}) =  \sum^{\quad \p}_{j, j^\prime, \mathbf{R}^\prime }  \frac { \mathfrak{t} (d_{j \mathbf{0}; j^\prime \mathbf{R}^\prime}) \,  } {\xi_0 } \, \left( \varphi^*_{j^\prime; (\mathbf{k}+\mathbf{q}) m^\prime} \varphi_{j; \mathbf{k} m} e^{i (\mathbf{k} + \mathbf{q}) \cdot \mathbf{d}_{j \mathbf{0}; j^\prime \mathbf{R}^\prime}} \right) \, \Upsilon_{j  j^\prime;  \mathbf{q} s} ( \mathbf{R}^\prime) \, ,
\label{el_ph_coupling_hop_eqn}
\eeq
and 
\beq
\mathfrak{g}^{(ce)}_{m^\prime m; s} (\mathbf{k}, \mathbf{q}) = -  \sum^{\quad \p}_{j, j^\prime, \mathbf{R}^\prime }  \frac { V_0 \delta n_{j^\prime} \,  } { d^2_{j \mathbf{0}; j^\prime \mathbf{R}^\prime} } \, \left( \varphi^*_{j; (\mathbf{k}+\mathbf{q}) m^\prime} \varphi_{j; \mathbf{k} m} \right) \, \Upsilon_{j  j^\prime;  \mathbf{q} s} (\mathbf{R}^\prime);
\label{el_ph_coupling_ce_eqn}
\eeq
where
\beq
\Upsilon_{j  j^\prime;  \mathbf{q} s} (\mathbf{R}^\prime)  \equiv  \sqrt{ \frac{\hbar}{2 \omega_{\mathbf{q} s}} } \sum_\alpha \frac {d^\alpha_{j \mathbf{0}; j^\prime \mathbf{R}^\prime}} {d_{j \mathbf{0}; j^\prime \mathbf{R}^\prime}} \left[ \frac { \eta^\alpha_{j; \mathbf{q} s}  } { \sqrt{M_j} }  -  \frac { \eta^\alpha_{j^\prime; \mathbf{q} s}   } { \sqrt{M_{j^\prime}} } e^{- i \mathbf{q} \cdot \mathbf{d}_{j \mathbf{0}; j^\prime \mathbf{R}^\prime}} \right] .
\label{Upsilon_defn_eqn}
\eeq
\end{widetext}
We note that the second contribution $\mathfrak{g}^{(ce)}$ above arises only if $\delta n_{j}$ is nonzero, i.e., due to the charge redistribution and the consequent formation of interface dipoles as discussed earlier. It would be absent in pure Au or Ag. 

We have calculated these matrix elements using our results for the S-SL  discussed above, and in this paper we restrict ourselves to reporting two measures involving different levels of averaging them. One is the Migdal-Eliashberg   ``$\alpha^2 F \,$'' spectral function, given within the ``double delta  approximation'' \cite{McMillan-68, Allen_Dynes-75}  by
\begin{multline}
\alpha^2F (\omega) \equiv  \frac {1} {(\mathcal{N})^2 \mathcal{D}(0)}  \sum_{\mathbf{k}, m, m^\prime } \sum_{\mathbf{q}, s} \lvert\mathfrak{g}_{m^\prime m; s} (\mathbf{k}, \mathbf{q})\rvert^2  \; \times \\ \delta (\varepsilon_{\mathbf{k} m} ) \; \delta (\varepsilon_{(\mathbf{k}+\mathbf{q}) \, m^\prime} ) \; \delta (\hbar \omega - \hbar \omega_{\mathbf{q} s} ) .
\label{alpha_squared_F_eqn}
\end{multline}
Here $ \mathcal{D}(0) $  is the one-electron DOS  at the chemical potential discussed earlier (cf., Eq. \ref {dos_tr_dos_eqn} ).  From this, the dimensionless effective {\em el-ph} coupling strength $\lambda^{(\mathcal{S})}$ and the effective log-averaged phonon frequency $\omega_{log}$,  both relevant for superconductivity (see Appendix D for details),  are calculated as
\bea
 \lambda^{(\mathcal{S})} &\equiv& 2 \int_0^\infty \, d\omega \; \frac {\alpha^2F (\omega)} {\omega} \, ; \; \nonumber  \\ \omega_{log} &\equiv& \exp { \left( \frac {2} {\lambda^{(S)}} \int_0^\infty \, d\omega \; \frac {\alpha^2F (\omega)} {\omega} \; \log {(\omega)} \right) };
\label{lambda_omega_log_eqn}
\eea

The {\em el-ph} coupling matrix elements also determine the transport properties of the system, in particular, the resistivity reported in ref. \cite{AG_el_ph_enh-2024}. The detailed expressions connecting the two within a Boltzman-transport theory within a relaxation time approximation (or the equivalent approximation starting from the Kubo formula for the conductivity) are given in Appendix D. As discussed there, with some additional approximations for averages around the Fermi level, and for temperatures of the order of the Debye temperature or larger, one can write
\beq
\rho(T) \approxeq \frac{1}{e^2} \left( \frac {m_{eff} } {n_{eff}} \right) \left(\frac {2 \pi \, k_B T } {\hbar} \right) \lambda^{(tr)}
\label{rho_eqn}
\eeq
Here $n_{eff}$ is the total electron occupancy per unit volume in the {\em partially filled bands} of the system (over the entire SBZ), and  $(m_{eff})^{-1}$ is the average of the diagonal components of the inverse effective mass tensor over these occupied states. Equivalently, $(n_{eff} / m_{eff})$ is also the transport DOS at the Fermi level [with corrections of order $(k_B T / \mathfrak{W}_0)$ at finite temperatures - see Appendix D for details, in particular Eq. (\ref{mean_sq_vel_to_neff_by_meff_eqn-SI}) ], results for which were discussed earlier, in the context of Fig. 6.  $\lambda^{(tr)}$ is another dimensionless {\em el-ph} coupling constant, given by Eq. \ref{lambda_tr_eqn-SI} in Appendix D,  whose calculation is somewhat more involved than that of $\lambda^{(\mathcal{S})}$, but we expect the two to be of the same order. For the purposes of this paper, and the discussion below, we make the simplifying approximation that  $\lambda^{(tr)} \approx \lambda^{(\mathcal{S})}$. Within the framework of our model calculations, the resistivity as well as its slope with respect to $T$ at high temperatures for the nano-structured superlattice is then expected to be enhanced compared to their values for pure gold by the factor 
\bea
 r_{enh} &\equiv&  \frac {\rho(T)}{\rho_0(T)}  \sim \left (  \frac {\lambda^{(\mathcal{S})} }  { \lambda^{(\mathcal{S})}_0} \right ) \times \left( \frac {n_{eff; 0}} {m_{eff; 0}} \frac {m_{eff}} {n_{eff}} \right) \nonumber \\ &=& \left (  \frac {\lambda^{(\mathcal{S})} }  { \lambda^{(\mathcal{S})}_0} \right ) \times \left( \frac {\mathcal{D}_0^{(tr)}{0}} { \mathcal{D}^{(tr)}(0)} \right)\; ,
 \label{r_enh_eqn} 
 \eea 
 where the subscript 0 refers, as before, to pure Au (corresponding to $\epsilon_0 = 0$ {\em and} all $M_j = M^{(Au)}$).

\begin{figure}
    \centering
    \includegraphics[width=0.48\textwidth]{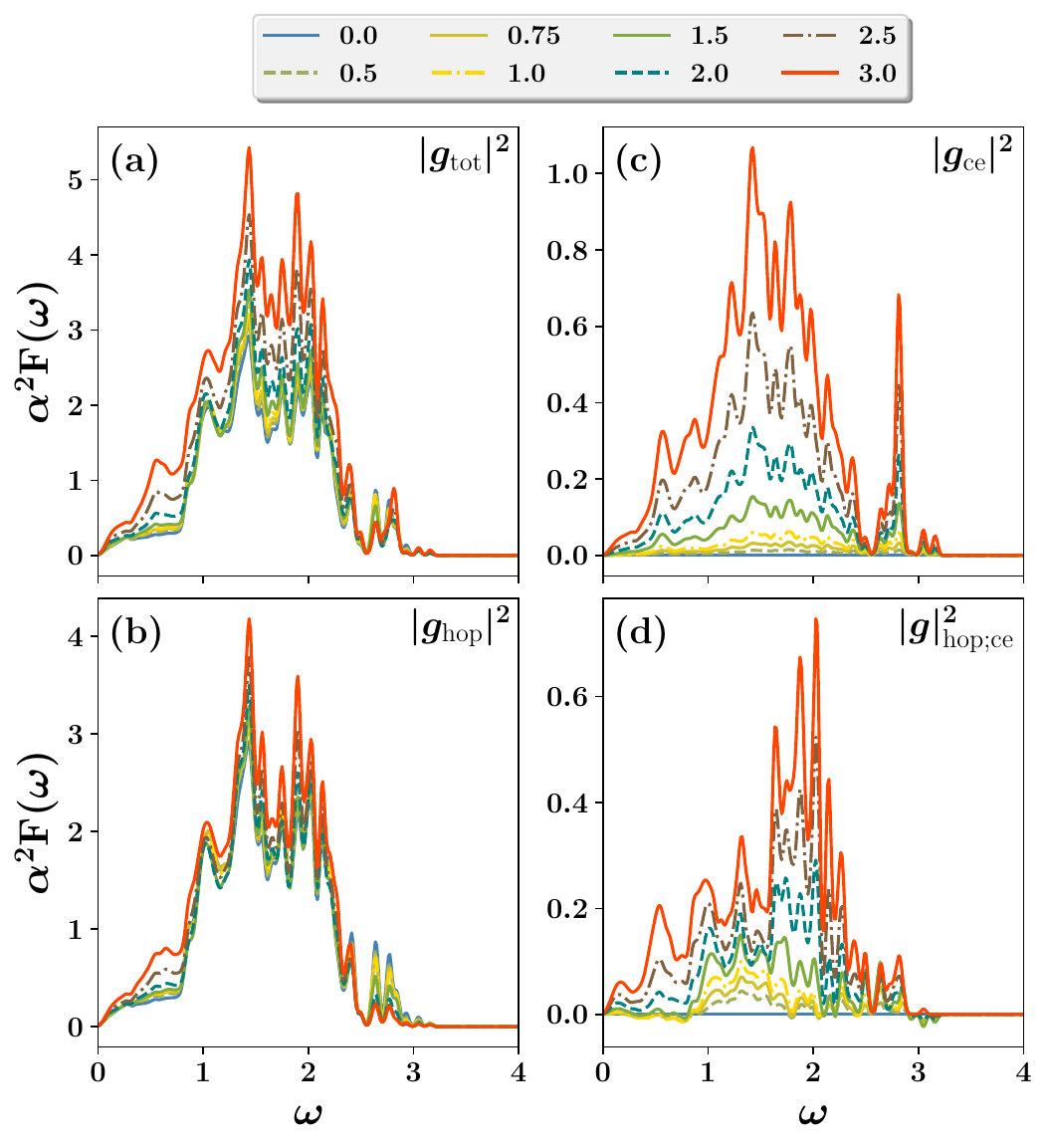}
    \caption { $\alpha^2 {F}(\omega)$ for various values of $\epsilon_0$, distinguishable by the colours indicated (in eV) at the top. The other parameters used are $U_0=2.0$ eV, $V_0/d_1 = 1.0$ eV, $\xi_0 = 1.0 d_1$ and $d_c = 8 d_1$. While panel (a) (label: $\lvert g_{\text{tot}} \rvert^2 $) shows the total $\alpha^2 {F}(\omega)$,  the other panels show the different components that contribute to it,  according to the choices for  $ \{ \mathcal{G} \} $ in the expression [compare Eq. (\ref{alpha_squared_F_eqn})] $ \sum \{ \mathcal{G} \} \delta(\varepsilon_{n,\mathbf{k}}) \delta(\varepsilon_{\mathbf{k+q},m})\delta(\hbar\omega - \hbar\omega_{\mathbf{q},s})$  as indicated below. 
Panel (b) (label: $\lvert g_{\text{hop}}\rvert^2 $): $\{ \mathcal{G} \} = \lvert \mathfrak{g}^{(hop)} \rvert^2$. 
Panel (c) (label: $\lvert g_{\text{ce}} \rvert^2 $): $\{ \mathcal{G} \} = \lvert \mathfrak{g}^{(ce)} \rvert^2$. 
Panel (d) (label: $\lvert g \rvert^2_{\text{hop;ce}} $):  $\{ \mathcal{G} \} = \lvert (\mathfrak{g}^ {(hop)} + \mathfrak{g}^{(ce)} ) \rvert^2 - \lvert \mathfrak{g}^{(hop)} \rvert^2 - \lvert \mathfrak{g}^{(ce)} \rvert^2 $. (For convenience, momentum and other labels associated with $\mathfrak{g}$ and  $\sum$ have been suppressed.) }
    \label{fig:fig8}
\end{figure}

\begin{figure}
    \centering
    \includegraphics[width=0.48\textwidth]{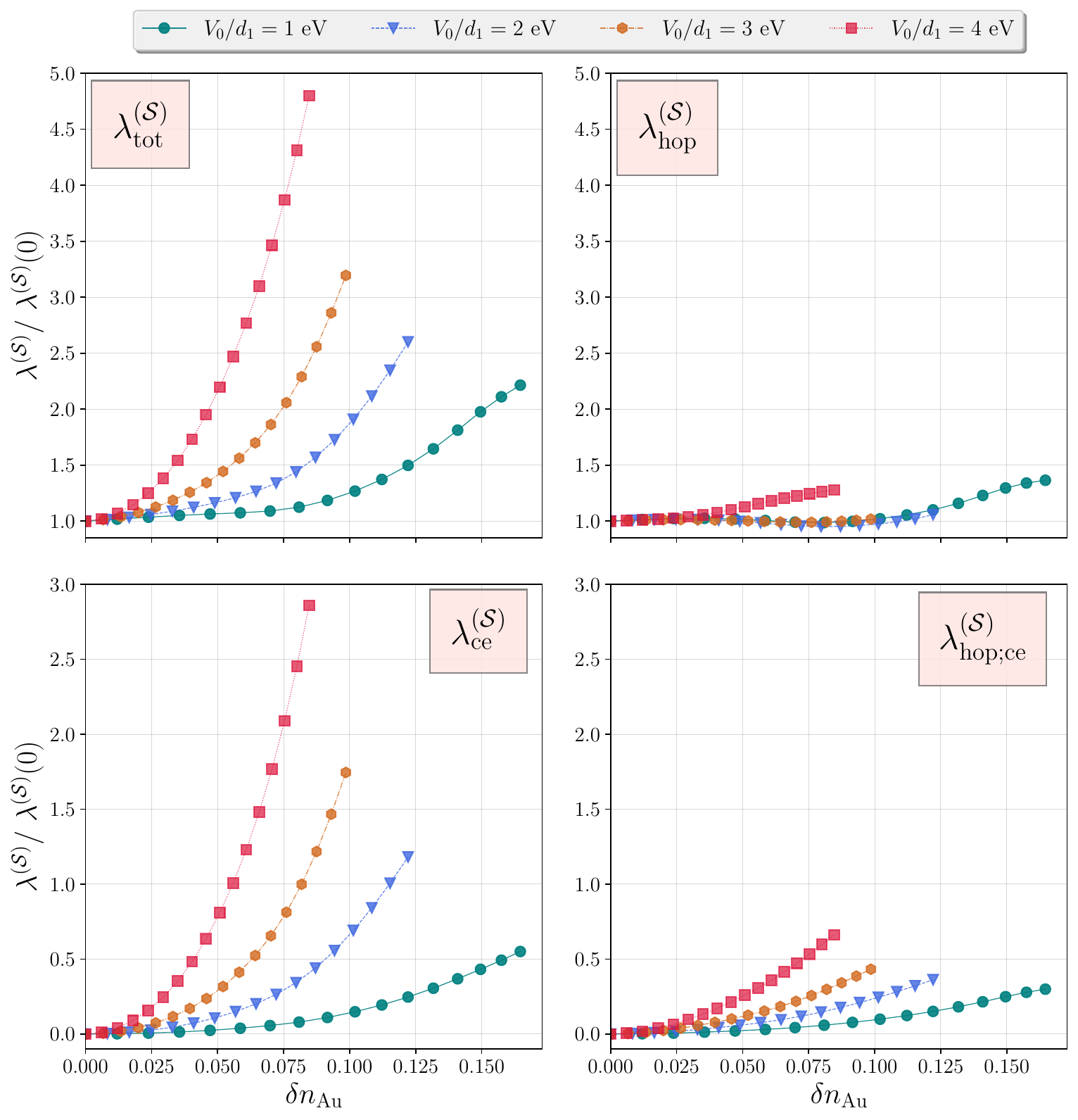}
    \caption{ Plots of the total as well as the different component contributions to the ratio 
$ \lambda^{(\mathcal{S})}$/ $\lambda^{(\mathcal{S})}_0$, versus $ \delta n_{Au} $, the average electron occupancy excess per atom on the Au atoms, shown for various values of $V_0/d_1$ in different colours as indicated at the top.  The different panels are in one to one correspondence with the panels in Fig. 8. The plots in the panel labelled  $\lambda^{(\mathcal{S})}_{ \text{tot}}$  correspond to the total contributions, arising from $\lvert (\mathfrak{g}^ {(hop)} + \mathfrak{g}^{(ce)} ) \rvert^2 $. The plots in the panels labelled  $\lambda^{(\mathcal{S})}_{ \text{hop}}$,  $\lambda^{(\mathcal{S})}_{ \text{ce}}$, and  $\lambda^{(\mathcal{S})}_{ \text{hop;ce}}$  contain, respectively, the contributions arising from $\lvert \mathfrak{g}^{(hop)} \rvert^2$,  $\lvert \mathfrak{g}^{(ce)} \rvert^2$, and   $\lvert (\mathfrak{g}^ {(hop)} + \mathfrak{g}^{(ce)} ) \rvert^2 - \lvert \mathfrak{g}^{(hop)} \rvert^2 - \lvert \mathfrak{g}^{(ce)} \rvert^2 $ .}
    \label{fig:fig9}
\end{figure}

The different panels of Fig. 8  show our results for the total as well as the separate component contributions to  $\alpha^2 F $ for a range of values of $\epsilon_0$ (in eV, shown in different colours as indicated at the top), with the other parameters being the same as mentioned earlier, in the context of Fig. 4, i.e, $U_0=2.0$ eV, $V_0/d_1 = 1.0$ eV, $\xi_0 = 1.0 d_1$ and $d_c = 8 d_1$.  Plots of  
$\alpha^2 F $ for other, larger values of $V_0/d_1$ are given in Appendix D. The different panels of Fig. 9 shows the total as well as the separate components of the contributions to  $\lambda^{(\mathcal{S})} / \lambda^{(\mathcal{S})}_0$ for a range of values of 
$\epsilon_0$ and for all the values of  $V_0/d_1$ we have explored, plotted now as a function of $\delta n_{Au}$, the {\em average} electron occupancy excess per atom on the Au sites (discussed earlier in the context of Fig. 5). Here $\lambda^{(\mathcal{S})}_0$ is obtained from the calculations carried out  for pure Au,  and therefore has only contributions from  $\mathfrak{g}^{(hop)}$. 

The increasing enhancement in both  $\alpha^2 F $ and $\lambda^{(\mathcal{S})} / \lambda^{(\mathcal{S})}_0$ with increasing values of $V_0/d_1$ and $\delta n_{Au}$ is clearly evident, as is to be expected from Eq.s (\ref{el_ph_coupling_ce_eqn}), (\ref{alpha_squared_F_eqn}) and (\ref{lambda_omega_log_eqn}). However, the overall excess contributions arising from the charge disproportionation is modest compared to the contribution from the hopping terms in the Hamiltonian, determined by $|{\mathfrak{g}}^{hop}|^2$, unless sufficiently large values of $V_0/d_1$ and $\delta n_{Au}$ are invoked, whence the dominant component of the excess contributions to $\lambda^{(\mathcal{S})} / \lambda^{(\mathcal{S})}_0$, governed by $|{\mathfrak{g}}^{ce}|^2$, increases rapidly, as $[V_0 \delta n_{Au} \xi_0 / (d_1^2 t_0)]^2$, as is clearly seen in the results presented in Fig.9. Note also that the excess contributions to $\alpha^2 F$ arising $|{\mathfrak{g}}^{ce}|^2$ have a significant weight in the nearly localised modes of the Ag atoms, and increasingly so for increasing  $V_0/d_1$ and $\delta n_{Au}$ (see Fig.s 14-16 in Appendix D), which is completely in line with the picture we have presented of the vibrational modes of Ag coupling strongly via Coulomb interactions to the interface dipoles.

As already mentioned, in the experimental work in ref. \cite{AG_el_ph_enh-2024},  numbers as large as  $\sim$ 60 for for $f_{Ag} \sim 0.25$ (and even larger numbers, $\sim$ 100, for $f_{Ag} \sim 0.5$) have been reported for the enhancement factor $r_{enh}$ defined above.  As seen in the results for the flattening of the bands  we have presented earlier from the DFT calculations for the 3D FCC-SL  as well as the Hartree self-consistent calculations for the S-SL as shown in Fig. 6, $r_{enh}$ already has a contribution in the range 2 - 3 coming from the reduction in $\mathcal{D}^{(tr)}(0)$.  Hence, to account for the observed values of $r_{enh}$ for $f_{Ag} \sim 0.25$  one needs an additional factor of 20 - 30 from the enhancement in $\lambda^{(\mathcal{S})}$.

From the plots for $\lambda^{(\mathcal{S})} / \lambda^{(\mathcal{S})}_0$ shown in Fig. 9, we see that for $V_0/d_1 = 3 t_0 = 3 eV$, for example\cite{V0-values}, the enhancement factor for $\delta n_{Au} =0.1$ is $\sim 3.2$. Assuming a quadratic scaling of the enhancement factor with $\delta n_{Au}$,  the system can clearly attain the required enhancement factors in $\lambda^{(\mathcal{S})} / \lambda^{(\mathcal{S})}_0$ if $\delta n_{Au}$ is larger, in the range 0.25 - 0.31. (For $V_0/d_1 = 4 $eV, the range required would be 0.15-0.18.)   Interestingly, ref. \cite{AG_el_ph_enh-2024} also reports core-level XPS data for Au 4f and Ag 3d, and the shifts of the peak positions suggest that for $f_{Ag} \sim 0.5$, $\delta n_{Au} \sim$  0.3 - 0.4.  Thus the mechanism we have proposed above, leading to new contributions that substantially enhance the electron-phonon coupling  together with at least a 2 - 3 fold reduction in the transport DOS at the Fermi level,  does contain the ingredients needed to account for the astoundingly large increases in the resistivity (slopes with respect to temperature) seen experimentally.  

However, there remains the question as to how such large charge transfers as considered above and as seen in the XPS spectra can possibly come about. An alluring explanation within the framework of what we have discussed so far is the possibility that the basis relaxation we have mentioned earlier can actually become very large, because a contraction of the distances between the oppositely charged interfacial Ag and Au atoms, and an expansion of the other Au-Au and Ag-Ag distances, is clearly energetically favourable as regards the Coulomb interactions.  This will lead to further enhancements of the interfacial charges, and so on, leading to a run-away instability that can only be arrested by the short-range repulsive and long-range attractive (an-harmonic) inter-atomic forces contained in $E_{core}$. The resulting self-consistent Hartree potentials could be so large and attractive on the Au sites near the interfaces as to split off nearly localised ``{\em polaronic bound states}'' with large enough binding energies that they become doubly occupied, constituting  {\em bound bipolarons}. Conversely, the Hartree potentials would be so large and repulsive on the Ag sites near the interfaces as to split off nearly localised anti-bound states, with energies large enough that they are nearly unoccupied. This mechanism could effectively drive $\delta n$ values for the Au sites near the interface close to 1, and for the Ag sites near the interface close to -1, leading to the large $\delta n_{Au}$ observed experimentally\cite{delta_n_Au}.  In a homogeneous system, when the {\em el-ph} interactions are strong enough to form bipolarons, they tend to do so everywhere, causing the system to go insulating \cite{Esterlis_et_al-2018,Murthy_et_al-2024}. We believe that in the present context the system is able to avoid this and stay metallic as the bipolaronic states are formed only near the interface.

The above bipolaronic mechanism has two additional attractive features. First, it will deplete carriers from the conduction band, leading to an additional source of the reduction of the conductivity or the enhancement of the resistivity, and hence of $r_{enh}$. Second, if the binding energy of the polaronic bound state (as measured from the Fermi level) is $\epsilon_p$, then that of the bipolaron would be $ \Delta \equiv (2 \epsilon_p - U_{eff})$ where $U_{eff}$ is the effective Coulomb repulsion energy cost for the double occupancy of the localised bound state. If  $ \Delta$ is small enough that thermal excitations can break up each bipolaron into a polaron and a conduction electron with probability $ \propto exp(-\Delta/ (k_B T)$, this would lead to a parallel activated channel of conduction, causing the high temperature resistivity to deviate from a linear temperature dependence. This seems indeed to be what is observed as per the resistivity data reported in ref. \cite{AG_el_ph_enh-2024}. The best fits to the data with a parallel channel of conduction have a parallel conductivity of activated form with $\Delta \approx 60 meV$ for $f_{AG} = 0.33$ (see section IV and Fig. S4 in the supplementary information of ref. \cite{AG_el_ph_enh-2024}).                                          

The large increase in $\lambda^{(\mathcal{S})}$ has obvious implications for superconductivity, and these are discussed in greater detail in Appendix D, together with the results for $\omega_{log}$. There we also present arguments to suggest that the bipolaron formation proposed above will be accompanied by an increase in $\omega_{log}$,  and this together with a likely breakdown of the Migdal-Eliashberg theory\cite{Esterlis_et_al-2018, Zhang_et_al-2023} could be the mechanism at play in the Au-Ag nanostructured systems with superconducting transition temperatures in the ``room temperature'' regime reported in refs. \cite {Au_Ag_unpub1, Au_Ag_unpub2, Saha_et_al-2022}. 

Needless to say, strong electron-phonon interactions have a significant impact on many other properties of the system as well, leading to several other measures that are possible, and relatively  easily accessible experimentally, such as thermal transport, Raman line-widths, ultrasound attenuation, etc.\cite{Giustino-2017}. It is a question of obvious interest as to how these other measures are affected by the enhanced electron-phonon interactions and other aspects of  bimetallic nano-structures discussed above. We hope to address them, as well as the challenging calculations allowing for bipolaron formation as suggested above, and the consequent, much harder, theoretical challenges of dealing with large {\em el-ph} interactions and a possible breakdown of Migdal-Eliashberg theory, in future work.

\section{Concluding comments}

In summary, we have shown in this paper that superlattices engineered by placing a periodic array of nano-clusters of a metal inside another dissimilar metal (preferably with heavier atoms and with a different work function) can lead to substantial enhancements of the electron-phonon interaction compared to that of the pure metals, with obvious consequent possibilities for high temperature superconductivity. The mechanism responsible is the charge transfer arising from the mismatch of the local potentials between the two species of metal atoms, thereby generating charge dipoles at all the interfaces between the two metals, which couple strongly, via Coulomb interactions, to the breathing and other phonon modes of the lighter atoms of the nano-clusters inside the ``cages'' formed by the matrix of the heavier metals. Within a Hartree mean-field theory framework, we have also presented the formalism for calculating the enhancements, and illustrated its use  in the context of Ag nano-clusters embedded periodically inside a Au matrix. The enhancements can be spectacularly large if the system parameters are such as to favour the formation of  bipolaronic bound states localised at the Au sites near the interface.

The mechanism we have proposed and the formalism we have presented above for the enhancement of electron-phonon interactions are clearly fairly general, and should be applicable to other combinations of nano-clusters and host matrix as well. Furthermore, we expect the enhancement to be robust, within limits, with respect to disorder in the placement of the centres of the nano-clusters as well in the distribution of their sizes. In addition, given that the enhancement is connected to the formation of interface charge dipoles, it stands to reason that it will be larger if the interface area increases, which can be achieved, again within limits, by reducing the sizes of the nano-clusters and increasing their proportion relative to the host matrix. These observations are all consistent with the trends reported in the experiments \cite{Saha_et_al-2022, AG_el_ph_enh-2024}. 

If the experience of Ag-Au is any indication, the difficulties are likely to be in the processes of making small-enough nano-clusters and embedding them roughly uniformly inside a host matrix, and the chemical (and physical) homogeneity and stability of the obtained systems. In this context, {\em we would like to suggest the promising possibility of using Fullerene C$_{60}$ molecules, where all the 60 C atoms are at the surface}, or even fragments of carbon nanotubes, as the nanoparticles instead of Ag or other metallic nanoparticles with their attendant difficulties of synthesis. The van der Waals diameter of C$_{60}$ is only 1.1 nanometers, and its reported work function of  ~ 4.71 eV as measured in poly-crystal films \cite{Shiraishi-2003}  is close to that of Ag. Furthermore, Carbon atoms are much lighter than Ag atoms, leading to much higher phonon frequencies.  Hence, if one can find ways of embedding C$_{60}$ molecules nearly uniformly in a Au or Pt matrix, the resulting system may well have similar, or hopefully even better, enhancement of the electron-phonon interactions, and with the much larger values of $\omega_{log}$, consequent high temperature superconductivity. 

Fullerenes are well known acceptors of electrons\cite{Illescas-2006}; hence, possibilities for embedding them in metals with {\em lower} work functions, of which there are many\cite{workfunction}, such as the Alkali metals (Na, K, etc.,) and the Alkaline earth metals (Mg, Ca, etc.,), are also worth exploring. Many of these metals have relatively low melting points, so a way to prepare the systems could be to mix the carbon nanoparticles into the molten metals.  If the carbon nanoparticles all get charged identically, as the theory discussed above suggests, electrostatic repulsion would keep them apart inside the liquid metal, leading essentially to a charge stabilised metallic nano-colloid, whose properties would be fascinating by themselves.  Cooling such a system slowly might well lead to the formation of crystalline superlattices akin to the idealised models discussed in this paper! The chemical or physical stability of these systems (eg., absence of phase separation, though the charge stabilisation mentioned above makes this very likely) is not easy to investigate theoretically, and experimental trials to determine the best combinations seem to be the best bet in the short run. We hope that our work will stimulate researchers in exploring these as well as other possibilities, using other combinations of metals, for engineered nano-structured bimetallic superlattices. 

\begin{acknowledgments}
We thank our colleague Arindam Ghosh and his group for sharing their as yet unpublished experimental results and for stimulating discussions. We thank the Supercomputer Education and Research Centre (SERC) at IISc for providing the computational resources. M.J. acknowledges the National Supercomputing Mission of the Department of Science and Technology, India, and Nano Mission of the Department of Science and Technology for financial support under Grants No. DST/NSM/R\&D\_HPC\_Applications$/2021/23$ and No. DST/NM/TUE/QM-10/2019 respectively. H. R. K. gratefully acknowledges support from the Indian National Science Academy under grant number No. INSA/SP/SS/2023/, the Science and Engineering Research Board of the Department of Science and Technology, India under grant No. SB/DF/005/2017/, and at ICTS from the Simons Foundation (grant No. 677895, R.G.).
\end{acknowledgments}

\appendix

\section{DFT calculations for the Au-Ag FCC-SL}

The DFT calculations we have reported were performed using the \texttt{Quantum ESPRESSO} package \cite{giannozzi2009quantum,giannozzi2017advanced} employing optimised norm-conserving Vanderbilt pseudo-potentials \cite{hamann2013optimized}. The exchange correlation was approximated using the Perdew-Burke-Ernzerhof (PBE) parametrisation of the generalised gradient approximation \cite{perdew1996generalized}. The wave-functions were expanded in a plane wave basis upto an energy cutoff of $50$ Ryd. The small Brillouin zone (SBZ) of the superlattice was sampled using a shifted $6\times6\times6$  $\mathbf{k}$-grid.\

The Wannier orbitals were constructed using the \texttt{Wannier90} package \cite{mostofi2014updated}. Based on the DFT bands, selectively localised Wannier functions with the centres constrained to each atomic centre (SLWF+C), were constructed with 9 orbitals per atom (Au: 5d, 6s, 6p and Ag: 4d, 5s, 5p), leading to 576 orbitals per supercell. The Lagrange multiplier for constrained site localisation was set at $1.0$, and the Wannierisation was performed on the $6\times 6 \times 6$ $\mathbf{k}$-grid. The DFT band structures and the Wannier interpolated band structures (which can be done on finer $\mathbf{k}$-grids as per requirement) have been shown in Fig. \ref{fig:fig2} in section II.

For the Wannier interpolated band structures, the band velocities are easily calculated as the expectation values of the $\mathbf{k}$-derivative of the band Hamiltonian (prior to diagonalization, which can be written down analytically) in the band wave-functions. The DOS (the transport DOS) at the chemical potential quoted in sections I and II were obtained by summing 1 (the squared velocities) multiplied by a Gaussian weight function of the band energies with a spread of ~ .025 eV around the Fermi level, and dividing by the number of points in the  $\mathbf{k}$-grid. 

\section{Details regarding the Hartree Solution of the semi-phenomenological TBM with long range Coulomb interactions, and additional results}

In the (restricted) Hartree approximation where we do not allow for the superlattice-translation or spin symmetry to be broken, $H_{ce}$ gets replaced by
\begin{multline}
    \boldsymbol{H}_{ce;H} = \sum_{j, \mathbf{R}} \Phi_j \, \boldsymbol{n}_{j \mathbf{R} }  - \bigg[ U_0 \sum_{j, \mathbf{R}} \frac {(\delta n_j)^2} {4}  \\ + \frac{1}{2} \; V_0 \sum^{\quad \p}_{j, \mathbf{R}; j^\prime, \mathbf{R}^\prime} \, \frac {\delta n_j \, \delta n_{j^\prime}} {d_{j \mathbf{R}; j^\prime \mathbf{R}^\prime}} \bigg] ;
\label{H_coul_hartree_eqn-SI}
\end{multline}
where, as mentioned earlier, $\delta n_j \equiv \langle (\boldsymbol{n}_{j \mathbf{R} } -1) \rangle$, the excess occupancy (of the Wannier orbital) at the site $j \mathbf{R}$, and
\beq
\Phi_j \equiv  \frac{1}{2} \, U_0 \, \delta n_j + V_0 \sum^{\quad \p}_{j^\prime, \mathbf{R}^\prime} \, \frac{\delta n_{j^\prime}} {d_{j \mathbf{R}; j^\prime \mathbf{R}^\prime}} \, ,
\label{Phi_j_eqn-SI}
\eeq
the effective Hartree potential at the site $j\mathbf{R}$,  are to be determined self consistently. Note that by assumption (of unbroken superlattice-translation symmetry) $\delta n_j$ and $\Phi_j$ are independent of $\mathbf{R}$. To diagonalise the total effective Hamiltonian within the Hartree approximation, $\boldsymbol{H}_{el;H} \equiv \boldsymbol{H}_{ke} + \boldsymbol{H}_{ce;H}$ we make the transformations
\bea
\boldsymbol{c}_{j \mathbf{R} \s} &=& \frac {1} {\sqrt{\mathcal{N}}} \sum_{\mathbf{k}} e^{i \mathbf{k} \cdot (\mathbf{R} + \mathbf{d}_j)} \tilde{\boldsymbol{c}}_{j \mathbf{k} \s} \nonumber \\ &=&  \frac {1} {\sqrt{\mathcal{N}}} \sum_{\mathbf{k}, m} e^{i \mathbf{k} \cdot (\mathbf{R} + \mathbf{d}_j)} \varphi_{j; \mathbf{k} m} \tilde{\boldsymbol{c}}_{\mathbf{k} m \s}
\label{c_ctilde_transform_eqn-SI}
\eea
where $\varphi_{j; \mathbf{k} m}$ are eigenvectors of the Fourier transform of the $N_c \times N_c$ (first quantised) one-electron Hamiltonian matrix with eigenvalues $\varepsilon_{\mathbf{k} m} \, $, both being labeled by the wave-vector $\mathbf{k}$  and the integer band index $m$ which distinguishes the $N_c$ different bands. i.e.,
\beq
\sum_{j^\prime} \tilde{h}_{j j^\prime} (\mathbf{k}) \varphi_{j^\prime; \mathbf{k} m} = \varepsilon_{\mathbf{k} m} \varphi_{j; \mathbf{k} m}
\label{vphi_veps_eigen_eqn-SI}
\eeq
with
\beq
\tilde{h}_{j j^\prime} (\mathbf{k}) \equiv (\epsilon_j -\mu + \Phi_j) \delta_{j j^\prime} - \sum^{\quad \p}_{\mathbf{R}^\prime}  \, \left[ e^{ - i \mathbf{k} \cdot \mathbf{d}_{j \mathbf{R}; j^\prime \mathbf{R}^\prime} } \, \mathfrak{t}_{j \mathbf{R}; j^\prime \mathbf{R}^\prime} \right] \, .
\label{k_space_ham_eqn-SI}
\eeq
Because of the periodic boundary conditions $\mathbf{k}$ takes $\mathcal{N}$ discrete values inside the superlattice Brillouin zone (SBZ) whose volume (or area in the 2-D case) in reciprocal space is $1/N_c$ of that of the  unit-cell Brilloun zone (UBZ) of the pure-lattice.  In terms of the operators $\tilde{\boldsymbol{c}}_{\mathbf{k} m \s}$,  $\boldsymbol{H}_{el;H}$ simplifies to
\begin{multline}
    \boldsymbol{H}_{el;H} = \sum_{\mathbf{k}, m, \s} \varepsilon_{\mathbf{k} m}  \tilde{\boldsymbol{c}}^\dag_{\mathbf{k} m \s} \tilde{\boldsymbol{c}}_{\mathbf{k} m \s}  - \bigg[ U_0 \sum_{j, \mathbf{R}} \frac {(\delta n_j)^2} {4}  \\ + \frac{V_0}{2} \sum^{\quad \prime}_{j, \mathbf{R}; j^\prime, \mathbf{R}^\prime} \frac {\delta n_j \, \delta n_{j^\prime}} {d_{j \mathbf{R}; j^\prime \mathbf{R}^\prime}} \bigg] \, .
\label{Hartree_H_el_diag_form_eqn-SI}
\end{multline}
For a system temperature $T$, the excess (or deficit if negative) occupancies $\{\delta n_j \}$ are determined in terms of the eigenvectors $\varphi_{j; \mathbf{k} m}$, and the thermal occupancies of the band levels $\varepsilon_{\mathbf{k} m}$ (which are already relative to the chemical potential) governed by the Fermi function  $n^-_F$ as
\bea
\delta n_j = \left( \frac {2} {\mathcal{N}} \sum_{\mathbf{k}, m} n^-_F(\varepsilon_{\mathbf{k} m}) |\varphi_{j; \mathbf{k} m}|^2 \right) - 1 ; \nonumber \\ \; n^-_F(\epsilon) \equiv \frac {1} {e^{ \beta \epsilon} + 1} , \, \ \ \ \ \  \beta \equiv 1/(k_B T) \, .
\label{nj_eqn-SI}
\eea
Eq.s~(\ref{Phi_j_eqn-SI}), (\ref{vphi_veps_eigen_eqn-SI}), (\ref{k_space_ham_eqn-SI}) and (\ref{nj_eqn-SI}) together with the half-filling constraint which requires $ \sum_j \delta n_j = 0 $ constitute the self-consistency conditions which we solve to determine the $N_c$ excess occupancies $\{\delta n_j \}$, the corresponding Hartree potentials $\{ \Phi_j \}$,  and the chemical potential $\mu$. All the Hartree self-consistent calculations we report in this paper have been done with the atoms in the {\em unrelaxed} basis positions $\{\mathbf{d}^0_j \}$, for which $\mathbf{d}_{j \mathbf{R}; j^\prime \mathbf{R}^\prime}$ and $d_{j \mathbf{R}; j^\prime \mathbf{R}^\prime}$ in the above Eq.s  are replaced by $\mathbf{d}^0_{j \mathbf{R}; j^\prime \mathbf{R}^\prime}$ and $d^0_{j \mathbf{R}; j^\prime \mathbf{R}^\prime}$ respectively.

\begin{figure*}
    \centering
        \includegraphics[width=0.91\textwidth]{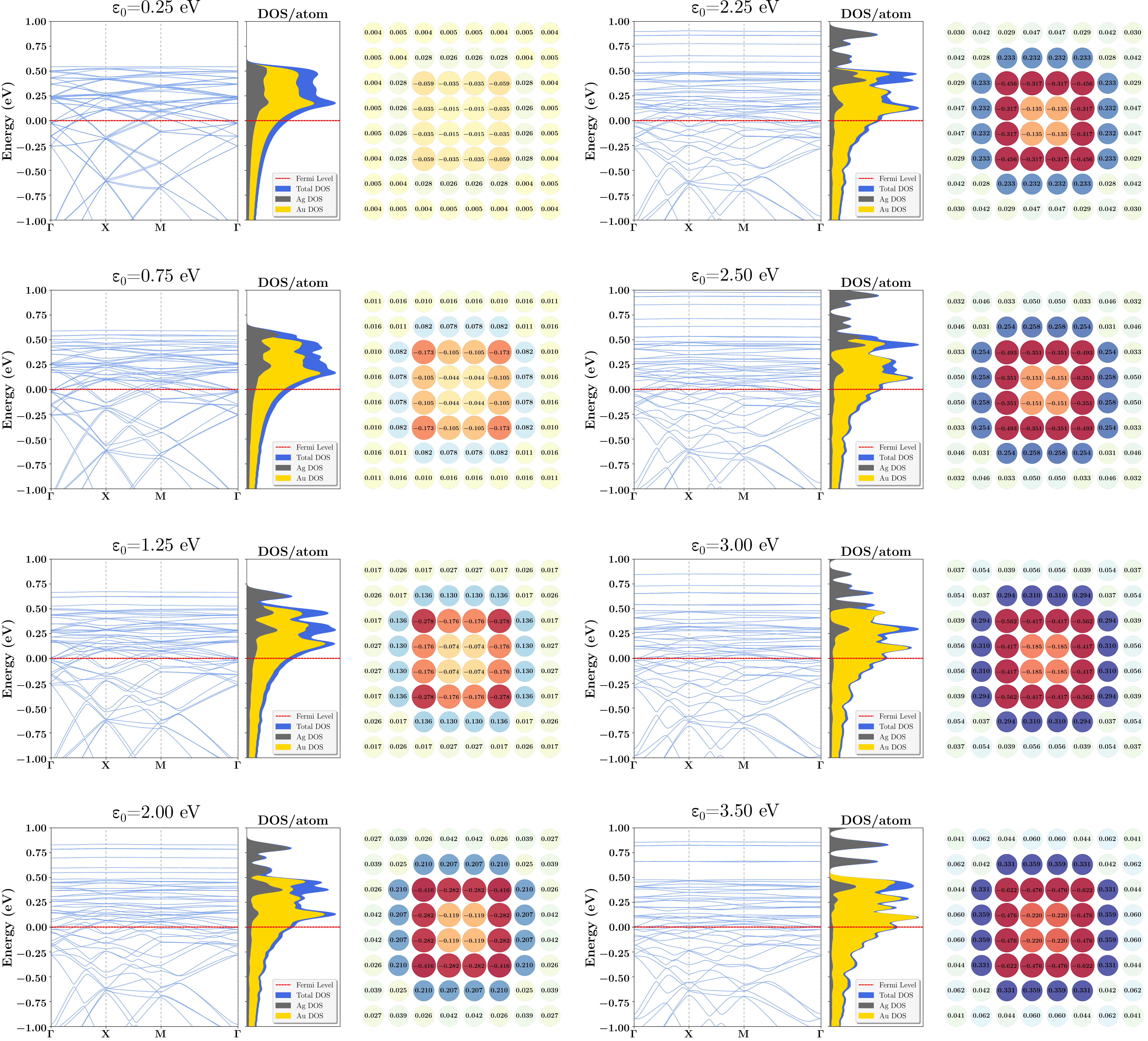}
    \caption{The band structure and the electron occupancy excess $(\delta n_j)$ at each atomic site for the S-SL model for various values of $\epsilon_0$, with $U_0=2.0, V_0/d_1 = 1.0$ (both in eV), $\xi_0 = 1.0 d_1$ and $d_c = 8 d_1$ being held fixed.  $(\delta n_j)$ is, depicted using colour-coding, and in addition, its value at each site is given within the corresponding circle.}
    \label{fig:fig10}
\end{figure*}

Both the diagonalization and the solving of the self-consist solutions have to be carried out numerically. The evaluations for $\{ \Phi_j \}$ [Eq. (\ref{Phi_j_eqn-SI})] involving the long range Coulomb interaction need the use of Ewald summation techniques \cite{Ewald-ref}.  Our calculations of the self-consistent densities and Hartree potentials were carried out using the root finding routine available in Scipy, \texttt{scipy.optimize.root}, on $20 \times 20$ or smaller $\mathbf{k}$-grids in the SBZ of the superlattice, with verified convergence with respect to increasing the grid size. The square point group symmetry (about the centre of the supercell) was taken advantage of to identify equivalent sites in the supercell that have the same values of $\delta n_j $ ( and also of $\Phi_j $), so that the number of unknown variables to be solved for by the root finding code got reduced. The self-consistent Hartree potentials were then used to carry out non-self-consistent calculations of the band structure, DOS, etc., on $24 \times 24$ $\mathbf{k}$-grids in the SBZ (equivalent to $192 \times 192$ $\mathbf{k}$-grids in the UBZ). The velocities needed for the calculation of the transport DOS were calculated using the relation:
\bea
 \mathbf{v}_{\mathbf{k} m} &=&  \sum_{j,j^{\prime}} \varphi^*_{j; \mathbf{k} m} \left[ \nabla_{\mathbf{k}} \tilde{h}_{j j^\prime} (\mathbf{k}) \right] \varphi_{j^\prime; \mathbf{k} m} \nonumber \\  
\nabla_{\mathbf{k}} \tilde{h}_{j j^\prime} (\mathbf{k})&=& -i \sum^{\quad \p}_{\mathbf{R}^\prime} \mathbf{d}_{j \mathbf{R}; j^\prime \mathbf{R}^\prime}  e^{ - i \mathbf{k} \cdot \mathbf{d}_{j \mathbf{R}; j^\prime \mathbf{R}^\prime} } \, \mathfrak{t}_{j \mathbf{R}; j^\prime \mathbf{R}^\prime}
\eea  
The DOS and the transport DOS given by Eq. (\ref{dos_tr_dos_eqn}) for which results were presented in Section II were calculated by summing over the discrete $\mathbf{k}$-grid and the band index $m$ with the delta functions in the equation being approximated as normalised Gaussian functions with a width $\sim 0.025$ eV.

 \begin{figure}
    \centering
    \includegraphics[width=0.48\textwidth]{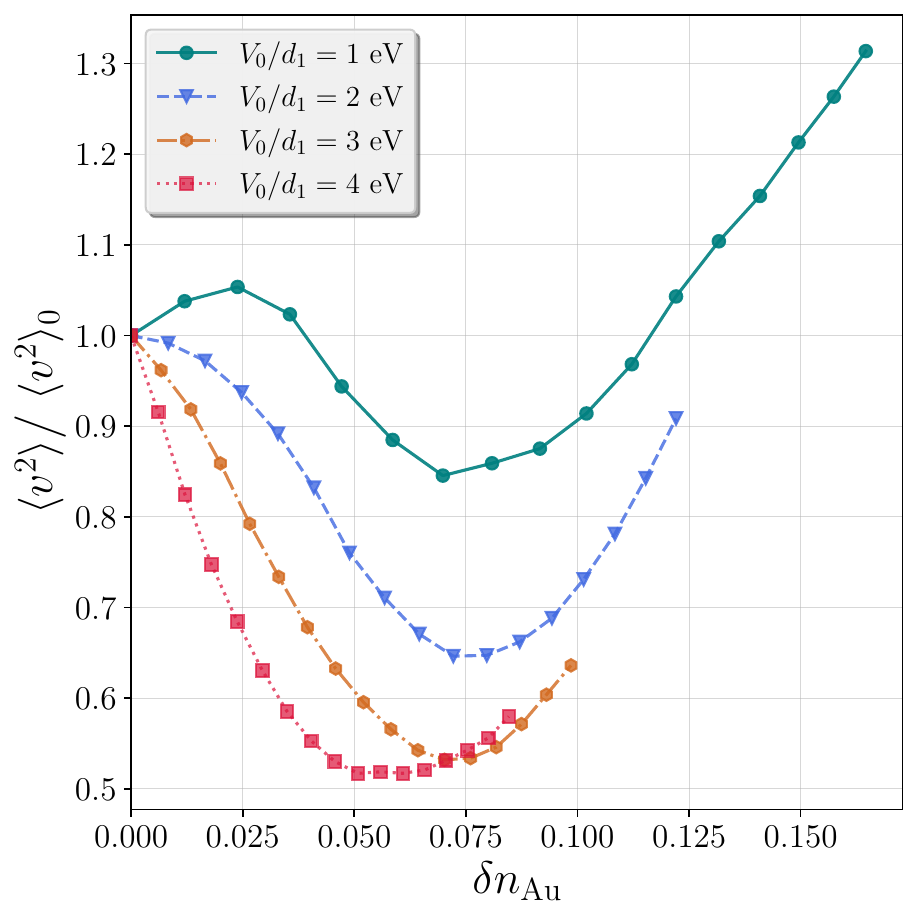}
    \caption{Plots of mean squared velocity ratio $ \langle v^2 \rangle /  \langle v^2 \rangle_0$ calculated by dividing  the transport DOS ratio, $\mathcal{D}^{(tr)}(0)/\mathcal{D}_0^{(tr)}(0)$ by the electronic DOS ratio $ \mathcal{D}(0) / \mathcal{D}_0(0)$,  for the S-SL system  as a function of  $\delta n_{Au}$ for several values of $V_0/d_1$. (The other parameters are held fixed at the values mentioned in the context of Fig. 6). As earlier, the subscript 0 refers to the pure Au case. }
    \label{fig:fig11}
\end{figure}

In case of pure gold or silver,  both $\delta n_j$ and $\Phi_j$ are self-consistently zero for all $j$, i.e., each site has exactly one electron, and the Hartree potential is zero. Then the band dispersions $\{ \varepsilon_{\mathbf{k} m} \}$ and the eigenvectors $\varphi_{j; \mathbf{k} m}$ for the superlattice are merely those obtainable by {\em folding} the single pure Au (or Ag) s-band with wave-vectors $\mathbf{k}_0$ in the pure-FCC UBZ down to the much smaller SBZ. i.e.,
\bea
m &\Leftrightarrow& \mathbf{G}_m ; \; \nonumber \\ \varepsilon^{(Au)}_{\mathbf{k} m} &=&  \varepsilon^{(Ag)}_{\mathbf{k} m} - \epsilon_0 = -\mu - \tilde{\mathfrak{t}}^{(0)} (\mathbf{k} + \mathbf{G}_m) \, ; \;  \nonumber \\ \varphi^{(0)}_{j; \mathbf{k} m} &=& \frac {1} {\sqrt{N_c}} e^{i \mathbf{G}_m \cdot \mathbf{d}^0_j}
\label{pure_Au_Ag_eps_vphi_eqn-SI}
\eea
where $\{ \mathbf{G}_m , m = 1, N_c \}$ are the $N_c$ smallest {\em superlattice} reciprocal lattice vectors (RLVs) which will fit inside the UBZ, and  $\tilde{\mathfrak{t}}^{(0)}$ denotes the pure-FCC or pure-SL  Fourier transform of the hopping amplitudes:
\bea
\tilde{\mathfrak{t}}^{(0)} (\mathbf{k}_0)  &\equiv& \sum^{\quad \p}_{\mathbf{R}^\prime , j^\prime}  \, \left[ e^{ - i \mathbf{k}_0 \cdot \mathbf{d}^0_{j \mathbf{R}; j^\prime \mathbf{R}^\prime} } \, \mathfrak{t} (d^0_{j \mathbf{R} ; j^\prime \mathbf{R}^\prime}) \right] \nonumber 
\\ &=& \sum^{\quad \p}_{\mathbf{R}_0^\prime}  \, \left[ e^{ - i \mathbf{k}_0 \cdot (\mathbf{R}_0 - \mathbf{R}_0^\prime)}  \, \mathfrak{t} (|\mathbf{R}_0 - \mathbf{R}_0^\prime|) \right]
\label{pure_Au_Ag_t_to_tilde_eqn-SI}
\eea
\begin{widetext}
\noindent For example, in the case when $d_c = 1.5 d_1$ whence only the first and second neighbour amplitudes, which we denote as $\mathfrak{t}, \mathfrak{t}^\prime$, are non zero, for the S-SL model, one has
\begin{equation}
    \tilde{\mathfrak{t}}^{(0)} (\mathbf{k}_0)  = 2 \mathfrak{t}  \, \big[\cos (k_0^x a_0) + \cos (k_0^y a_0)\big]  + 4 \mathfrak{t}^\prime \, \cos (k_0^x a_0)  \cos (k_0^y a_0)
\label{pure_Au_Ag_band_square_eqn-SI}
\end{equation}
and for the FCC-SL model, 
\begin{multline}
    \tilde{\mathfrak{t}}^{(0)} (\mathbf{k}_0)  =  4 \mathfrak{t} \, [ \cos (k_0^x a_0 /2)  \cos (k_0^y a_0/2) + \cos (k_0^y a_0 /2)  \cos (k_0^z a_0 /2)  + \cos (k_0^z a_0)  \cos (k_0^x a_0/2) ] \\
+ 2 \mathfrak{t}^\prime \, [\cos (k_0^x a_0 ) + \cos (k_0^y a_0) + \cos (k_0^z a_0 )]
\label{pure_Au_Ag_band_FCC_eqn-SI}
\end{multline}
\end{widetext}
However, as discussed in section II, when clusters of Ag atoms are embedded in the Au matrix, the repulsive local potential $\epsilon_0$ on the Ag sites causes these folded bands to hybridise and split, and also build up of a deficit of electrons on the Ag sites, and a consequent excess of electrons on the Au sites to maintain over all charge neutrality.  i.e., $\delta n_j > 0$ for the Au sites and $< 0$ for the Ag sites. The onsite Coulomb repulsion $U_0$ tends to suppresses this charge disproportionation, whereas the long range Coulomb repulsion controlled by $V_0$, while suppressing it overall somewhat, promotes a redistribution of the charge such that it is concentrated on the sites at the Ag-Au interface; i.e., in the self-consistent solution the Ag sites at the surface of the Ag cluster have the most negative  $\delta n_j$ values and the Au sites closest to this surface the largest positive $\delta n_j$ values, thus leading to the formation of interface dipoles. Two representative figures of the above (Fig.s 3 and 4) were shown in section II. Fig. 10 shows several more such figures illustrating the evolution of the band-structure, DOS and the charge distribution as $\epsilon_0$ is varied, all with $U_0 = 2.0$ eV, $V_0/d_1= 1.0$ eV, $\xi_0 = 1.0 d_1$ and $d_c = 8 d_1$  being held fixed. Fig 11 presents plots for the mean squared velocity ratio $ \langle v^2 \rangle /  \langle v^2 \rangle_0$  calculated from the ratios of the transport DOS and the regular DOS [cf., Eq. (\ref{dos_tr_dos_eqn})] for the  range of values of $\delta n_{Au}$ and $V_0/d_1$ that was discussed in section II in the context of Fig.s 5 and 6.                                                                                                      
\section{Details regarding the calculations of phonons in the Nano-structured Superlattices}

Given the  harmonic terms in the  expansion for the dependence of $E_{core}$, the core energy term in Eq. (\ref{model_ham_eqn}),  on the displacement of the atoms from $\{ \mathbf{R}+\mathbf{d}^0_j \}$ to $\{ \mathbf{R}+\mathbf{d}^0_j + \mathbf{u}_{j \mathbf{R}} \}$ in terms of spring-constant tensors or force-constant-tensors as specified in Eq. (\ref{E_core_expn_eqn}), the Hamiltonian for the lattice vibrations or phonons can be written as
\begin{multline}
\boldsymbol{H}_{ph} = \sum_{\mathbf{R}, j, \alpha} \frac { P^\alpha_{j \mathbf{R}} P^\alpha_{j \mathbf{R}}} { 2 M_j } \\ + \; \frac{1}{2} \, \sum_{\substack{\mathbf{R}, j, \alpha \\ \mathbf{R}^\prime, j^\prime, \alpha^\prime}}  
\kappa^{\alpha \alpha^\prime}_{j j^\prime } (\mathbf{R}-\mathbf{R}^\prime)  \; u^{\alpha}_{j \mathbf{R}}  \; u^{\alpha^\prime}_{j^\prime \mathbf{R}^\prime},
\label{phonon_Hamiltonian_eqn-SI} 
\end{multline}
where $M_j$ is the mass of the atom at the site $(j \mathbf{R})$, $P^\alpha_{j \mathbf{R}}$ and $u^{\alpha}_{j \mathbf{R}}$ are canonically conjugate momentum and coordinate variables (operators) classically (quantum-mechanically).

As is well known \cite{Ashcroft_Mermin-76}, the phonon Hamiltonian for a D-dimensional case can be completely decoupled into $D N_c$ independent harmonic oscillators, referred to as normal modes or phonon modes, characterised by wave vectors $\{\mathbf{q}\}$, an integer mode number label $s$ which takes values from 1 to $D N_c$, frequencies  $\{ \omega_{\mathbf{q} s} \} $, and polarisation vector components $\{ \eta^{\alpha}_{j; \mathbf{q} s} \}$ via the following canonical transformations to new variables $\zeta_{\mathbf{q} s}$ and $\pi_{\mathbf{q} s}$ corresponding respectively to the normal mode coordinates and momenta: 
\bea
u^{\alpha}_{j \mathbf{R}} &=& \frac {1} {\sqrt{\mathcal{N} M_j}} \sum_{\mathbf{q}, s} \zeta_{\mathbf{q} s} \eta^{\alpha}_{j;\mathbf{q} s} e^{i \mathbf{q} \cdot (\mathbf{R} + \mathbf{d}^0_j)} ; \; \nonumber \\ P^{\alpha}_{j \mathbf{R}} &=& \sqrt {\frac {M_j} {\mathcal{N}} } \sum_{\mathbf{q}, s} \pi_{\mathbf{q} s} \eta^{\alpha}_{j;\mathbf{q} s} e^{i \mathbf{q} \cdot (\mathbf{R} + \mathbf{d}^0_j)}.
\label{u_jR_to_zeta_qs_eqn-SI}
\eea
Here $\{ \omega_{\mathbf{q} s} \} $ and $\{ \eta^{\alpha}_{j; \mathbf{q} s} \}$ are determined as solutions of the eigenvalue equation
\beq
\sum_{\alpha^\prime, j^\prime}  D^{\alpha \alpha^\prime}_{j j^\prime } (\mathbf{q}) \; \eta^{ \alpha^\prime}_{j^\prime; \mathbf{q} s} = (\omega_{\mathbf{q} s})^2 \;  \eta^{\alpha}_{j; \mathbf{q} s}
\label{dyn_matrix_eigen_eqn-SI}
\eeq
where the $DN_c \times DN_c$ {\em dynamical matrix} $D^{\alpha \alpha^\prime}_{j j^\prime } (\mathbf{q})$ is given by
\bea
D^{\alpha \alpha^\prime}_{j j^\prime } (\mathbf{q})  &=&  \frac{1} { \sqrt{M_j M_{j^\prime}} } \sum_{\mathbf{R}^\prime}  \left[    e^{ i \mathbf{q} \cdot \mathbf{d}_{j \mathbf{R}; j^\prime \mathbf{R}^\prime} } \;  \kappa^{\alpha \alpha^\prime}_{j j^\prime} (\mathbf{R}-\mathbf{R}^\prime) \right] \, \nonumber \\ &\equiv& \, \frac{1} { \sqrt{M_j M_{j^\prime}} } \, \tilde{\kappa}^{\alpha \alpha^\prime}_{j j^\prime} (\mathbf{q}) \, .
\label{kappa_to_dyn_matrix_eqn-SI}
\eea
Superlattice translational symmetry ensures that the right-hand-side (RHS) of this equation does not depend on the vector $\mathbf{R}$ ; hence one can set $\mathbf{R} = \mathbf{0}$ in the equation without any loss of generality. In terms of the new variables, the phonon hamiltonian takes the simple form
\bea
\boldsymbol{H}_{ph} &=& \sum_{\mathbf{q}, s}  \left[ \frac{1}{2} \boldsymbol{\pi}_{-\mathbf{q} s} \boldsymbol{\pi}_{\mathbf{q} s} + \; \frac{1}{2} \omega^2_{\mathbf{q} s} \boldsymbol{\zeta}_{-\mathbf{q} s} \boldsymbol{\zeta}_{\mathbf{q} s} \right] \; \\ &=& \sum_{\mathbf{q}, s} \left( \frac{1}{2} + \boldsymbol{a}^\dag_{\mathbf{q} s} \boldsymbol{a}_{\mathbf{q} s} \right) \,  \hbar \omega_{\mathbf{q} s} \, ,
\label{H_ph_diag_form_eqn-SI}
\eea
where the second form is obtained in the quantum case upon expressing the normal mode coordinates and momenta in terms of phonon creation and destruction operators $\boldsymbol{a}^\dag_{\mathbf{q} s}$ and $\boldsymbol{a}_{\mathbf{q} s}$ respectively in the standard way, as
\bea
\boldsymbol{\zeta}_{\mathbf{q} s} &=& \sqrt{ \frac{\hbar}{2 \omega_{\mathbf{q} s}} } \, (\boldsymbol{a}_{\mathbf{q} s} + \boldsymbol{a}^\dag_{-\mathbf{q} s}) \, ; \; \nonumber \\ \boldsymbol{\pi}_{\mathbf{q} s} &=& \frac{1}{i} \sqrt{ \frac{\hbar \omega_{\mathbf{q} s}}{2} }\, (\boldsymbol{a}_{\mathbf{q} s} - \boldsymbol{a}^\dag_{-\mathbf{q} s}).
\label{zeta_to_a_adag_eqn-SI}
\eea

In the phenomenological  {\em axially symmetric} force-constant modelling of lattice dynamics mentioned in section III, the contribution to the harmonic energy cost for displacements involving the pair of atoms at sites  $j \mathbf{R}$ and $j^\prime \mathbf{R}^\prime$   involving only  the bond-stretching and bond-bending spring constants $K^{(s)}_{j j^\prime} (\mathbf{R}-\mathbf{R}^\prime)$ and $K^{(b)}_{j j^\prime} (\mathbf{R}-\mathbf{R}^\prime)$ has the form
\begin{multline}
\frac{1}{2} K^{(s)}_{j j^\prime}(\mathbf{R}-\mathbf{R}^\prime) \left[ \frac{(\mathbf{d}^0_{j \mathbf{R}; j^\prime \mathbf{R}^\prime} \cdot \mathbf{u}_{j \mathbf{R}; j^\prime \mathbf{R}^\prime})^2} {|\mathbf{d}^0_{j \mathbf{R}; j^\prime \mathbf{R}^\prime}|^2} \right]
\\ + \frac{1}{2}  K^{(b)}_{j j^\prime}(\mathbf{R}-\mathbf{R}^\prime) \left[ |\mathbf{u}_{j \mathbf{R}; j^\prime \mathbf{R}^\prime}|^2 - \frac{(\mathbf{d}^0_{j \mathbf{R}; j^\prime \mathbf{R}^\prime} \cdot \mathbf{u}_{j \mathbf{R}; j^\prime \mathbf{R}^\prime})^2} {|\mathbf{d}^0_{j \mathbf{R}; j^\prime \mathbf{R}^\prime}|^2} \right] 
\label{stretch_bend_energy_cost_eqn-SI}
\end{multline}
Correspondingly, we get 
\begin{multline}
    K^{\alpha \alpha^\prime}_{j j^\prime } (\mathbf{R}-\mathbf{R}^\prime)  =  2 K^{(s)}_{j j^\prime}(\mathbf{R}-\mathbf{R}^\prime) \frac{ d^{0,\alpha}_{j \mathbf{R}; j^\prime \mathbf{R}^\prime} d^{0, \alpha^\prime} _{j \mathbf{R}; j^\prime \mathbf{R}^\prime}} {|\mathbf{d}^0_{j \mathbf{R}; j^\prime \mathbf{R}^\prime}|^2} \\ + 2 K^{(b)}_{j j^\prime}(\mathbf{R}-\mathbf{R}^\prime) \left[   \delta_{\alpha, \alpha^\prime} -  \frac{ d^{0,\alpha}_{j \mathbf{R}; j^\prime \mathbf{R}^\prime} d^{0, \alpha^\prime} _{j \mathbf{R}; j^\prime \mathbf{R}^\prime}} {|\mathbf{d}^0_{j \mathbf{R}; j^\prime \mathbf{R}^\prime}|^2} \right] ,
\label{stretch_bend_force_const_eqn-SI}
\end{multline}
as stated in Eq. \ref{stretch_bend_force_const_eqn} in section III. Using these, one can proceed as discussed above to construct the force constant tensor and the dynamical matrix, and numerically diagonalising the latter, one can determine all the $D N_c$ phonon frequencies and polarisation vectors over a grid of $\mathbf{q}$. 

\begin{figure}
    \centering
    \includegraphics[width=0.48\textwidth]{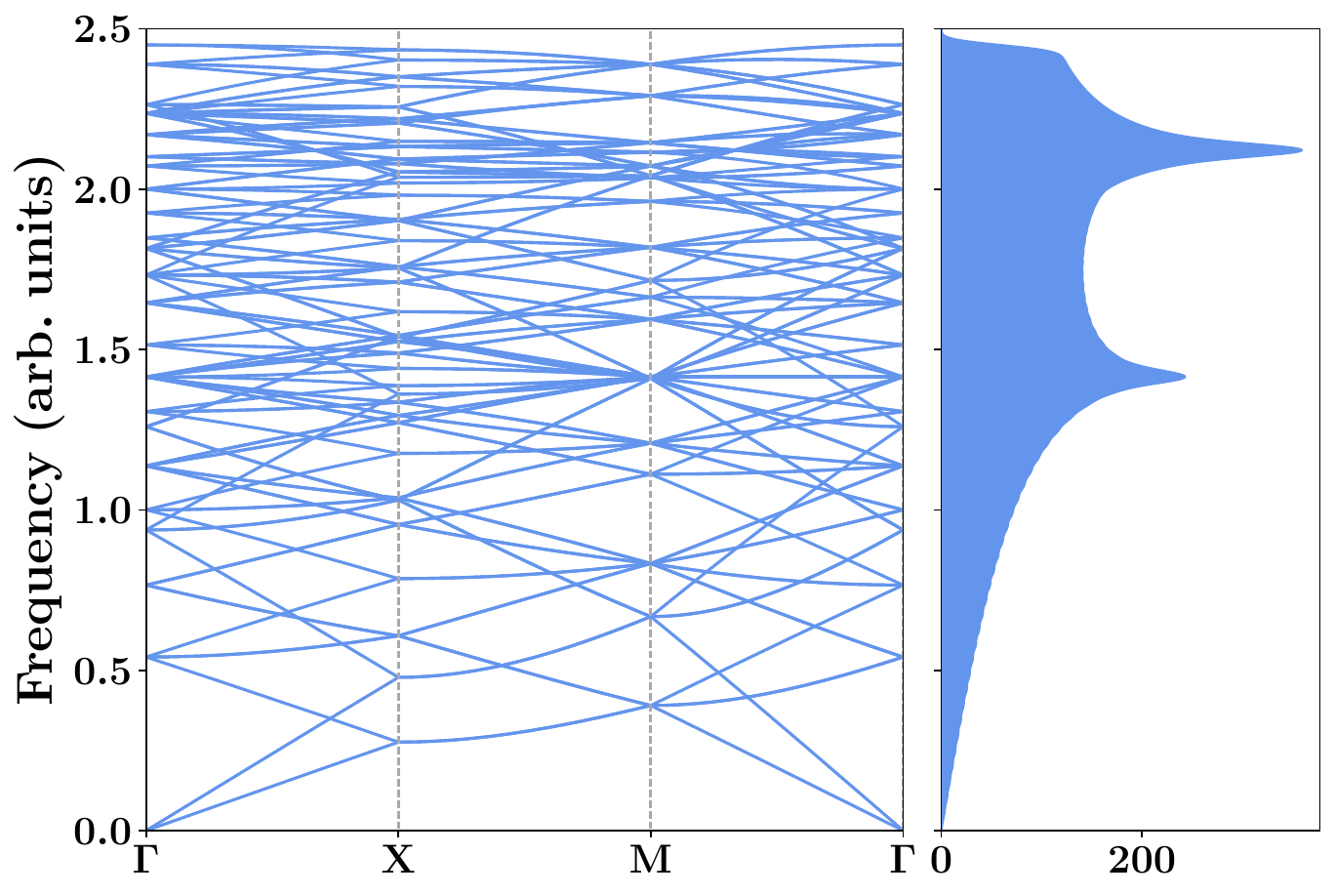}
    \caption{Phonon dispersion of pure Au in the FCM model on the S-SL system that we have considered, with only the first neighbour and second neighbour stretching spring constants, $K_1^{(s)}$ and $K_2^{(s)}$, assumed to be  non zero, and in units where $K_1^{(s)}$ and $K_2^{(s)}$  have been set equal to 1.0 and 0.5 respectively, and $M^{(Au)}$, the mass of the Au atom, has also been set equal to 1.0. The values of the phonon frequencies in real units can be obtained by multiplication with the appropriate value of $\sqrt{(K_1^{(s)}/M^{(Au)})}$\cite{real-units}. }
    \label{fig:fig12}
\end{figure}

\begin{figure*}
    \centering
    \includegraphics[width=0.88\textwidth]{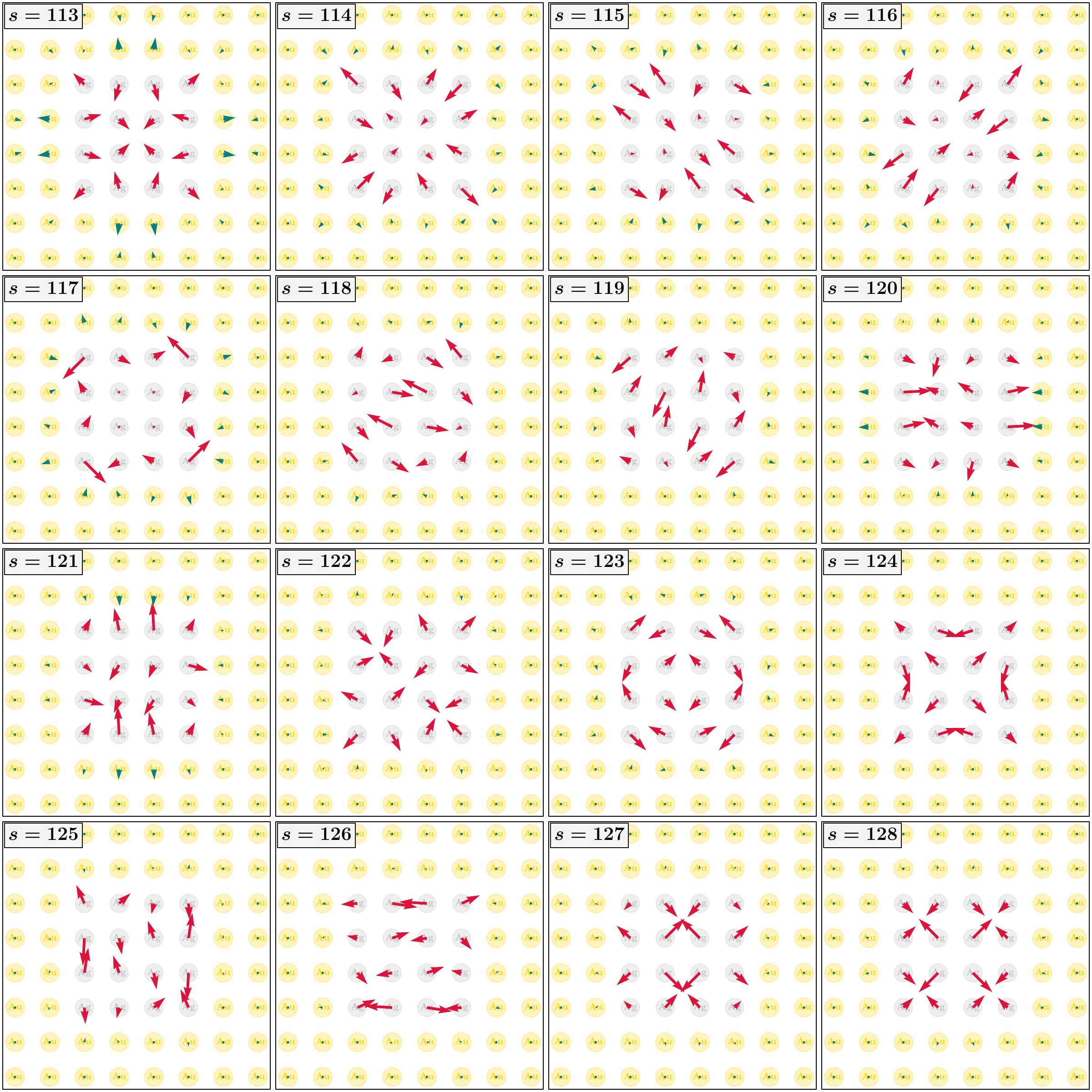}
    \caption{Polarization vectors of the highest 16 optical $\Gamma$ point phonon modes in the FCM model that we have considered. For each mode, the arrows  indicate the directions of vibrations of the different atoms, and their lengths are proportional to their relative amplitudes. Note that for all of these, essentially only the Ag atom vibrations are involved. }
    \label{fig:fig13}
\end{figure*}

The phonon calculations we have reported in this paper were carried out on a $24  \times 24$  $\mathbf{q}$-grid. Furthermore, as mentioned in section III, we assumed that only the first and second neighbour stretching spring constants, $K_1^{(s)}$ and $K_2^{(s)}$, are non zero, and used units where $K_1^{(s)} = 1.0$, $K_2^{(s)} = 0.5$,   $M^{(Au)} = 1.0$ and $M^{(Ag)} = 0.54$. The values of the phonon frequencies in real units can be obtained by multiplication with the appropriate value of $\sqrt{(K_1^{(s)}/M^{(Au)})}$\cite{real-units}. The phonon DOS (per supercell), $\mathcal{D}_{ph}(\omega) \equiv \mathcal{N}^{-1} \sum_{\mathbf{q}, s} \delta(\omega - \omega_{\mathbf{q} s})$, was evaluated by using a frequency grid with a spacing of 0.0125 and summing over the $ 2 \times 64 \times \mathcal{N}$ modes with the delta function above approximated  as  a normalised Gaussian function with a spread of 0.025.

In case of pure silver or gold\cite{Drexel-72, Lynn_et_al-73}, the phonon calculations are simplest when when the systems are viewed as pure-FCC lattices with one atom per unit cell, with wave vectors $\mathbf{q}_0$  ranging over the UBZ. Then we obtain 3 {\em acoustic} branches of phonons with labels $s_0 = 1, 2, 3$ as all have linear dispersions for small $|\mathbf{q}_0|$. This  follows from the fact that the basis labels $j$ and $j^\prime$ are now redundant, so that the dynamical matrix is just a  $3 \times 3$ matrix; and furthermore, because of the inversion symmetry of the pure-FCC lattice it can be re-expressed as
\bea
D_{(0)}^{\alpha \alpha^\prime} (\mathbf{q}_0)  & =  & \frac{1} { M } \sum_{\mathbf{R}_0}  \left[    e^{ i \mathbf{q}_0 \cdot \mathbf{R}_0  } \;  \kappa_0^{\alpha \alpha^\prime} (\mathbf{R}_0) \right] \, \equiv \, \frac{1} {M}  \tilde{\kappa}_0^{\alpha \alpha^\prime} (\mathbf{q}_0) \, ; \nonumber \\
\tilde{\kappa}_{(0)}^{\alpha \alpha^\prime} (\mathbf{q}_0) & = &  \sum_{\mathbf{R}_0}  \left[ 1 -   \cos ( \mathbf{q}_0 \cdot \mathbf{R}_0 )  \right] \;  K_0^{\alpha \alpha^\prime} (\mathbf{R}_0)  \nonumber \\ &=& \sum_{\mathbf{R}_0}   2   \sin^2  \left(  \frac {\mathbf{q}_0 \cdot \mathbf{R}_0 } { 2 } \right) \;  K_0^{\alpha \alpha^\prime} (\mathbf{R}_0)  ,
\label{pure_Au_Ag_K_to_dyn_matrix_eqn-SI}
\eea
where $M$ is $M^{(Au)}$ or $M^{(Ag)}$ as appropriate. We denote the corresponding phonon frequencies as $\omega^{(0)}_{\mathbf{q}_0 s_0}$ and the polarisation vector components as $\eta^{(0) \, \alpha}_{\mathbf{q}_0 s_0}$ . 

In the simplified axially symmetric model with just the two first-neighbour spring constants $K^{(s)}_1$ and $K^{(b)}_1$ being non zero, the sum over $\mathbf{R}_0$ in Eq. (\ref{pure_Au_Ag_K_to_dyn_matrix_eqn-SI}) gets constrained, in the pure-FCC case, to be over just the 12 first-neighbour vectors of the site at the origin: $$\{ \mathbf{R}_0^{(1)} \} = \{ a_0 (\pm \frac{1}{2}, \pm \frac{1}{2}, 0) ,   a_0 (0, \pm \frac{1}{2}, \pm \frac{1}{2}) ,  a_0 (\pm \frac{1}{2}, 0, \pm \frac{1}{2}) \} . $$  The corresponding  $3 \times 3  \; \mathbb{K}_0$  matrices  are hence given by
\begin{multline}
K_{(0)}^{\alpha \alpha^\prime} (\mathbf{R}^{(1)}_0)  =  2 K^{(s)}_1  \frac{ R_0^{(1) \alpha} \, R_0^{(1) \alpha^\prime}} {|\mathbf{R}_0^{(1)}|^2}  \\+ 2 K^{(b)}_1 \left[   \delta_{\alpha, \alpha^\prime} -   \frac{ R_0^{(1) \alpha} \, R_0^{(1) \alpha^\prime}} {|\mathbf{R}_0^{(1)}|^2}  \right] ,
\label{pure_FCC_force_const_eqn-SI}
\end{multline}
the same for both Au and Ag. Hence the phonon dispersions are essentially the same for both, except for an over all scale factor of $M^{-\frac{1}{2}}$.  In case of the square lattice with lattice constant $a_0$ and only in-plane displacements permitted, the sum over $\mathbf{R}_0$ in Eq. (\ref{pure_Au_Ag_K_to_dyn_matrix_eqn-SI}) is constrained to just the 4 first-neighbour sites $\{ \mathbf{R}_0^{(1)} \} = \{ a_0 (\pm 1, \pm 1) \} $, with the corresponding  $\mathbb{K}_0$  now being $2 \times 2 $  matrices, again given by Eq. (\ref{pure_FCC_force_const_eqn-SI}), leading to 2 acoustic branches.

If the calculations for pure Au or Ag are artificially done in the superlattice with $N_c$ sites per supercell, analogously to the case of the electronic structure, one will merely get {\em folded} versions of the above, via the mapping
\bea
(\mathbf{R}+ \mathbf{d}^0_j)  & \Leftrightarrow & \mathbf{R}_0 ; \; \nonumber \\   \kappa^{\alpha \alpha^\prime}_{j j^\prime } (\mathbf{R}-\mathbf{R}^\prime)  &\Leftrightarrow& \kappa_0^{\alpha \alpha^\prime} ( (\mathbf{R}+ \mathbf{d}^0_j) - (\mathbf{R}^\prime+ \mathbf{d}^0_{j^\prime}) ); \; \nonumber \\ s &\Leftrightarrow& (s_0, \mathbf{G}_\ell) ;  \nonumber \\
 (\mathbf{q} \, s) & \Leftrightarrow & ((\mathbf{q}+\mathbf{G}_\ell) \, s_0);  \; \nonumber \\ \omega_{\mathbf{q} s}  &\Leftrightarrow&  \omega^{(0)}_{(\mathbf{q} + \mathbf{G}_\ell) \, s_0} ; \; \nonumber \\ \eta^\alpha_{j; \mathbf{q} s} &\Leftrightarrow&   \frac {1} {\sqrt{N_c}} \, e^{ i \mathbf{G}_\ell \cdot \mathbf{d}^0_j} \, \eta^{(0) \alpha}_{ (\mathbf{q} + \mathbf{G}_\ell) \, s_0} \, ;
\label{pure_Au_Ag_omega_eta_eqn-SI}
\eea
where, as defined before, $\{ \mathbf{G}_\ell , \ell = 1, N_c \}$ are the $N_c$ smallest superlattice RLVs which span the UBZ. 

Fig. 12 shows these folded phonon bands for the pure Au S-SL system, these being the folded versions of the 2 acoustic branches one would have obtained for the pure square lattice as discussed above. The ones for pure Ag will be essentially the same as these, except for their frequencies getting enhanced by a factor of  $\sqrt{(M^{(Au)} / M^{(Ag)})} = 1.36$. 

However, in case of the superlattices with nano-clusters of  Ag in an Au matrix, the results are drastically different. Apart from the hybridisation and splitting of the folded Au phonon branches, as  discussed in the context of Fig. 7 in section III, one finds clear evidence of nearly localised higher frequency phonon modes comprised primarily of Ag atoms rattling inside the ``cages''  formed by the the heavier Gold  atoms. This followed from the partial DOS provided in Fig. 7, as was pointed out in section III.  Fig. 13 provides additional evidence illustrating this by depicting the polarisation vectors of the top 16 Phonon modes at the $\Gamma$ point, which are almost entirely comprised of vibrations of Ag atoms, and have very little participation from the Au atoms, except near the interface.

\section{Details regarding Electron-Phonon interaction effects in the nano-structured Superlattice}

We show in this appendix that, as stated in section IV, the dependence of the hopping amplitude in Eq. (\ref{H_hop_eqn}) and the long range Coulomb energy in Eq. (\ref{H_ce_eqn}) on the positions of the atoms leads to three characteristic consequences {\em due to the nano-structuring}.
\begin{enumerate}
\item {There is lattice relaxation leading to the displacement of atoms from $\mathbf{d}^0_j$ to new equilibrium positions $\mathbf{d}_j$ within the supercell;}
\item {additional contributions to the spring constants of the system arise, especially at the Au-Ag interface, leading generally to their stiffening; and}
\item {{\em new contributions} to the electron-phonon interactions arise, leading to their substantial enhancement, the focus of this paper. }
\end{enumerate}

We note that both the hopping and the Coulomb interaction Hamiltonians in our model depend on the atomic coordinates only via the set of distances $\{ d_{j \mathbf{R}; j^\prime \mathbf{R}^\prime} \}$ between  pairs of atoms. In the presence of the displacements  $ \{ \mathbf{u}_{j \mathbf{R}} \}$ defined above, we have, to second order in $\mathbf{u}$,

\bea
\delta d_{j \mathbf{R}; j^\prime \mathbf{R}^\prime}  & = &  \delta^{(1)} d_{j \mathbf{R}; j^\prime \mathbf{R}^\prime} + \delta^{(2)} d_{j \mathbf{R}; j^\prime \mathbf{R}^\prime} \, ; \; \nonumber \\ \delta^{(1)} d_{j \mathbf{R}; j^\prime \mathbf{R}^\prime} &\equiv& \left[ \frac{(\mathbf{d}_{j \mathbf{R}; j^\prime \mathbf{R}^\prime} \cdot \mathbf{u}_{j, \mathbf{R}; j^\prime, \mathbf{R}^\prime})} {d_{j \mathbf{R}; j^\prime \mathbf{R}^\prime}} \right] \, ; \nonumber \\
\delta^{(2)} d_{j \mathbf{R}; j^\prime \mathbf{R}^\prime} & \equiv &  \frac{1}{2}  \left[ \frac {|\mathbf{u}_{j, \mathbf{R}; j^\prime, \mathbf{R}^\prime}|^2}{d_{j \mathbf{R}; j^\prime \mathbf{R}^\prime}}  - \frac{(\mathbf{d}_{j \mathbf{R}; j^\prime \mathbf{R}^\prime} \cdot \mathbf{u}_{j, \mathbf{R}; j^\prime, \mathbf{R}^\prime})^2} {(d_{j \mathbf{R}; j^\prime \mathbf{R}^\prime})^3} \right] . \notag \\ & &
\label{delta_d_eqn-SI}
\eea
Hence we get, correct to second order in $\mathbf{u}$,
\begin{multline}
\delta \boldsymbol{H}_{hop}  =   \sum^{\quad \p}_{j, \mathbf{R}; j^\prime, \mathbf{R}^\prime ; \s}  \mathfrak{t} (d_{j \mathbf{R}; j^\prime \mathbf{R}^\prime})  \, \boldsymbol{c}_{j \mathbf{R} \s}^\dag \boldsymbol{c}_{j^\prime \mathbf{R}^\prime \s} \\ \left[ \frac {\delta d_{j \mathbf{R}; j^\prime \mathbf{R}^\prime}} {\xi_0} - \frac{1}{2}  \frac{(\mathbf{d}_{j \mathbf{R}; j^\prime \mathbf{R}^\prime} \cdot \mathbf{u}_{j, \mathbf{R}; j^\prime, \mathbf{R}^\prime})^2} { \xi_0^2 \, (d_{j \mathbf{R}; j^\prime \mathbf{R}^\prime})^2} \right]
\label{del_H_hop_eqn-SI}
\end{multline}
and
\begin{multline}
\delta \boldsymbol{H}_{ce}  =  - \frac {V_0}{2} \sum^{\quad \p}_{j, \mathbf{R}; j^\prime, \mathbf{R}^\prime} \,  \frac{(\boldsymbol{n}_{j \mathbf{R} } -1)  \,   (\boldsymbol{n}_{j^\prime \mathbf{R}^\prime} - 1)} {d_{j \mathbf{R}; j^\prime \mathbf{R}^\prime}}  \, \\ \left[ \frac {\delta d_{j \mathbf{R}; j^\prime \mathbf{R}^\prime}} {d_{j \mathbf{R}; j^\prime \mathbf{R}^\prime}} -  \frac{1}{3}  \frac{(\mathbf{d}_{j \mathbf{R}; j^\prime \mathbf{R}^\prime} \cdot \mathbf{u}_{j, \mathbf{R}; j^\prime, \mathbf{R}^\prime})^2} { (d_{j \mathbf{R}; j^\prime \mathbf{R}^\prime})^4} \right]
\label{del_H_ce_eqn-SI}
\end{multline}

\subsection{Force balance and basis relaxation}

Taking the expectation value of the the terms linear in $\mathbf{u}$ above with respect to the electronic ground state (within the Hartree approximation that we are restricting ourselves to) gives us, via the Hellman-Feynman theorem, the first order changes in the electronic energy, and hence the forces. We have,
\bea
& &\delta^{(1)} \langle  \boldsymbol{H}_{hop} \rangle  = \langle \delta^{(1)} \boldsymbol{H}_{hop} \rangle = \nonumber \\ & &  \sum^{\quad \p}_{\substack{j, \mathbf{R} \\ j^\prime, \mathbf{R}^\prime }}  \mathfrak{t} (d_{j \mathbf{R}; j^\prime \mathbf{R}^\prime})  \; C_{j \mathbf{R}; j^\prime \mathbf{R}^\prime} \, \left[ \frac{(\mathbf{d}_{j \mathbf{R}; j^\prime \mathbf{R}^\prime} \cdot \mathbf{u}_{j, \mathbf{R}; j^\prime, \mathbf{R}^\prime})} {\xi_0 \, d_{j \mathbf{R}; j^\prime \mathbf{R}^\prime}} \right] \, ; \notag \\ & &
\label{del1_E_hop_eqn-SI}
\eea
where
\bea
 C_{j \mathbf{R}; j^\prime \mathbf{R}^\prime} &\equiv&  \sum_\s \langle \boldsymbol{c}_{j \mathbf{R} \s}^\dag \boldsymbol{c}_{j^\prime \mathbf{R}^\prime \s} \rangle  \nonumber \\ & =& \frac {2} {\mathcal{N}} \sum_{\mathbf{k}, m} n^-_F(\varepsilon_{\mathbf{k} m}) \,\varphi^*_{j; \mathbf{k} m} \varphi_{j^\prime; \mathbf{k} m} \, e^{- i \mathbf{k} \cdot \mathbf{d}_{j \mathbf{R}; j^\prime \mathbf{R}^\prime}} \notag \\ & &
\label{C_jR_jpRp_eqn-SI}
\eea
Similarly,
\bea
\delta^{(1)} &\langle&  \boldsymbol{H}_{ce} \rangle  = \langle \delta^{(1)} \boldsymbol{H}_{ce} \rangle  = \nonumber \\ &-& \frac {V_0}{2} \sum^{\quad \p}_{j, \mathbf{R}; j^\prime, \mathbf{R}^\prime}  \frac{\delta n_j  \delta n_{j^\prime}} {d_{j \mathbf{R}; j^\prime \mathbf{R}^\prime}}   \left[ \frac {(\mathbf{d}_{j \mathbf{R}; j^\prime \mathbf{R}^\prime} \cdot \mathbf{u}_{j, \mathbf{R}; j^\prime, \mathbf{R}^\prime})} {(d_{j \mathbf{R}; j^\prime \mathbf{R}^\prime})^2}  \right] \notag \\ & &
\label{del1_E_ce_eqn-SI}
\eea

We note again that, when the Hartree self-consistent calculations for the Ag-Au superlattice are done with the atoms in the {\em unrelaxed} basis positions  $\{ \mathbf{d}^0_j \}$, $\mathbf{d}_{j \mathbf{R}; j^\prime \mathbf{R}^\prime}$ and $d_{j \mathbf{R}; j^\prime \mathbf{R}^\prime}$ in Eq.s (\ref{del1_E_hop_eqn-SI}) - (\ref{del1_E_ce_eqn-SI}) above need to be replaced by $\mathbf{d}^0_{j \mathbf{R}; j^\prime \mathbf{R}^\prime}$ and $d^0_{j \mathbf{R}; j^\prime \mathbf{R}^\prime}$ respectively.

In the case of pure Au or Ag, it is straightforward to see that both of these expectation values $\langle \delta^{(1)} \boldsymbol{H}_{hop} \rangle ^{(0)}$ and $\langle \delta^{(1)} \boldsymbol{H}_{ce} \rangle ^{(0)}$ vanish, consistent with the pure-lattice being in force balance characteristic of equilibrium. The vanishing of $\langle \delta^{(1)} \boldsymbol{H}_{ce} \rangle ^{(0)}$ is obvious because $\delta n_j = 0$  in this case. The vanishing of $\langle \delta^{(1)} \boldsymbol{H}_{hop} \rangle ^{(0)}$ is subtler; it arises from the fact that for any given $j \mathbf{R}$, $C^{(0)}_{j \mathbf{R}; j^\prime \mathbf{R}^\prime}$ in this case will be the same for all $j^\prime \mathbf{R}^\prime $ related by the point group symmetry of the pure-lattice {\em about} $j \mathbf{R}$. Hence the coefficient of $u^\alpha_{j \mathbf{R}}$ in Eq.~(\ref{del1_E_hop_eqn-SI}) involves sums of  $d^{0 \alpha}_{j \mathbf{R}; j^\prime \mathbf{R}^\prime}$ over these point group symmetry related sites, which vanishes.

However, in the case of our primary interest, i.e., supercells with Ag clusters embedded in a Au matrix, both the above expectation values are non-vanishing, signalling a force imbalance if the basis atoms remain at $\{ \mathbf{d}^0_j \}$. Force balance is restored by the {\em relaxation} of the basis via the displacement of  the basis atoms to $\{ \mathbf{d}_j \}$ such that the forces generated from  $\langle \delta^{(1)} \boldsymbol{H}_{hop} \rangle $ and $\langle \delta^{(1)} \boldsymbol{H}_{ce} \rangle$ are precisely cancelled by the harmonic forces generated due the displacements via the potential energy term of $H_{ph}$ in Eq.(\ref{phonon_Hamiltonian_eqn-SI}). It is easy to see using Eq.s~(\ref{del1_E_hop_eqn-SI}) and (\ref{del1_E_ce_eqn-SI}) that the displacements $\{(\mathbf{d}_j-\mathbf{d}^0_j)\}$ corresponding to the relaxation are determined by the (vector) equation
\begin{multline}
\sum_{j^\prime, \alpha^\prime}  \tilde{\kappa}^{\alpha \alpha^\prime}_{j j^\prime} (\mathbf{0}) \,  (d^{\alpha^\prime}_{j^\prime} - d^{0 \alpha^\prime}_{j^\prime}) \, = \,  \sum^{\quad \p}_{j^\prime,  \mathbf{R}^\prime} \bigg[  V_0 \frac{\delta n_j  \, \delta n_{j^\prime}} {(d_{j \mathbf{R}; j^\prime \mathbf{R}^\prime})^2} \\ - \mathfrak{t} (d_{j \mathbf{R}; j^\prime \mathbf{R}^\prime}) \frac{( C_{j \mathbf{R}; j^\prime \mathbf{R}^\prime} + C_{j^\prime \mathbf{R}^\prime ; j \mathbf{R}}) } {\xi_0}  \bigg] \,  \frac{d^{\alpha} _{j \mathbf{R}; j^\prime \mathbf{R}^\prime} } {d_{j \mathbf{R}; j^\prime \mathbf{R}^\prime}}  \, ,
\label{lat_relax_eqn-SI}
\end{multline}
where the matrix $\tilde{\kappa}$ was defined in Eq. (\ref{kappa_to_dyn_matrix_eqn-SI}). If the displacements due to the relaxation are small, one can replace $d_{j \mathbf{R}; j^\prime \mathbf{R}^\prime}$ and $d^{\alpha}_{j \mathbf{R}; j^\prime \mathbf{R}^\prime}$ on the RHS by $d^0_{j \mathbf{R}; j^\prime \mathbf{R}^\prime}$ and $d^{0 \alpha}_{j \mathbf{R}; j^\prime \mathbf{R}^\prime}$, whence the above equation can be solved by a simple matrix inversion. If not, one can iterate the procedure (i.e., change $d_{j \mathbf{R}; j^\prime \mathbf{R}^\prime}$ and $d^{\alpha}_{j \mathbf{R}; j^\prime \mathbf{R}^\prime}$ on the RHS iteratively based on the solution from the previous step) until convergence to self consistence, as long as the displacements are still small enough that {\em anharmonic} contributions to the LHS (i.e., anharmonic forces arising from $E_{core}$) are not significant. If these simplifications are not valid, the problem of finding the final relaxed structure gets more challenging, requiring some modelling of the anharmonic components of  $E_{core}$ (such as using an appropriate force field model) , and in addition, the task of minimising the Hartree energy plus the core energy with respect to the $DN_c$ variables  $\{ d^{\alpha}_j \}$. 

\subsection{New contributions to spring constants}

Next, we consider the terms in the electronic energy which are of second order in the displacements $\mathbf{u}$. Two contributions are readily obtained by taking the expectation values of the second order terms in $\delta \boldsymbol{H}_{hop}$ and $\delta \boldsymbol{H}_{ce}$ above. We get
\begin{multline}
\langle \delta^{(2)}  \boldsymbol{H}_{hop} \rangle =   \sum^{\quad \prime}_{j, \mathbf{R}; j^\prime, \mathbf{R}^\prime }  \mathfrak{t} (d_{j \mathbf{R}; j^\prime \mathbf{R}^\prime})  \, C_{j \mathbf{R} ; j^\prime \mathbf{R}^\prime} \bigg[ \frac {\delta^{(2)} d_{j \mathbf{R}; j^\prime \mathbf{R}^\prime}} {\xi_0}   \\ - \frac{1}{2}  \frac{(\mathbf{d}_{j \mathbf{R}; j^\prime \mathbf{R}^\prime} \cdot \mathbf{u}_{j, \mathbf{R}; j^\prime, \mathbf{R}^\prime})^2} { \xi_0^2 \, (d_{j \mathbf{R}; j^\prime \mathbf{R}^\prime})^2} \bigg] , 
\label{del2_H_hop_eqn-SI}
\end{multline}
with $ \delta^{(2)} d_{j \mathbf{R}; j^\prime \mathbf{R}^\prime}$ as specified in Eq. (\ref {delta_d_eqn-SI}). Similarly, 
\begin{multline}
\langle \delta^{(2)}  \boldsymbol{H}_{ce} \rangle =  - \frac {V_0}{2} \sum^{\quad \prime}_{j, \mathbf{R}; j^\prime, \mathbf{R}^\prime} \,  \frac{\delta n_j  \, \delta n_{j^\prime}} {d_{j \mathbf{R}; j^\prime \mathbf{R}^\prime}}  \, \bigg[ \frac {\delta^{(2)} d_{j \mathbf{R}; j^\prime \mathbf{R}^\prime}} {d_{j \mathbf{R}; j^\prime \mathbf{R}^\prime}} \\ -  \frac{1}{3}  \frac{(\mathbf{d}_{j \mathbf{R}; j^\prime \mathbf{R}^\prime} \cdot \mathbf{u}_{j, \mathbf{R}; j^\prime, \mathbf{R}^\prime})^2} { (d_{j \mathbf{R}; j^\prime \mathbf{R}^\prime})^4} \bigg]
\label{del2_H_ce_eqn-SI}
\end{multline}
There are two additional contributions to the electronic energy to second order in $\mathbf{u}$, which come from the changes in $\langle \delta^{(1)} \boldsymbol{H}_{hop} \rangle $ (cf.,  Eq. (\ref{del1_E_hop_eqn-SI}) induced via first order changes  in $C_{j \mathbf{R} ; j^\prime \mathbf{R}^\prime}$, and in $\langle \delta^{(1)} \boldsymbol{H}_{ce} \rangle$ (cf.,  Eq. (\ref{del1_E_ce_eqn-SI}) induced via first order changes in $(\delta n_j  \, \delta n_{j^\prime})$ ;  i.e.,  
\begin{multline}
\delta^{(2)} E_{hop}  =  \langle \delta^{(2)} \boldsymbol{H}_{hop} \rangle + \frac{1}{2}  \sum^{\quad \prime}_{j, \mathbf{R}; j^\prime, \mathbf{R}^\prime }  \mathfrak{t} (d_{j \mathbf{R}; j^\prime \mathbf{R}^\prime})  \, \times \\  \left[ \frac{\delta^{(1)} d_{j \mathbf{R}; j^\prime \mathbf{R}^\prime}} {\xi_0} \right] \;  \delta^{(1)} C_{j \mathbf{R}; j^\prime \mathbf{R}^\prime} \,;
\label{del2_E_hop_eqn-SI}
\end{multline}
and 
\begin{multline}
\delta^{(2)} E_{ce}   = \langle \delta^{(2)} \boldsymbol{H}_{ce} \rangle  -  \frac {V_0}{2} \sum^{\quad \prime}_{j, \mathbf{R}; j^\prime, \mathbf{R}^\prime} \, \, \left[ \frac {\delta^{(1)} d_{j \mathbf{R}; j^\prime \mathbf{R}^\prime}} {(d_{j \mathbf{R}; j^\prime \mathbf{R}^\prime})^2}  \right]  \;  \times \\ \left[ \delta n_{j^\prime} \, \delta^{(1)} (\delta n_j ) \,  \right] .
\label{del2_E_ce_eqn-SI}
\end{multline}

The first order changes  $\delta^{(1)} C_{j \mathbf{R}; j^\prime \mathbf{R}^\prime}$ and $ \delta^{(1)} (\delta n_j) $  induced by the first order hamiltonian  $\delta^{(1)} \boldsymbol{H} \equiv \delta^{(1)} \boldsymbol{H}_{hop} + \delta^{(1)}\boldsymbol{H}_{ce} $  are straightforwardly calculable using linear response theory. Within the Hartree approximation we are using, we can write, retaining only the {\em Hartree component} of  $\delta^{(1)}\boldsymbol{H}_{ce}$ ,
\begin{multline}
\delta^{(1)} \boldsymbol{H}_H  =   \sum^{\quad \prime}_{j_1, \mathbf{R}_1; j_1^\prime, \mathbf{R}_1^\prime} \bigg[ \frac{ \mathfrak{t} (d_{j_1 \mathbf{R}_1; j_1^\prime \mathbf{R}_1^\prime}) }{\xi_0} \, \sum_\s \boldsymbol{c}_{j_1^\prime \mathbf{R}_1^\prime \s}^\dag \boldsymbol{c}_{j_1 \mathbf{R}_1 \s}  \\ - \frac{ V_0  \, \delta n_{j_1^\prime} } {d^2_{j_1 \mathbf{R}_1; j_1^\prime \mathbf{R}_1^\prime}} \boldsymbol{n}_{j_1 \mathbf{R}_1 }  \bigg]   \bigg[ \frac {(\mathbf{d}_{j_1 \mathbf{R}_1; j_1^\prime \mathbf{R}_1^\prime} \cdot \mathbf{u}_{j_1, \mathbf{R}_1; j_1^\prime, \mathbf{R}_1^\prime})} {d_{j_1 \mathbf{R}_1; j_1^\prime \mathbf{R}_1^\prime}}  \bigg] 
\label{del1_H_el_ph_int_eqn-SI}
\end{multline}
where, for convenience, we have interchanged the order of the  $j_1 \mathbf{R}_1$ and $j_1^\prime \mathbf{R}_1^\prime$ labels on the electron operators in the hopping part.
From linear response theory, we get, to first order in $\delta^{(1)} \boldsymbol{H}_H$ ,
\begin{widetext}
    \begin{align}
    \delta^{(1)} C_{j \mathbf{R}; j^\prime \mathbf{R}^\prime}  = -  \frac {2} {\mathcal{N}^2 } \sum_{\substack{\mathbf{k}, m \\ \mathbf{k}^\prime, m^\prime }}  \, 
    \sum^{\quad \p}_{\substack{j_1, \mathbf{R}_1 \\ j_1^\prime, \mathbf{R}_1^\prime}}
    &\bigg[\frac{ \mathfrak{t} (d_{j_1 \mathbf{R}_1; j_1^\prime \mathbf{R}_1^\prime}) }{\xi_0} 
    \bigg( \varphi^*_{j_1^\prime; \mathbf{k}^\prime m^\prime} \varphi_{j_1; \mathbf{k} m}  e^{ i  \mathbf{k}^\prime \cdot \mathbf{d}_{j^\prime \mathbf{R}^\prime; j_1^\prime \mathbf{R}_1^\prime}  } \bigg) \nonumber 
    \\ &- \frac{ V_0  \, \delta n_{j_1^\prime} } {d^2_{j_1 \mathbf{R}_1; j_1^\prime \mathbf{R}_1^\prime}} \bigg( \varphi^*_{j_1; \mathbf{k}^\prime m^\prime} \varphi_{j_1; \mathbf{k} m} \, e^{i  \mathbf{k}^\prime \cdot \mathbf{d}_{j^\prime \mathbf{R}^\prime; j_1 \mathbf{R}_1} } \bigg) \bigg] 
    \left( \varphi^*_{j; \mathbf{k} m} \varphi_{j^\prime; \mathbf{k}^\prime m^\prime} \,  e^{ i \mathbf{k} \cdot \mathbf{d}_{j_1 \mathbf{R}_1; j \mathbf{R}}  }  \right) \nonumber \\ & \ \ \ \ \ \ \ \ \ \ \ \ \ \ \ \ \ \ \ \ \ \ \ \ \ \  \ \ \ \  \ \times
    \left(  \frac { n^-_F(\varepsilon_{\mathbf{k} m}) - n^-_F(\varepsilon_{\mathbf{k}^\prime m^\prime})} { \varepsilon_{\mathbf{k} m} - \varepsilon_{\mathbf{k}^\prime m^\prime} } \right) \left[ \frac {(\mathbf{d}_{j_1 \mathbf{R}_1; j_1^\prime \mathbf{R}_1^\prime} \cdot \mathbf{u}_{j_1, \mathbf{R}_1; j_1^\prime, \mathbf{R}_1^\prime})} {d_{j_1 \mathbf{R}_1; j_1^\prime \mathbf{R}_1^\prime}}  \right]  
\label{del1_C_jR_jpRp_eqn-SI}
    \end{align}
Similarly, 
\begin{align}
    \delta^{(1)} \delta n_{j }  = \delta^{(1)} n_j =  -  \frac {2} {\mathcal{N}^2 } \sum_{\substack{\mathbf{k}, m \\ \mathbf{k}^\prime, m^\prime }}  \, \sum^{\quad \p}_{\substack{j_1, \mathbf{R}_1 \\  j_1^\prime, \mathbf{R}_1^\prime}} &\bigg[ \frac{ \mathfrak{t} (d_{j_1 \mathbf{R}_1; j_1^\prime \mathbf{R}_1^\prime}) }{\xi_0} \,  \left( \varphi^*_{j_1^\prime; \mathbf{k}^\prime m^\prime} \varphi_{j_1; \mathbf{k} m}  e^{ i  \mathbf{k}^\prime \cdot \mathbf{d}_{j  \mathbf{R}; j_1^\prime \mathbf{R}_1^\prime}  } \right)    \nonumber \\  & -   \frac{ V_0  \, \delta n_{j_1^\prime} } {d^2_{j_1 \mathbf{R}_1; j_1^\prime \mathbf{R}_1^\prime}} \left( \varphi^*_{j_1; \mathbf{k}^\prime m^\prime} \varphi_{j_1; \mathbf{k} m} \, e^{i  \mathbf{k}^\prime \cdot \mathbf{d}_{j \mathbf{R}; j_1 \mathbf{R}_1} } \right) \, \bigg]  \left( \varphi^*_{j; \mathbf{k} m} \varphi_{j; \mathbf{k}^\prime m^\prime} \,  e^{ i \mathbf{k} \cdot \mathbf{d}_{j_1 \mathbf{R}_1; j \mathbf{R}}  }  \right) \nonumber \\  & \ \ \ \ \ \ \ \ \ \ \ \ \ \ \ \ \ \ \ \ \ \ \  \ \ \ \  \ \times \left(  \frac { n^-_F(\varepsilon_{\mathbf{k} m}) - n^-_F(\varepsilon_{\mathbf{k}^\prime m^\prime})} { \varepsilon_{\mathbf{k} m} - \varepsilon_{\mathbf{k}^\prime m^\prime} } \right) \left[ \frac {(\mathbf{d}_{j_1 \mathbf{R}_1; j_1^\prime \mathbf{R}_1^\prime} \cdot \mathbf{u}_{j_1, \mathbf{R}_1; j_1^\prime, \mathbf{R}_1^\prime})} {d_{j_1 \mathbf{R}_1; j_1^\prime \mathbf{R}_1^\prime}}  \right]
    \label{del1_delta_nj_eqn-SI}
\end{align}
\end{widetext}
Thus we have 4 new contributions to the spring constants of the system. The two contributions in $\delta^{(2)} E_{hop}$ would clearly be present even in pure gold, and basically represent the contributions to the inter-atomic forces arising from the energetics of the conduction electrons. The other two, coming from $\delta^{(2)} E_{hop}$ are present only when $\delta n_{j }$ are nonzero, i.e, arise due to the nano-structuring involving dis-similar metals and the resulting charge disproportionation. We note also that the first  term  in each of these, i.e., $\langle \delta^{(2)} \boldsymbol{H}_{hop} \rangle$ and $\langle \delta^{(2)} \boldsymbol{H}_{ce} \rangle$,  lead to spring constants which are of the axially symmetric form, where as the second term in each, which involve linear response coefficients, lead to spring constants  which are not of this form.

\subsection{New contributions to the {\em el-ph} interactions}

Finally, and most importantly, with respect to the relaxed (or unrelaxed) lattice,  $\delta \boldsymbol{H}_{hop}$ (cf., Eq. \ref{del_H_hop_eqn-SI})  and  $\delta \boldsymbol{H}_{ce}$ (cf., Eq. \ref{del_H_ce_eqn-SI}) are sources of electron-phonon interactions. The simplest of these, involving the scattering of a {\em single} electron together with the absorption or emission of a {\em single} phonon, comes from   $\delta^{(1)} \boldsymbol{H}_{hop}$ and  the {\em Hartree component} of  $ \delta^{(1)}\boldsymbol{H}_{ce} $ , and is exactly given by $\delta^{(1)} \boldsymbol{H}_H$ above, since $\mathbf{u}$ is linear in the phonon creation and destruction operators [cf., Eq.s (\ref{u_jR_to_zeta_qs_eqn-SI}) and (\ref{zeta_to_a_adag_eqn-SI})].  The Fock or exchange component of $\delta^{(1)} \boldsymbol{H}$ is also of this type, but amounts to a renormalisation of the hopping term in $\delta^{(1)} \boldsymbol{H}_H$.   The non-Hartree-Fock component of $\delta^{(1)} \boldsymbol{H}_{ce}$ corresponds to processes involving the scattering of two electrons together with the absorption or emission of a single phonon, and in this paper we will not discuss these or the other more complicated electron-phonon interactions, including those involving two phonon processes that arise from $\delta^{(2)} \boldsymbol{H}_{hop}$ and $\delta^{(2)} \boldsymbol{H}_{ce}$ .

Eq. (\ref{del1_H_el_ph_int_eqn-SI}) represents the electron-phonon interactions in terms of site localised electron and phonon displacement operators. it is more conventional, and useful, to express them in terms of the electron operators that diagonalise $\boldsymbol{H}_{el;H}$ (Eq. (\ref{Hartree_H_el_diag_form_eqn-SI})) and the phonon operators that diagonalise $\boldsymbol{H}_{ph}$ (Eq. (\ref{H_ph_diag_form_eqn-SI})). Using Eq.s (\ref{c_ctilde_transform_eqn-SI}), (\ref{u_jR_to_zeta_qs_eqn-SI}) and (\ref{zeta_to_a_adag_eqn-SI}), and simplifying the lattice sums involved using superlattice translational invariance which leads to superlattice (crystal) momentum conservation, it is straightforward to verify that {\em both} the terms in the electron-phonon interaction hamiltonian above can be reduced to the standard form: 
\begin{multline}
\boldsymbol{H}_{el-ph} = \frac {1} {\sqrt{\mathcal{N}}} \sum_{\mathbf{k}, m, m^\prime, \s} \sum_{\mathbf{q}, s}  \mathfrak{g}_{m^\prime m; s} (\mathbf{k}, \mathbf{q}) \, \times 
    \\ \tilde{\boldsymbol{c}}^\dag_{(\mathbf{k}+\mathbf{q})  m^\prime \s} \tilde{\boldsymbol{c}}_{\mathbf{k} m \s} \,(\boldsymbol{a}_{\mathbf{q} s} + \boldsymbol{a}^\dag_{-\mathbf{q} s}) ,
\label{el_ph_ham_std_form_eqn-SI}
\end{multline}

where the electron-phonon coupling matrix element $\mathfrak{g}_{m^\prime m; s} (\mathbf{k}, \mathbf{q})$ is the sum of two distinct contributions, as asserted in the section IV, namely 
\begin{widetext}
\beq
\mathfrak{g}^{(hop)}_{m^\prime m; s} (\mathbf{k}, \mathbf{q}) =  \sum^{\quad \p}_{j, j^\prime, \mathbf{R}^\prime }  \frac { \mathfrak{t} (d_{j \mathbf{R}; j^\prime \mathbf{R}^\prime}) \,  } {\xi_0 } \, \left( \varphi^*_{j^\prime; (\mathbf{k}+\mathbf{q}) m^\prime} \varphi_{j; \mathbf{k} m} e^{i (\mathbf{k} + \mathbf{q}) \cdot \mathbf{d}_{j \mathbf{R}; j^\prime \mathbf{R}^\prime}} \right) \, \Upsilon_{j  j^\prime;  \mathbf{q} s} (\mathbf{R} - \mathbf{R}^\prime) \, ,
\label{el_ph_coupling_hop_eqn-SI}
\eeq
and 
\beq
\mathfrak{g}^{(ce)}_{m^\prime m; s} (\mathbf{k}, \mathbf{q}) = -  \sum^{\quad \p}_{j, j^\prime, \mathbf{R}^\prime }  \frac { V_0 \delta n_{j^\prime} \,  } { d^2_{j \mathbf{R}; j^\prime \mathbf{R}^\prime} } \, \left( \varphi^*_{j; (\mathbf{k}+\mathbf{q}) m^\prime} \varphi_{j; \mathbf{k} m} \right) \, \Upsilon_{j  j^\prime;  \mathbf{q} s} (\mathbf{R} - \mathbf{R}^\prime);
\label{el_ph_coupling_ce_eqn-SI}
\eeq
where we have set 
\beq
\Upsilon_{j  j^\prime;  \mathbf{q} s} (\mathbf{R} - \mathbf{R}^\prime)  \equiv  \sqrt{ \frac{\hbar}{2 \omega_{\mathbf{q} s}} } \sum_\alpha \frac {d^\alpha_{j \mathbf{R}; j^\prime \mathbf{R}^\prime}} {d_{j \mathbf{R}; j^\prime \mathbf{R}^\prime}} \left[ \frac { \eta^\alpha_{j; \mathbf{q} s}  } { \sqrt{M_j} }  -  \frac { \eta^\alpha_{j^\prime; \mathbf{q} s}   } { \sqrt{M_{j^\prime}} } e^{- i \mathbf{q} \cdot \mathbf{d}_{j \mathbf{R}; j^\prime \mathbf{R}^\prime}} \right] .
\label{Upsilon_defn_eqn-SI}
\eeq
\end{widetext}
Once again, we note that the superlattice translation invariance ensures that the sums over $\mathbf{R}^\prime$ in Eq.s (\ref{el_ph_coupling_hop_eqn-SI}) and (\ref{el_ph_coupling_ce_eqn-SI})  lead to results that are independent of $\mathbf{R}$, whence we can choose $\mathbf{R} = \mathbf{0}$ without loss of generality. Furthermore, we note that $\Upsilon$ has dimensions of length which ensures that both the above contributions to $\mathfrak{g}$ have dimensions of energy. 

In case of pure Au or Ag, as we have discussed earlier,  $\delta n_{j^\prime} = 0$, and the labels $j, j^\prime, m, m^\prime$  are redundant, and we can also set $\varphi = 1$,  if we use the pure lattice vectors $\mathbf{R}_0$ and wave-vectors $\mathbf{k}_0, \mathbf{q}_0$ within the UBZ in all the expressions. Accordingly, the only contribution to 
$\mathfrak{g}$ is from the hopping term, and  because of our simplifying assumptions that $\mathfrak{t}$ and the force constants are the same for both, we have 
\begin{multline}
\mathfrak{g}^{(0)}_{s_0} (\mathbf{k}_0, \mathbf{q}_0) =  \sum^{\quad \p}_{\mathbf{R}^\prime_0 }  \frac { \mathfrak{t} (|\mathbf{R}_0 - \mathbf{R}^\prime_0|)} {\xi_0 } \, \left( e^{i (\mathbf{k}_0 + \mathbf{q}_0) \cdot (\mathbf{R}_0 - \mathbf{R}^\prime_0)} \right) \times \, \\ \Upsilon^{(0)}_{\mathbf{q}_0 s_0} (\mathbf{R}_0 - \mathbf{R}^\prime_0) \, 
\label{pure_Au_Ag_el_ph_coupling_eqn-SI}
\end{multline}
with 
\begin{multline}
\Upsilon^{(0)}_{\mathbf{q}_0 s_0} (\mathbf{R}_0 - \mathbf{R}^\prime_0)  \equiv  \sqrt{ \frac {\hbar} {2 M \omega^{(0)}_{\mathbf{q}_0 s_0} } }  \sum_\alpha \frac { (\mathbf{R}_0 - \mathbf{R}^\prime_0)^\alpha \, \eta^{(0) \alpha}_{\mathbf{q}_0 s_0} } {|\mathbf{R}_0 - \mathbf{R}^\prime_0|}  \times   \\ \left( 1 -  e^{- i \mathbf{q}_0 \cdot (\mathbf{R}_0 - \mathbf{R}^\prime_0)} \right) .
\label{Upsilon_pure_Au_Ag_eqn-SI}
\end{multline}
where $M$ is $M^{(Au)}$ or $M^{(Ag)}$ as appropriate. Within the framework of the simplifying assumptions we have made, as mentioned earlier, $\omega^{(0)}_{\mathbf{q}_0 s_0}$ scales as $M^{-\frac{1}{2}}$, hence the electron-phonon coupling matrix elements of Au and Ag are the same except for an overall multiplication factor of $M^{-1/4}$.  Using the mapping in Eq.s (\ref{pure_Au_Ag_eps_vphi_eqn-SI}) and  (\ref{pure_Au_Ag_omega_eta_eqn-SI}) it is straightforward to verify that essentially the same results for $\mathfrak{g}^{(0)}$ are recovered even when the pure Au or pure Ag calculations are done using the superlattice rather than the pure-lattice. Specifically, one finds (with $ m \Leftrightarrow \mathbf{G}_m; \; m^\prime \Leftrightarrow \mathbf{G}_{m^\prime}; \; s \Leftrightarrow s_0, \mathbf{G}_\ell $ as before)
\beq
\mathfrak{g}_{m^\prime m; s} (\mathbf{k}, \mathbf{q}) = \frac {\delta_{ (\mathbf{G}_m + \mathbf{G}_\ell) \, , \mathbf{G}_{m^\prime} }} {\sqrt{N_c}} \;  \;  \mathfrak{g}^{(0)}_{s_0} ( (\mathbf{k} + \mathbf{G}_m), (\mathbf{q} + \mathbf{G}_\ell) ) 
\label{pure_Au_Ag_superlattice_el_ph_coupling_eqn-SI}
\eeq

The electron-phonon coupling matrix elements $\mathfrak{g}$ discussed above constitute much too detailed measures  of the electron phonon interaction, and furthermore, they are not directly observable. It is therefore more convenient to use averaged measures that relate more directly to observable properties. We discuss two such commonly used sets of measures below, connected with superconductivity and resistivity respectively. Both were presented and discussed in section IV; here we re-discuss them at greater length, including providing details regarding our calculations.

The first set is connected with the superconductivity that results from the effective retarded attraction between electrons induced by the electron-phonon interaction.  Within the ``strong-coupling Migdal-Eliashberg theory'' of superconductivity \cite{McMillan-68, Allen_Dynes-75} , the crucial measure involved is the ``$\alpha^2 F \,$'' spectral function, given within the ``double delta approximation'' by 
\begin{multline}
\alpha^2F (\omega) \equiv  \frac {1} {(\mathcal{N})^2 \mathcal{D}(0)}  \sum_{\mathbf{k}, m, m^\prime } \sum_{\mathbf{q}, s} \lvert \mathfrak{g}_{m^\prime m; s} (\mathbf{k}, \mathbf{q})\rvert^2 \times \\  \delta (\varepsilon_{\mathbf{k} m} ) \; \delta (\varepsilon_{(\mathbf{k}+\mathbf{q}) \, m^\prime} ) \; \delta (\hbar \omega - \hbar \omega_{\mathbf{q} s} ) .
\label{alpha_squared_F_eqn-SI}
\end{multline}
Here $ \mathcal{D}(0) $ is the one-electron DOS (per spin per supercell) at the chemical potential defined and discussed earlier, in section II [cf., Eq. (\ref{dos_tr_dos_eqn}) and Fig. 6]. From this, the dimensionless effective electron-phonon coupling strength relevant for superconductivity  $\lambda^{(\mathcal{S})}$ and the effective log-averaged phonon frequency $\omega_{log}$  are calculated as
\bea
 \lambda^{(\mathcal{S})} &\equiv& 2 \int_0^\infty \, d\omega \; \frac {\alpha^2F (\omega)} {\omega} \,  \; \nonumber \\ \omega_{log} &\equiv& \exp { \left( \frac {2} {\lambda^{(\mathcal{S})}} \int_0^\infty \, d\omega \; \frac {\alpha^2F (\omega)} {\omega} \; \log {(\omega)} \right) }
\label{lambda_omega_log_eqn-SI}
\eea
The superconducting transition temperature $T_c$ within the strong coupling theory is then calculable using the McMillan-Allen-Dynes formula  \cite{McMillan-68, Allen_Dynes-75} which can be written as
\beq
 T_c =   \frac { \hbar \omega_{log} } { 1.2 k_B } \exp { \left(  - \frac {1.04 (1 + \lambda^{(\mathcal{S})})/ ( 1 + 0.62 \lambda^{(\mathcal{S})})} { \lambda^{(\mathcal{S})} / ( 1 + 0.62 \lambda^{(\mathcal{S})}) - \mu^*  } \right) }
\label{Tc_eqn-SI}
\eeq
Here $\mu^*$ is a dimensionless ``Coulomb pseudo-potential'' which is a measure of the screened Coulomb interactions averaged over the Fermi surface taking into account the effects of retardation via a renormalisation (from the plasma frequency scale down to the phonon frequency scale) \cite{Morel_Anderson-62}. It is rarely calculated, and is instead often taken as an adjustable parameter. 
\begin{figure}
    \centering
    \includegraphics[width=0.48\textwidth]{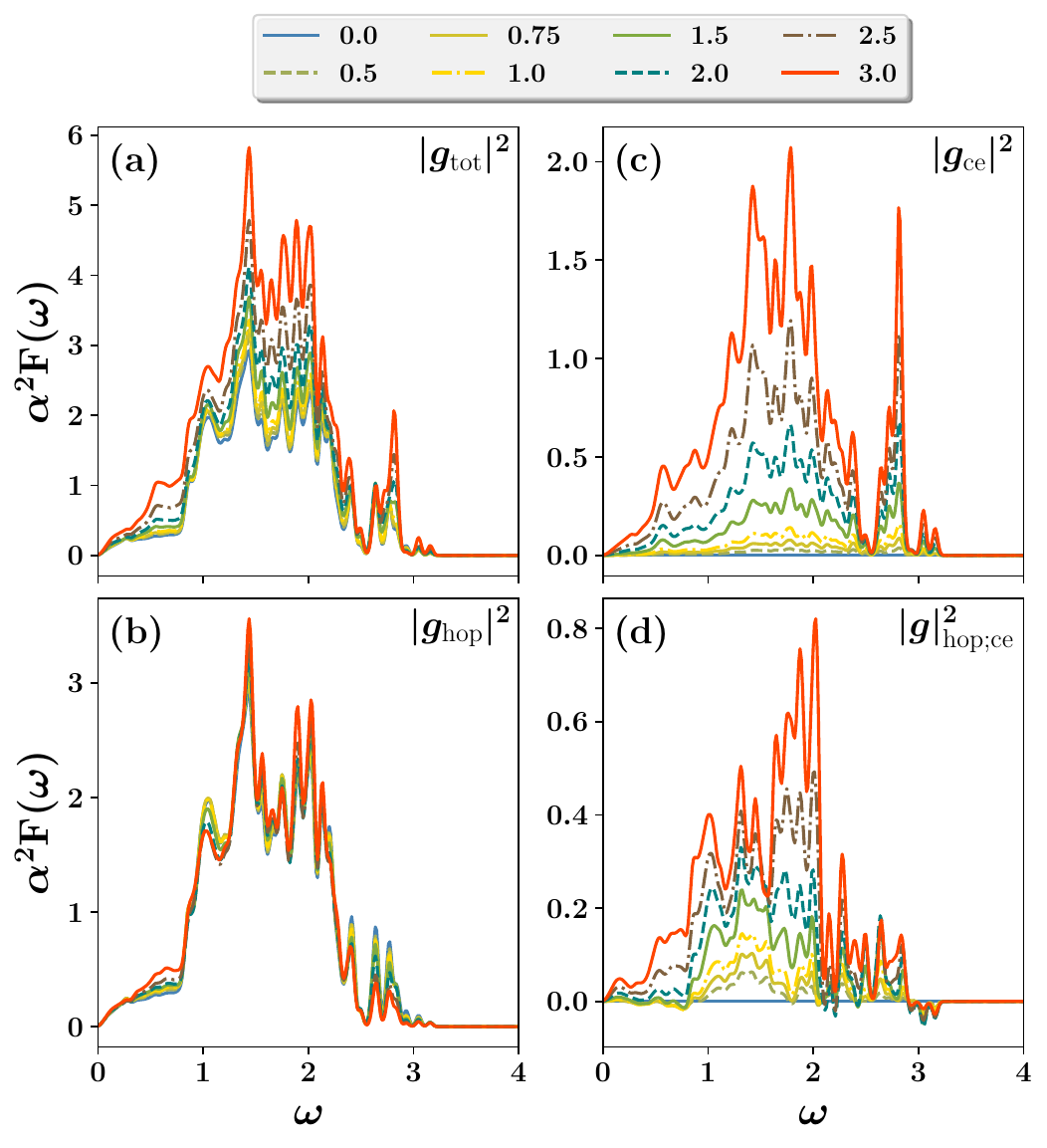}
    \caption { $\alpha^2 {F}(\omega)$ for various values of $\epsilon_0$, distinguishable by the colours indicated (in eV) at the top, and for $V_0/d_1 = 2.0$ eV. The other parameters used are $U_0=2.0$ eV,  $\xi_0 = 1.0 d_1$ and $d_c = 8 d_1$. While panel (a) (label: $\lvert g_{\text{tot}} \rvert^2 $) shows the total $\alpha^2 {F}(\omega)$,  the other panels show the different components that contribute to it,  exactly as in the case of Fig. 8 in section IV - see caption of Fig.8 for details. }
    \label{fig:fig14}
\end{figure}

Our calculated results for the different components of  $\alpha^2 F \,$ for various values of $\epsilon_0$, the work-function or local potential mismatch, for $U_0 = 2.0$ eV, $V_0/d_1 = 1.0$ eV, $\xi_0 = 1.0 d_1$ and $d_c = 8 d_1$ were presented and discussed in Sec.  IV of the paper (cf., Fig. 8). Fig.s 14, 15 and 16 show similar data for larger $V_0/d_1$ values, of 2, 3, and 4 eV respectively. The trends mentioned in Sec. IV,  of the rapidly increasing enhancement of  $\alpha^2 F $ with increasing values of $V_0/d_1$ and $\epsilon_0$, arising from $|{\mathfrak{g}}^{ce}|^2$ and with increasing weight in the nearly localised modes of the Ag atoms,  is clearly evident. All of the features seen in these figures are completely in line with the picture we have presented, of the vibrational modes of Ag nano-clusters coupling strongly via Coulomb interactions to the interface-dipoles that form at the Ag-Au interfaces. The calculations required for the plots in these  figures were done using a  $ 24 \times 24 $ grids within the SBZ of the superlattice for both $\mathbf{k}$ and $\mathbf{q}$, and over a discrete $\omega$ grid with spacing $0.0125$. The 3 delta functions in Eq. \ref{alpha_squared_F_eqn-SI} were approximated by normalised Gaussian functions with a width of 0.025 in the appropriate units. 

\begin{figure}
    \centering
    \includegraphics[width=0.48\textwidth]{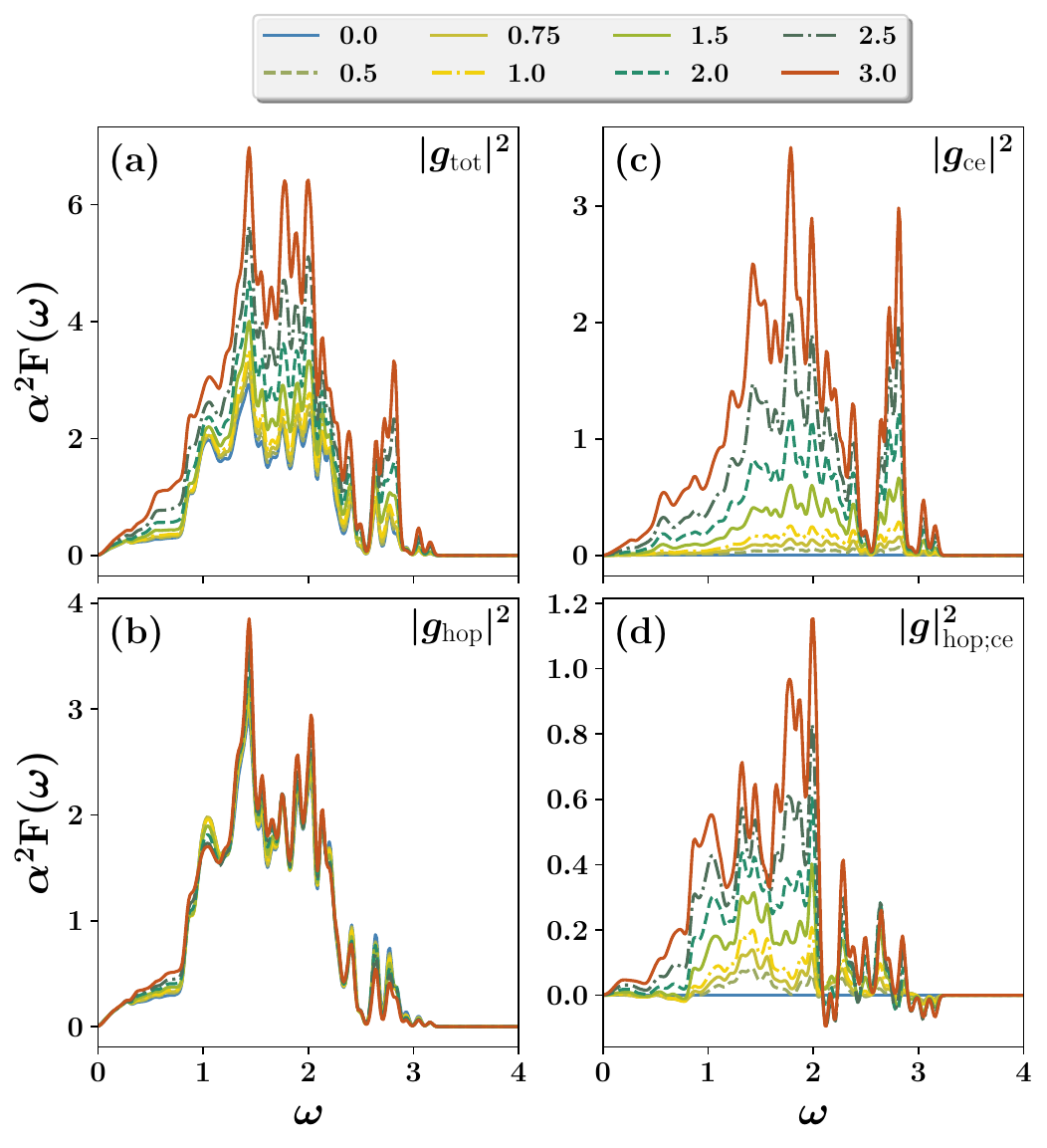}
    \caption { $\alpha^2 {F}(\omega)$ for various values of $\epsilon_0$, distinguishable by the colours indicated (in eV) at the top, and for $V_0/d_1 = 3.0$ eV. The other parameters used are $U_0=2.0$ eV,  $\xi_0 = 1.0 d_1$ and $d_c = 8 d_1$. While panel (a) (label: $\lvert g_{\text{tot}} \rvert^2 $) shows the total $\alpha^2 {F}(\omega)$,  the other panels show the different components that contribute to it,  exactly as in the case of Fig. 8 in section IV - see caption of Fig.8 for details. }
    \label{fig:fig15}
\end{figure}

\begin{figure}
    \centering
    \includegraphics[width=0.48\textwidth]{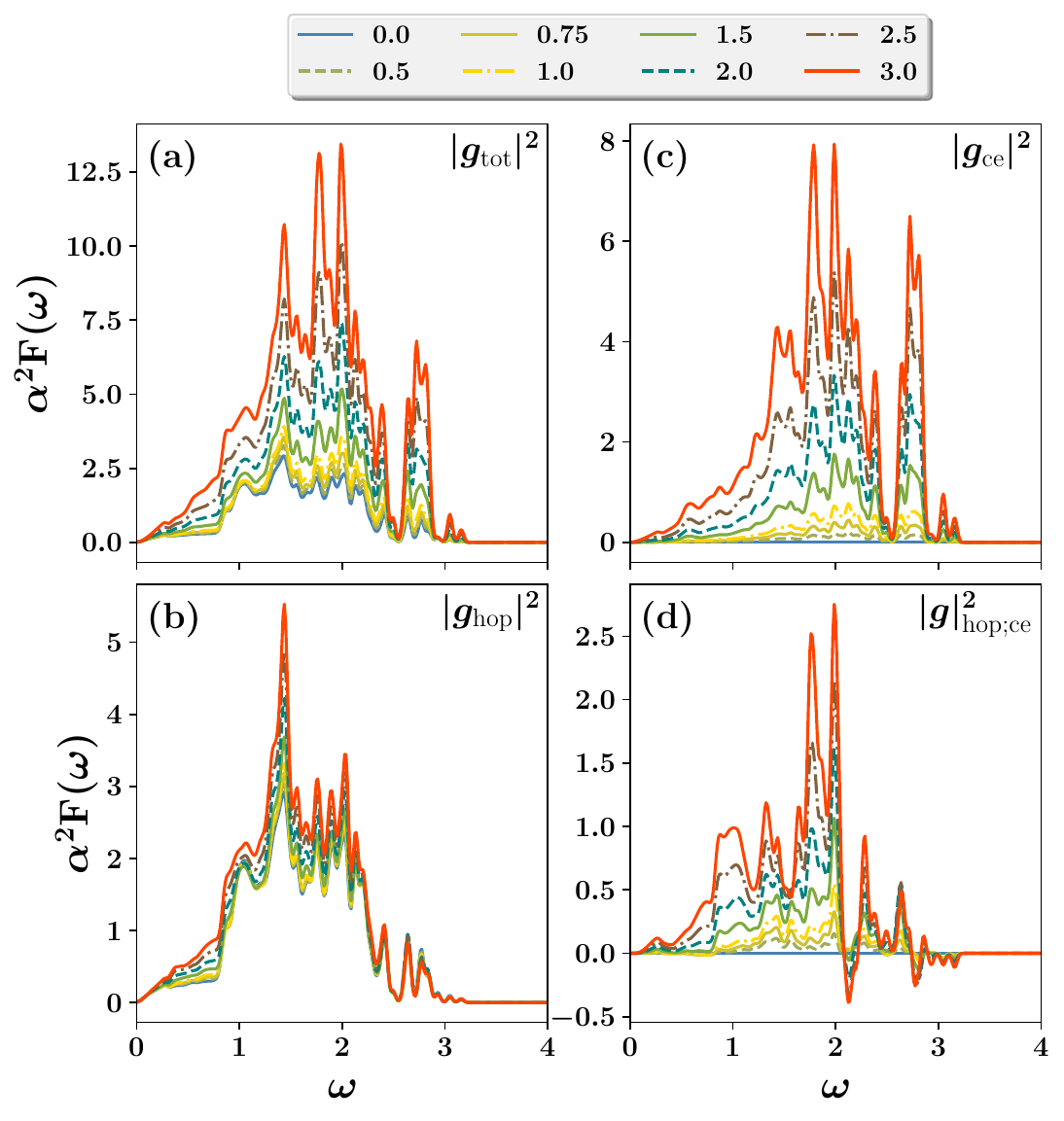}
    \caption { $\alpha^2 {F}(\omega)$ for various values of $\epsilon_0$, distinguishable by the colours indicated (in eV) at the top, and for $V_0/d_1 = 4.0$ eV. The other parameters used are $U_0=2.0$ eV,  $\xi_0 = 1.0 d_1$ and $d_c = 8 d_1$. While panel (a) (label: $\lvert g_{\text{tot}} \rvert^2 $) shows the total $\alpha^2 {F}(\omega)$,  the other panels show the different components that contribute to it,  exactly as in the case of Fig. 8 in section IV - see caption of Fig.8 for details. }
    \label{fig:fig16}
\end{figure}

Fig. 9 in section IV showed the total as well as the separate components of the contributions to  $\lambda^{(\mathcal{S})} / \lambda^{(\mathcal{S})}_0$ for a range of values of $\epsilon_0$ as well as of  $V_0/d_1$, plotted as a function of $\delta n_{Au}$, the {\em average} electron occupancy excess per atom on the Au sites. 
$\lambda^{(\mathcal{S})}$ for each parameter set was computed from the corresponding  
$\alpha^2 F \,$ data and implementing the first of the integrals in Eq. \ref{lambda_omega_log_eqn-SI}. $\lambda^{(\mathcal{S})}_0$ was obtained from similar calculations carried out  for pure Au,  and therefore with contributions only from 
$\mathfrak{g}^{(hop)}$. Fig.17 shows plots of $\omega_{log}$ versus  $\delta n_{Au}$ again obtained from the $\alpha^2 F \,$ data by implementing the second of the integrals in Eq. \ref{lambda_omega_log_eqn-SI}. All the integrals over $\omega$ were carried out using Simpson's rule and the data over the discrete $\omega$ grid mentioned above. Note that unlike $\lambda^{(\mathcal{S})}$, which increases monotonically and rapidly with increasing values of both  $V_0/d_1$ and $\delta n_{Au}$, $\omega_{log}$ has a more complicated dependence; while it does increase with $V_0/d_1$ for a fixed $\delta n_{Au}$, for a fixed $V_0/d_1$ it increases with increasing $\delta n_{Au}$ initially, but for larger values of the latter, it starts decreasing because $\lambda^{(\mathcal{S})}$, the denominator of the exponent in the expression for $\omega_{log}$ in Eq. \ref{lambda_omega_log_eqn-SI} increases faster than the integral over $\omega$ in its numerator. 

\begin{figure}
    \centering
    \includegraphics[width=0.48\textwidth]{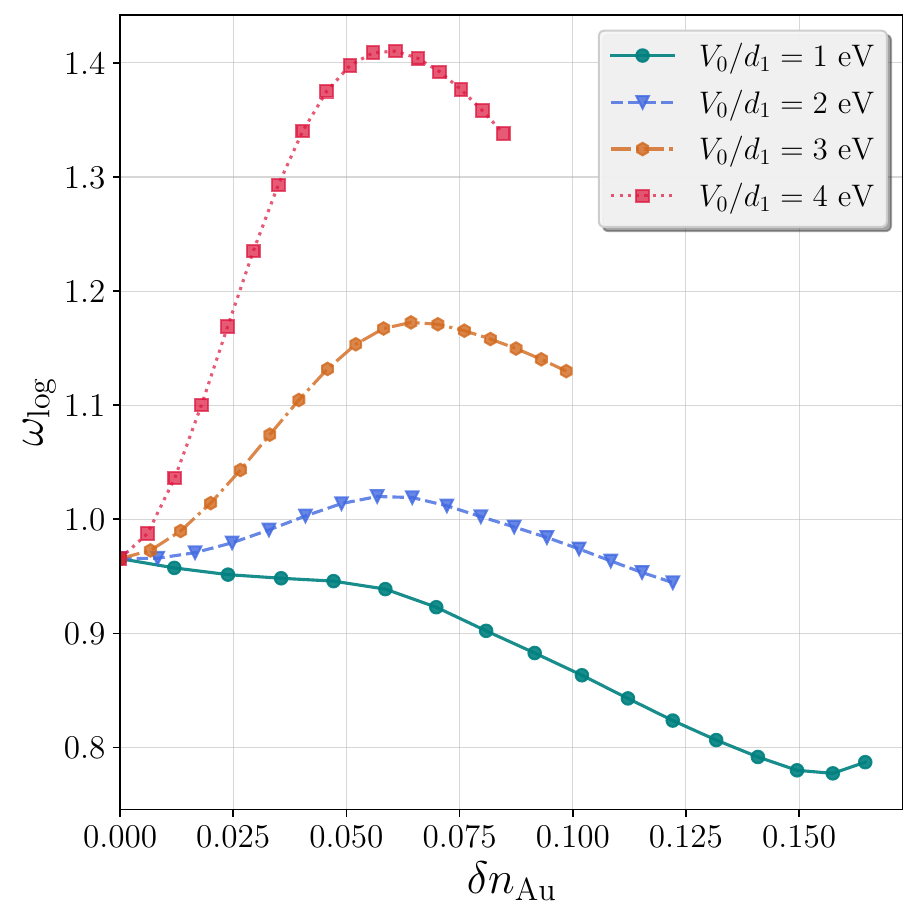}
    \caption {  $\omega_{log}$ computed from the second integral in Eq. \ref{lambda_omega_log_eqn-SI} for a range of values of $\epsilon_0$ as well as of  $V_0/d_1$, plotted versus $\delta n_{Au}$ .The other parameters used are $U_0=2.0$ eV,  $\xi_0 = 1.0 d_1$ and $d_c = 8 d_1$ as earlier.  }
    \label{fig:fig17}
\end{figure}

It is instructive to examine the enhancements in the superconducting $T_c$ that the above discussed enhancements in $\omega_{log}$ and $\lambda^{(\mathcal{S})} / \lambda^{(\mathcal{S})}_0$ can lead to. For this purpose, we note that for a reasonable choice of $K_1^{(s)}$, such that the maximum phonon frequency for the pure square lattice matches the maximal experimentally observed phonon frequency of Au, the optimal $\omega_{log}$ values in dimensionless units, in the range from 1.2  - 1.4 (cf., Fig. 17), would correspond to the range 107 - 125 K in temperature units\cite{real-units}. Furthermore, given that pure Au is known not to show superconductivity down to the lowest temperatures for which it has been checked, it is reasonable to assume that its Coulomb pseudo-potential $\mu^*$ is big enough to cause the denominator of the exponent in Eq. \ref{Tc_eqn-SI} to vanish. Then, using the estimates from the literature \cite{Allen-2000,AG_el_ph_enh-2024} that $\lambda^{(\mathcal{S})}_0 \simeq 0.15$ for Au, we  obtain $\mu^*  \simeq 0.137$. Using this value for $\mu^*$, and the above mentioned ranges of $\omega_{log}$, we obtain $T_c$ values in the range 21.7 - 25.4 K for $\lambda^{(\mathcal{S})} = 3.0$ (corresponding to   $\lambda^{(\mathcal{S})} / \lambda^{(\mathcal{S})}_0 = 20$), and in the range 25.4 -  29.7 K for $\lambda^{(\mathcal{S})} = 4.5$ (corresponding to  $\lambda^{(\mathcal{S})} / \lambda^{(\mathcal{S})}_0 = 30$). 

The above numbers represent relatively modest superconducting transition temperatures, nowhere near room temperature, despite the relatively large values of  $\lambda^{(\mathcal{S})}$. This is because the argument of the exponential in Eq. \ref{Tc_eqn-SI} is a monotonic increasing function of (increasing) 
$\lambda^{(\mathcal{S})}$ that saturates at [$ - 1.6774 / (1.6129 - \mu^*)] = -1.1367$ as  
$\lambda^{(\mathcal{S})} \rightarrow \infty$. Hence we get $T_c \leq 0.2674 (\hbar \omega_{log})/k_B $. For $ \omega_{log} = $ 125 K, this bound yields $T_c \leq $ 33.4 K \cite{Tc-bound}. Needless to say, the validity of the Migdal-Eliashberg theory for such large values of $\lambda^{(\mathcal{S})}$ is debatable. Nevertheless, the above discussion underscores the importance of increased values of $\omega_{log}$ for achieving higher-$T_c$ superconductivity. In this context, the changes in the spring constants arising from the relaxation effects that we have discussed earlier in this paper, especially the large relaxations connected with bipolaron formation and the consequent large stiffening of the interfacial Au-Ag spring constants,  can lead to significant enhancements in $\omega_{log}$. This effect likely plays a crucial role in the ``room temperature superconductivity'' reported in refs. \cite {Au_Ag_unpub1, Au_Ag_unpub2, Saha_et_al-2022}. 

\subsection{Resistivity enhancement due to enhanced  {\em el-ph} interactions}

In this subsection we discuss the enhancement of the DC resistivity of our model system that results from the enhanced electron-phonon interactions in greater detail than our presentation in section IV. For the regime of temperatures of the order of or larger than typical phonon frequencies which is of our interest,  within the framework of the Boltzmann transport theory treated in the relaxation-time approximation, or from the Kubo formula and diagrammatic perturbation theory, the DC conductivity is determined by the equation
\bea
\sigma^{\alpha \,  \alpha^\prime} =   \frac {e^2} {\mathcal{V}}  \sum_{\mathbf{k}, m}  v^\alpha_{\mathbf{k} m} &v^{\alpha^\prime}_{\mathbf{k} m}&  \left( - \frac { \del n^-_F( \varepsilon_{\mathbf{k}  m} ) } {\del \varepsilon_{\mathbf{k} m} } \right) \; \tau^{(tr)}_{\mathbf{k} m} \, ; \; \nonumber \\ v^\alpha_{\mathbf{k} m} &\equiv& \left( \frac {1}{\hbar} \frac { \del \varepsilon_{\mathbf{k}  m}  } { \del  k^\alpha } \right).
\label{DC_conductivity_eqn-SI}
\eea
Here $v^\alpha_{\mathbf{k} m} $ is the $\alpha^{th}$ component of the velocity of the electron in the band  $(\mathbf{k} m)$ that we have discussed earlier,  and $\tau^{(tr)}_{\mathbf{k} m}$ is the {\em transport lifetime} arising from the electron-phonon interaction, given (to leading order in $\mathfrak{g}$ and in $ \hbar \omega_{\mathbf{q} s} / (k_B T )$ )  by 
\begin{widetext}
\bea
\frac{1} {\tau^{(tr)}_{\mathbf{k} m} } & =&  \frac {2 \pi k_B T} {\hbar} \,  \;  \frac {1} {\mathcal{N}}  \sum_{m^\prime} \sum_{\mathbf{q}, s} |\mathfrak{g}_{m^\prime m; s} (\mathbf{k}, \mathbf{q})|^2  \; n^-_F (\varepsilon_{\mathbf{k} m} ) \; (1- n^-_F (\varepsilon_{(\mathbf{k}+\mathbf{q}) \, m^\prime} ) )  \nonumber \\ & & \ \ \ \ \ \ \times   \frac { [ \delta ( \varepsilon_{(\mathbf{k}+\mathbf{q}) \, m^\prime} - \varepsilon_{\mathbf{k} m} - \hbar \omega_{\mathbf{q} s} ) + \delta ( \varepsilon_{(\mathbf{k}+\mathbf{q}) \, m^\prime} - \varepsilon_{\mathbf{k} m} + \hbar \omega_{-\mathbf{q} s} ) ] } { \hbar \omega_{\mathbf{q} s} } \left( 1 - \frac {\mathbf{v}_{\mathbf{k} m} \cdot \mathbf{v}_{(\mathbf{k}+\mathbf{q}) \, m^\prime}} {|\mathbf{v}_{\mathbf{k} m}| |\mathbf{v}_{(\mathbf{k}+\mathbf{q}) \, m^\prime}|} \right)  
\label{tau_k_m_eqn-SI}
\eea
\end{widetext}
The two delta functions in the above expression correspond respectively to phonon absorption and phonon emission processes leading to the scattering of an electron from the Bloch state $\mathbf{k} m$ to $(\mathbf{k}+\mathbf{q}) \, m^\prime$, and the last factor (which would be not present in the otherwise identical expression for the inverse of the {\em single particle lifetime}) is connected with the fact that forward scattering does not degrade currents and hence does not contribute to the transport scattering rate.

The evaluation of the transport lifetime and the conductivity is a little more involved than that of $\alpha^2F$ and $\lambda^{(\mathcal{S})}$. For the purposes of this paper, we confine ourselves to an approximate evaluation of $\sigma$, the average diagonal component of the conductivity tensor, as follows. We recall that for temperatures  much less than the typical electronic energy scales, $ ( - \del n^-_F( \varepsilon_{\mathbf{k}  m} )  / \del \varepsilon_{\mathbf{k} m} )$ as a function of  $\varepsilon_{\mathbf{k} m} $ is sharply peaked at $\varepsilon_{\mathbf{k} m} = 0$ with a width $\sim k_B T$, approaching $\delta (\varepsilon_{\mathbf{k}  m}) $ as $T$ goes to zero. We can regard 
[$( \mathcal{N} \mathcal{D}_{eff} (0) )^ {-1} ( - \del n^-_F( \varepsilon_{\mathbf{k}  m} )  / \del \varepsilon_{\mathbf{k} m} ) $], where $  \mathcal{D}_{eff} (0) \equiv \mathcal{N} ^ {-1}  \sum_{\mathbf{k} m}  ( - \del n^-_F( \varepsilon_{\mathbf{k}  m} )  / \del \varepsilon_{\mathbf{k} m} )$,  as a probability distribution.  Denoting averages with respect to it, which are essentially averages over bands within an energy  window $\sim k_B T$ around the Fermi level, as $ \langle ... \rangle_{eff}$, we can write, by approximating averages of products as products of averages, 
\bea
\sigma  &\approx&  e^2  \left( \frac { \mathcal{N} \mathcal{D}_{eff} (0)\langle v^2 \rangle_{eff} }{\mathcal{V}}  \right) \; \left( 1 / \langle \frac{1}{ \tau_{\mathbf{k} m}} \rangle_{eff} \right) \, \nonumber \\ &\equiv& e^2  \left( \frac { \mathcal{N} \mathcal{D}^{(tr)}_{eff} (0)} {\mathcal{V}}   \right)  \tau_{eff} \equiv  \, e^2  \left( \frac{n_{eff}}{m_{eff}} \right) \, \tau_{eff}  \,  \notag \\ & &
\label{sigma_drude_form_eqn-SI}
\eea
We note that $\mathcal{D}_{eff} (0)$, $\mathcal{D}^{(tr)}_{eff} (0)$  and $\langle v^2 \rangle_{eff}$ are identical to the quantities $\mathcal{D}(0)$, $\mathcal{D}^{(tr)}(0)$  and $\langle v^2 \rangle$ we have discussed earlier [cf., Eq. (\ref{dos_tr_dos_eqn})] in the zero temperature limit, and differ from the latter at finite temperatures only by corrections of order ($k_B T / \mathfrak{W}_0)$. In fact, the approximations we have used to calculate the latter makes their correspondence even closer. For the purposes of this paper, we ignore the distinction between the two sets.

The identification of $ ( \mathcal{N} \mathcal{D}^{(tr)}_{eff} (0))/\mathcal{V}$ with $(n_{eff} / m_{eff})$, where $n_{eff}$ is the total electron occupancy {\em in the partially filled bands} of the system (over the entire SBZ) per unit volume, and  $(m_{eff})^{-1}$ is the average of the diagonal components of the inverse effective mass tensor over these occupied states, comes about as follows:
\begin{widetext}
\bea
 \frac { \mathcal{N} \mathcal{D}^{(tr)}_{eff} (0)} {\mathcal{V} }    & \equiv & \frac {1} {\mathcal{V} D} \sum_{ \mathbf{k} m} \mathbf{v}_{\mathbf{k} m} \cdot \mathbf{v}_{\mathbf{k} m}  \left( - \frac{\del n^-_F( \varepsilon_{\mathbf{k}  m} }  { \del \varepsilon_{\mathbf{k} m}} \right) = - \frac {1} {\mathcal{V} D \hbar} \sum_{ \mathbf{k} m} \mathbf{v}_{\mathbf{k} m} \cdot \left(  \frac {\del} {\del \mathbf{k}}  n_F^- (\varepsilon_{\mathbf{k}  m}) \right)  \nonumber \\
  & = & \frac {1} {\mathcal{V} D \hbar} \sum_{ \mathbf{k} m \alpha}  \left( \frac {\del} {\del \mathbf{k}^\alpha} \mathbf{v}^\alpha_{\mathbf{k} m} \right)   n_F^- (\varepsilon_{\mathbf{k}  m}) = \frac {1} {\mathcal{V} D } \sum_{ \mathbf{k} m \alpha}  [ \mathbf{M}^{-1} ( \mathbf{k} m) ] _{\alpha \alpha}  n_F^- (\varepsilon_{\mathbf{k}  m})  \nonumber \\  
  & = & \left( \frac { \sum^{pob}_{ \mathbf{k} m \alpha} n_F^- (\varepsilon_{\mathbf{k}  m})} {\mathcal{V} } \right)   \left( \frac {1}{D} \frac { \sum^{pob}_{ \mathbf{k} m \alpha}  [ \mathbf{M}^{-1} ( \mathbf{k} m) ] _{\alpha \alpha}  n_F^- (\varepsilon_{\mathbf{k}  m}) }{ \sum^{pob}_{ \mathbf{k} m \alpha} n_F^- (\varepsilon_{\mathbf{k}  m} ) } \right)   \nonumber \\
  & \equiv &  \left( \frac{N_{eff}}{\mathcal{V}} \right)  \left( \frac{1}{m_{eff}} \right) \equiv \frac{n_{eff}}{m_{eff}}
  \label{mean_sq_vel_to_neff_by_meff_eqn-SI}
\eea
\end{widetext}
Here $\sum^{pob}$ in the third line indicates that the sum over the bands involves only the  {\em partially occupied bands}, since for completely filled bands, for which $n_F^- (\varepsilon_{\mathbf{k}  m}) = 1$ for all $\mathbf{k}$, the sum over $\mathbf{k}$ of the inverse effective mass tensor $ [ \mathbf{M}^{-1} ( \mathbf{k} m) ] _{\alpha \alpha}$, becomes a sum over the entire SBZ of a $k_{\alpha}$ derivative of a periodic function in the SBZ, and consequently vanishes\cite{Ashcroft_Mermin-76}. [The integrated part in the step involving integration by parts in going from the first to the second line of Eq. (\ref{mean_sq_vel_to_neff_by_meff_eqn-SI}) vanishes for the same reason.] 

Eq. (\ref{sigma_drude_form_eqn-SI}), together with the identification   $\langle { \tau_{\mathbf{k} m}}^{-1} \rangle_{eff}  \equiv (\tau_{eff})^{-1}  \equiv  ( 2 \pi \, k_B T \, \lambda^{(tr)} ) / \hbar $, leads to the expression for the resistivity [cf., Eq. (\ref{rho_eqn})] discussed in section IV. Furthermore. the above definition of $\lambda^{(tr)}$ together with Eq. (\ref{tau_k_m_eqn-SI}) implies that  
\begin{widetext}
\bea
\lambda^{(tr)}  \, &=&   \,  \frac {1} {\mathcal{N}^2  \mathcal{D}_{eff} (0) }  \sum_{\mathbf{k}, m}  \left( - \frac{\del n^-_F( \varepsilon_{\mathbf{k}  m} }  { \del \varepsilon_{\mathbf{k} m}} \right)  \sum_{m^\prime} \sum_{\mathbf{q}, s} |\mathfrak{g}_{m^\prime m; s} (\mathbf{k}, \mathbf{q})|^2  \; n^-_F (\varepsilon_{\mathbf{k} m} ) \; [1- n^-_F (\varepsilon_{(\mathbf{k}+\mathbf{q}) \, m^\prime} ) ]  \nonumber \\ 
& & \ \ \ \ \ \ \ \times   \frac { [ \delta ( \varepsilon_{(\mathbf{k}+\mathbf{q}) \, m^\prime} - \varepsilon_{\mathbf{k} m} - \hbar \omega_{\mathbf{q} s} ) + \delta ( \varepsilon_{(\mathbf{k}+\mathbf{q}) \, m^\prime} - \varepsilon_{\mathbf{k} m} + \hbar \omega_{-\mathbf{q} s} ) ] } { \hbar \omega_{\mathbf{q} s} } \left( 1 - \frac {\mathbf{v}_{\mathbf{k} m} \cdot \mathbf{v}_{(\mathbf{k}+\mathbf{q}) \, m^\prime}} {|\mathbf{v}_{\mathbf{k} m}| |\mathbf{v}_{(\mathbf{k}+\mathbf{q}) \, m^\prime}|} \right)  
\label{lambda_tr_eqn-SI}
\eea
\end{widetext}
For phonon energies and $k_B T$ much smaller than the electronic energy scales, the Fermi functions and the energy conserving delta functions in Eq. (\ref{lambda_tr_eqn-SI}) ensure that both $ \varepsilon_{\mathbf{k} m} $ and  $ \varepsilon_{(\mathbf{k}+\mathbf{q}) \, m}$ get constrained to be close to the Fermi level. Hence we expect $\lambda^{(tr)}$ to be  roughly equal to $\lambda^{(\mathcal{S})}$, as assumed in the discussion presented in section IV. 

In future work, we hope to  carry out the calculations of $\sigma$ as well as $\lambda^{(tr)}$ directly and at finite temperatures {\em without making any of the above approximations}, whence we can also check the validity of the above approximations.

\bibliographystyle{apsrev4-2}
\bibliography{./references}

\begin{thebibliography}{37}
\bibitem [{vel()}]{vel-reduct}%
  \BibitemOpen
  \href@noop {} {}\bibinfo {note} {For the definition of the transport DOS, and for more details, see the discussion in the context of Fig. 6 where we discuss the similar behaviour encountered in the semi-phenomenological TBM for the S-SL, and also Appendices A and B.}\BibitemShut {Stop}%
\bibitem [{Har()}]{Hartree-wo-relax}%
  \BibitemOpen
  \href@noop {} {}\bibinfo {note} {We note that, when the Hartree self-consistent calculations for the Ag-Au superlattice are done with the atoms in the {\em unrelaxed} basis positions $\{\mathbf{d}^0_j \}$, as is the case with most of the results we report in this paper, $\mathbf{d}_{j \mathbf{R}; j^\prime \mathbf{R}^\prime}$ and $d_{j \mathbf{R}; j^\prime \mathbf{R}^\prime}$ in the above equations need to be replaced by $\mathbf{d}^0_{j \mathbf{R}; j^\prime \mathbf{R}^\prime}$ and $d^0_{j \mathbf{R}; j^\prime \mathbf{R}^\prime}$ respectively.}\BibitemShut {Stop}%
\bibitem [{no-()}]{no-R-dep}%
  \BibitemOpen
  \href@noop {} {}\bibinfo {note} {The lack of dependence of the right-hand-side (RHS) of Eq. \ref{kappa_to_dyn_matrix_eqn} on the vector $\mathbf{R}$ is guaranteed by superlattice translational symmetry, and one can set $\mathbf{R} = \mathbf{0}$ in the equation without any loss of generality.}\BibitemShut {Stop}%
\bibitem [{rea()}]{real-units}%
  \BibitemOpen
  \href@noop {} {}\bibinfo {note} {For example, we can choose $K_1^{(s)}$ such that the maximum frequency obtained for the pure-Au S-SL lattice, which is about 2.45 in the dimensionless units we have been using (see Fig.12 in appendix C), matches the experimentally observed maximal frequency for Au. The maximal phonon frequency of Au obtained from neutron scattering is about 4.6 THz \cite{Lynn_et_al-73}, i.e., about 152 cm$^{-1}$ or 18.8 meV or 219 K, in various other units. Hence, for such a choice, the conversion factor $\sqrt{(K_1^{(s)}/M^{(Au)})}$ would be about 89.4 K in temperature units. Matching the maximal frequency obtained from the same force constant model implemented for the pure-Au FCC lattice would give a slightly different conversion factor.}\BibitemShut {Stop}%
\bibitem [{del()}]{delta_n_Au}%
  \BibitemOpen
  \href@noop {} {}\bibinfo {note} {For example, in the S-SL case that we have been discussing, if every one of the 16 Au atoms at the interface in each supercell had an excess electron occupancy of +1, that alone would make a contribution of 0.33 to $\delta n_{Au}$. The precise numbers will depend on the Ag cluster sizes and on $f_{Ag}$, with the contribution being smaller for larger Ag clusters given a fixed $f_{Ag}$.}\BibitemShut {Stop}%
\end{thebibliography}

\end{document}